\documentclass[11pt,a4paper]{article}
 
\usepackage{amsmath,amsthm,amssymb}
\usepackage{graphics,graphicx}
\usepackage{epsfig}
\usepackage{multicol}
\usepackage{color}
\usepackage{mathrsfs}
\makeatletter
\@addtoreset{equation}{section}
\makeatother

\usepackage{centernot}

\usepackage{braket}

\usepackage{mathbbol}
\usepackage{cite}

\usepackage{mathabx}
\usepackage{datetime}

\begin{document}
\begin{center}
\LARGE{\bf How many degrees of freedom describe a quantum N-particle state?}
\end{center}

\begin{center}
{Matthew J. Lake,}${}^{a,b}$\footnote{matthew.lake@ubbcluj.ro}\large{and Marek Miller}${}^{c}$\footnote{mlm@math.ku.dk}
\end{center}
\begin{center}
\emph{${}^a$ Department of Physics, Babe\c s-Bolyai University, Mihail Kog\u alniceanu Street 1, 400084 Cluj-Napoca, Romania} \\
\emph{${}^b$ Physics Department, California State University, Fresno, California 93740, USA} \\
\emph{${}^c$ Quantum for Life Centre, University of Copenhagen, Universitetsparken 5, DK-2100 Copenhagen, Denmark}
\vspace{0.1cm}
\end{center}

\begin{center}
\today
\end{center}

\abstract{
In Newtonian spacetime, the canonical description of a classical $N$-particle system requires $3N$ degrees of freedom. 
Not all of these are physical, however, since the conservation of the net momentum implies that only $3(N-1)$ accelerations are independent. 
Hence, three constraints can be used to eliminate the unphysical centre-of-mass variables, at the level of the Lagrangian, leaving only the subset of observable displacements and momenta, which are relational. 
Imposing the constraints does not change the dynamics of these variables, at the classical level, and is analogous to a gauge-fixing procedure, which removes redundancy in the description of the system. 
In classical physics, therefore, the number of physical degrees of freedom equals the number of independent relational degrees of freedom. 
Here, we show that this is not the case in quantum mechanics. 
While an operator-analogue of the classical net momentum exists, it cannot be used to impose constraints that restrict the degrees of freedom in the theory, without a loss of physical information. 
This means that all $3N$ canonical degrees of freedom are physical, even though only $3(N-1)$ of them are relational. 
We explore the physical consequences of the non-relational variables and show that they give rise to generalised uncertainty relations (GURs), for the relational quantities that define the quantum reference frame (QRF). 
Hence, it is shown that the non-relational degrees of freedom refer to the frame itself and that the non-Heisenberg terms in the GURs define its Galilean-invariant spreads, in both real space and momentum space. 
The implications of this result for recent work on relational models, including the ``perspective neutral'' framework for QRFs, are discussed. 
Its implications for the wider relational program, and, in particular, the relevance of the latter to quantum gravity research, are also critically assessed.
}

\tableofcontents

\section*{Synopsis} \label{Synopsis}

A closed system of $N$ non-relativistic particles in Newtonian spacetime is described canonically by $3N$ configuration variables and a regular Lagrangian. 
Conservation of the total momentum, $\mathbf{P} := \sum_{J=1}^{N}\mathbf{p}_J$, implies that only $3(N-1)$ accelerations are independent: the centre-of-mass coordinates $X^i(t) = v^i t + x^i_0$, with $v^i$ and $x^i_0$ constant, are ignorable and are not Galilean-invariant. 
In the ``perspective-neutral'' formalism, originally proposed in \cite{Vanrietvelde:2018pgb}, it is claimed that subtracting the centre-of-mass energy from the canonical Lagrangian is equivalent imposing the primary first-class constraints $P_i \approx 0$, yielding a singular Lagrangian in which three canonical degrees of freedom must be removed by a ``classical gauge fixing''. 
It is further claimed that these degrees of freedom can be chosen at will, and, hence, that the conditions $x_I^i \approx 0$ can be imposed, for some $I \in \left\{1,2,\dots,N\right\}$, in addition to the constraint ${\bf P} \approx {\bf 0}$. 
We disagree with this, and argue that only the centre-of-mass variables can be ``gauge fixed'' in this way, yielding $X^i(t) \approx x^i_0$ for all $t$. 
Nonetheless, we agree with \cite{Vanrietvelde:2018pgb} regarding the number of physical degrees of freedom in a classical $N$-particle system, which is equal to $3(N-1)$. 
We then ask whether this counting survives quantisation, and give a negative answer, which contradicts the conclusions of the ``relational'' framework \cite{Giacomini:2017zju,Vanrietvelde:2018pgb}. 
In our view, imposing analogous constraints on the total momentum operator, $\widehat{\mathbf{P}} := \sum_{J=1}^{N}\hat{\mathbf{p}}_J$, removes physical information. 
Hence, all $3N$ canonical degrees of freedom of the quantum system are physical, although only $3(N-1)$ of them are relational.

Our critique of the relational QRF models rests on two objections. 
First, these models insist that Galilean symmetries hold in the frame of an accelerated reference particle, which violates canonical Galilean invariance (Secs. \ref{Sec.3} and \ref{Sec.X}). 
Second, they insist that the classically ignorable degrees of freedom remain unphysical after quantisation, which violates wave-particle duality (Secs. \ref{Sec.3.2} and \ref{Sec.4}). 
The remainder of this synopsis develops these two points.

The relational QRF models -- the formalism of Giacomini, Castro-Ruiz and Brukner (GCB) \cite{Giacomini:2017zju}, the perspective-neutral framework \cite{Vanrietvelde:2018pgb}, and the work built on them -- begin from the following premise: the degrees of freedom of the ``reference system'' are unphysical, so that the physical state space of a closed $N$-particle system must contain only $3(N-1)$ (or fewer) relational variables \cite{Vanrietvelde:2018pgb,Vanrietvelde:2018dit}. 
Transformations between the perspectives of different particles are then maps between such reduced descriptions. 
The perspective-neutral framework \cite{Vanrietvelde:2018pgb} presents this premise as a consequence of imposing translation invariance, and, guided by a ``gravity-inspired symmetry principle'', reaches the $3N \rightarrow 3(N-1)$ reduction by three routes (Secs. \ref{Sec.2}-\ref{Sec.3}). 
Let $A$ be one of three interacting particles, $A$, $B$ and $C$, described from an abstract inertial frame $O$. 
Reduced quantisation places $A$ at the origin, sets the total momentum to zero before quantising, and treats what remains as the description `seen' by $A$. 
However, if $A$ is interacting, these two conditions cannot hold simultaneously (Sec. \ref{Sec.3.1}). 
`Canonical' quantisation, {\` a} la \cite{Vanrietvelde:2018pgb}, imposes the same two conditions but claims to handle the constraints systematically via the Dirac bracket procedure. 
Dirac quantisation, {\` a} la \cite{Vanrietvelde:2018pgb}, quantises all nine canonical degrees of freedom then admits only states with zero total momentum, $\widehat{\mathbf{P}}|\Psi\rangle = {\bf 0}$, as ``physical''. 
This requires a redefinition of the canonical inner product. 
On this basis, the wider relational program claims to extend diffeomorphisms to the quantum regime \cite{delaHamette:2022cka,Kabel:2024lzr}, to give a quantum extension of the equivalence principle \cite{Giacomini:2020ahk,Giacomini:2021aof,Cepollaro:2021ccc}, to unify canonical, reduced and Dirac quantisation within a single framework \cite{Hohn:2018iwn,Hohn:2018toe,delaHamette:2021oex,Hoehn:2019fsy}, and to define spin operationally in the particle's rest frame \cite{Giacomini:2018gxh,Mikusch:2021kro}. 
Each of these claims rests on some form of relationally-reduced description, and is therefore open to both our objections.

The first objection concerns the frame of an accelerated reference particle. 
Rewriting the Hamiltonian in terms of the centre of mass and of positions relative to $A$ does not take us into $A$'s frame; the description remains that of $O$. 
Because $B$ and $C$ exert forces on $A$, the frame that follows $A$ is non-inertial, and in it every particle experiences a pseudo-force proportional to its mass. 
No frame therefore satisfies both $\mathbf{x}_A \approx \mathbf{0}$ and $\mathbf{P} \approx \mathbf{0}$ when $A$ is accelerated. 
The relational models take no account of these pseudo-forces: they read $O$'s description, rewritten in relational variables, as the one `seen' by $A$, and hold $A$'s accelerated frame to be physically equivalent to an inertial frame. 
The reference particle becomes an ``unmoved mover'', which violates the law of action and reaction. 
We reject this: Galilean symmetry relates only frames in uniform relative motion, and, in Newtonian spacetime, acceleration is absolute. Discarding the centre-of-mass degrees of freedom and discarding those of an accelerated particle are then physically distinct operations, unless the particle is isolated.
For an interacting particle the distinction disappears in the limit $m_A \to \infty$, provided the forces on $A$ do not grow with $m_A$: its acceleration then vanishes and its frame reduces to an inertial one (Sec. \ref{Sec.4}). The Newtonian treatment of gravity does not meet this proviso, since gravitational and inertial mass are equal, so that the gravitational `acceleration' of $A$ does not depend on $m_A$.

The second objection concerns the centre of mass, whose position $\mathbf{X}$ is ignorable, with conserved conjugate momentum $\mathbf{P}$. 
Classically, the centre of mass moves uniformly relative to $O$ with some velocity $\mathbf{v}$; although no classical physics depends on $\mathbf{v}$, the conditions to be quantised are $\mathbf{P} = M\mathbf{v}$, not $\mathbf{P} = \mathbf{0}$. 
Their quantum counterparts, $\widehat{\mathbf{P}} = M\hat{\mathbf{v}}$ with $\hat{\mathbf{v}}$ a constant of motion, impose no restriction on the states. 
Imposing, instead, $\widehat{\mathbf{P}}|\Psi\rangle = 0$ admits only states that cannot be canonically normalised, since the spectrum of $\widehat{\mathbf{P}}$ is continuous. 
The perspective-neutral framework renormalises these states with a modified inner product, and regards the result, a centre of mass spread over all space with sharply zero momentum, as an expression of gauge invariance \cite{Vanrietvelde:2018pgb}. 
In our view this discards the spread of the centre of mass without physical justification. 
Under an active Galilean boost the Hamiltonian and the time derivative in the Schr\"odinger equation acquire the same extra terms, so the equation is invariant only if the Hamiltonian retains the kinetic energy of the centre of mass, $\widehat{P}^2/2M$ (Sec. \ref{Sec.X}). 
The ``physical Schr\"odinger equation'' of the perspective-neutral framework, with the internal energy alone, is therefore not boost-invariant: the centre-of-mass degrees of freedom cannot be removed without violating Galilean invariance. 
Recast as a gauge theory, as in \cite{Vanrietvelde:2018pgb}, Galilean-invariant mechanics remains equivalent to the standard classical theory, by construction, but its quantisation -- whether reduced, `canonical' or Dirac -- does not recover canonical quantum theory.

Section \ref{Sec.4} argues that the discarded part of the canonical state is physical. 
The positions and momenta of the individual particles, and of $\widehat{\mathbf{X}}$ and $\widehat{\mathbf{P}}$, are not Galilean-invariant, but their variances are, and are therefore measurable; we call such quantities \emph{detectables}. 
Consider three particles in a state that factorises into single-particle states in $O$'s frame, and, along the $i^{\rm th}$ Cartesian axis, the position of $B$ relative to $A$, $\hat{x}^i_{B|A} = \hat{x}^i_B - \hat{x}^i_A$; the momentum of $B$ `seen' by $A$, along the same axis, is then $\hat{p}_{B:Ai} = \hat{p}_{Bi} - r\,\hat{p}_{Ai}$, with $r = m_B/m_A$. 
With $\sigma_A^i$ and $\tilde{\sigma}_{Ai}$ the position and momentum spreads of $A$, the standard Heisenberg relations for both $A$ and $B$ yield the GUR
\begin{equation*}
(\Delta x^i_{B|A})^2(\Delta p_{B:Aj})^2 \geq \Big(\frac{\hbar}{2}\Big)^2(\delta^{i}{}_{j})^2 + r^2(\tilde{\sigma}_{Aj})^2(\Delta x^i_{B|A})^2 + (\sigma^i_A)^2(\Delta p_{B:Aj})^2 - r^2(\sigma^i_A)^2(\tilde{\sigma}_{Aj})^2 \, , 
\end{equation*}
for the Galilean-invariant relational variables. 
Its non-Heisenberg terms are the frame's own spreads, which sum to at least $(\hbar/2)^2(2r + r^2)$. 
For separable states with $N \geq 3$ these spreads can be computed directly from Galilean-invariant data; for $N = 2$, or for non-separable states, they cannot. 
Since the canonical basis is unitarily equivalent to the centre-of-mass and relational basis, the three-particle state requires nine physical degrees of freedom, not six; for $N$ particles, $3N$. 
QRFs imply GURs.

In Newtonian mechanics and in general relativity alike, inertial effects mark the departure from geodesic motion in the spacetime structure, and the distinction between inertial and accelerated frames is physical. 
The gravitational claims of the program listed above, together with ``quantum general covariance'' \cite{Hoehn-Presentation:CERN-2019,Hoehn-Presentation:HK-2020,Hoehn-Presentation-fxqi-2021}, rest on QRF transformations that excise the degrees of freedom of the reference system, or make no distinction between inertial and accelerated frames. 
Each must answer, in its Newtonian limit, whether it treats the frame's acceleration as gauge, and whether it retains the frame's own spreads. 
The second question is quantitative: a reference system of finite mass registers the rest of the system through a GUR whose floor is fixed by the mass ratio, and coincides with an ideal frame only in the limit of infinite mass.

Our statements on observability concern closed systems. 
A classical measurement apparatus breaks Galilean invariance, and positions measured relative to it are observables of the open system that excludes it. 
Including the apparatus in the quantum description restores the invariance, but makes the apparatus a quantum reference frame, and raises the question that both objections address: how are the degrees of freedom of such a frame are to be treated? 
Our criticisms concern the answer given by the present QRF models, not every possible relational theory.

\section{Introduction} \label{Sec.1}

Work on QRFs dates back to the 1960's when the term ``quantum reference frame'' was first introduced \cite{Aharonov:1967zza,Aharonov:1967zz}. 
From the beginning, this work had a strongly relational flavour, as it concerned the properties of frames of reference that are {\bf embodied} as material systems -- and, in particular, their {\bf relations} with other material subsystems. 

The original QRF frameworks, however, by no means abandoned the idea of an {\bf external} or `objective' view of the system(s) in question \cite{Aharonov:1984zz,Rovelli:1990pi,quant-ph/0602069v2,Bartlett:2006tzx,arXiv:0711.0043v2,Angelo:2011,Palmer:2013zza,Vedral-1:1603.04583,Vedral-2:1803.03523v1,Busch:2016rui,Loveridge:2016tnh,Loveridge:2017pcv,Carette:2023wpz}. 
In short, while it was admitted that no physical observer could inhabit such a frame, it was also recognised that a closed system could be described, mathematically, using an {\bf abstract coordinate system}. 
Such a coordinate system is defined as a particular parameterisation of the spacetime manifold, whose coordinate origin need not be identified with the trajectory of any physical body \cite{Frankel:1997ec,Nakahara:2003nw}. 
The key question for these early models was, therefore, how to ``jump'' from the objective, God's eye view of the abstract coordinates, to the partial, subjective view of an embodied physical observer?  

The phrasing of this question presupposes that an objective mathematical description can first be obtained, which causes obvious epistemological difficulties. 
If all physical observers are non-ideal, how can we begin from the perfect viewpoint of an abstract coordinate system? 
To answer this requires us to grapple with metaphysical questions, which `respectable' physicists often prefer to avoid, but which are essential for clarifying, and justifying, the foundational assumptions of our field \cite{Maudlin:2007}. 
We do not propose to answer it fully, here, but provide instead a tentative answer, which we believe to have been tacitly and implicitly assumed in the earliest work on QRFs \cite{Aharonov:1967zza,Aharonov:1967zz,Aharonov:1984zz}. 

Our ideas follow roughly those put forward by Schr{\" o}dinger in his popular but insightful essay ``What is Life?'' \cite{Schrodinger:1944}. 
Therein, he argued that a living detector -- such as a human eye, for example -- must be sufficiently macroscopic for its internal state to be uninfluenced by the quantum fluctuations of its environment. 
It is clear that similar logic can be employed to define an {\bf ideal frame of reference}, whether this be embodied by a biological or a non-biological system. 
It is then a short step to an abstract coordinate system: since the internal degrees of freedom of the ideal detector may be considered negligible, its perspective is {\bf effectively objective}. 
 
Moreover, we may assume that all existing observations -- which rely, ultimately, on human sense-impressions, no matter how delicate or complex the intermediate technological apparatus \cite{Russell:1931} -- have been made from the perspective of such a frame.
\footnote{We do not rely on human consciousness to ``collapse the wave function'' \cite{Tomaz:2025eid}. Our point is simply that, whatever technical apparatus is used to perform the measurement, it must be coupled to an ancilla system that provides a `read out', which the human eye, ear, or sense of touch, is capable of perceiving \cite{Russell:1931}. At least this part of our measuring equipment, therefore, must be as big and heavy as our sense organs \cite{Schrodinger:1944}.} 
This implies that the laws of science, as currently formulated, are `objective' to within the limits of experimental precision. 

With regard to the quantum regime, this position suggests that we have effectively `lucked onto' the correct laws of physics, by virtue of our innate construction as macroscopic objects, vis-{\` a}-vis the quantum realm. 
Yet this also suggests that luck had nothing to do with it -- our macroscopic nature being a necessary precondition of our very existence, as persistent, non-randomised entities \cite{Schrodinger:1944}. 

This is, therefore, the view that we take. 
It allows us to assume that the non-relativistic Schr{\" o}dinger equation, and the Hilbert space states which satisfy it, yield an objective description of the quantum Universe, as defined formally within a Newtonian spacetime manifold \cite{SEP:Newtonian_Space-time}. 
In other words, it allows us to take canonical quantum theory -- within its regime of validity \cite{Bargmann:1954gh,Levy-Leblond:1963qdx,Levy-Leblond:1967eic,Horzela:1991pa,Giulini:1995te,Csillag:2003} -- as an objective description of the physical world, from the perspective of an abstract, ideal frame \cite{Frankel:1997ec,Nakahara:2003nw}. 

We can then return to our previous question and ask what the Universe would `look like' from the perspective of a microscopic system, whose internal quantum mechanical degrees of freedom are non-negligible, and which therefore influence its `perception' of the external world. 
The relevant ``jump'' is from a classical reference frame (CRF), from which a closed quantum system can be viewed objectively, and in its entirety, to the QRF associated with one of its constituent subsystems. 
This view is {\bf relational}, in the sense that the relative displacements and velocities between particles are regarded as the only physically meaningful position- and velocity-type variables, respectively. 

Recently, however, a more extreme form of relationalism has become popular in the QRF literature -- which, on account of this view, has undergone something of a resurgence 
\cite{Giacomini:2017zju,Castro-Ruiz:2019nnl,delaHamette:2020dyi,Giacomini:2021gei,Castro-Ruiz:2021vnq,Vanrietvelde:2018pgb,Hohn:2018iwn,Hohn:2018toe,delaHamette:2021oex,Krumm:2020fws,Hoehn:2019fsy,AliAhmad:2021adn,Vanrietvelde:2018dit,Hoehn:2023ehz,Hoehn:2023axh,DeVuyst:2025ezt}. 
In these new models, the very possibility of describing a closed system from an external viewpoint is denied and spacetime is reduced, {\` a} la Aristotle \cite{Aristotle,Aristotle:SEP}, Descartes \cite{Descartes,Descartes:SEP}, Leibniz \cite{Leibniz:Arthur,Leibniz:SEP}, and, ultimately, Mach \cite{Mach's_Principle:Book,Barbour:Mach's_Principle}, to a system of relations between material bodies \cite{Absolute_vs_Relational_Space:Classical_SEP,Absolute_vs_Relational_Space:Relativistic_SEP}. 
The only permissible frames are those associated with material particles \cite{Giacomini:2017zju,Vanrietvelde:2018pgb}, or with the centres of mass of groups of particles \cite{Krumm:2020fws}, and the Newtonian spacetime manifold is effectively dispensed with. 
All dynamical variables, including accelerations, are defined only {\bf relative} to such embodied frames of reference. 
The degrees of freedom associated with the so-called ``reference system'' are regarded as unobservable, and therefore unphysical, and are also dispensed with \cite{Giacomini:2017zju,delaHamette:2020dyi,Hoehn:2019fsy,Vanrietvelde:2018pgb,Krumm:2020fws}. 

We see two problems with the insistence that acceleration is relational  and its acceptance, largely unquestioned, by a large section of the ``quantum foundations'' research community. 
\footnote{Of the papers that, to date, have referenced the original work by Giacomini, Castro-Ruiz and Brukner (GCB) \cite{Giacomini:2017zju}, and the work in which the perspective-neutral framework was first introduced \cite{Vanrietvelde:2018pgb}, a great many are follow-up studies, that make use of the same basic formalisms to extend and extrapolate these models. (Source: INSPIRE website, circa $1^{\rm st}$ July 2026.) We are not aware of any other papers, with the exception \cite{Lake:2023nua,Lake:2026zrk} and the present work, that challenge the fundamental assumptions on which these models are based.} 
The {\bf first problem} is the necessity of explaining the physical distinction between inertial and accelerated motion. 
In canonical theories -- including the formalisms of Newtonian mechanics \cite{Kibble&Berkshire:2004,McCall:2011,Arnold:1978,Marsden&Ratiu:1999}, canonical quantum mechanics \cite{Rae:2002,Isham:1995,Dirac:1958,vonNeumann:1955}, special relativity \cite{French:1966}, canonical quantum field theory (QFT) \cite{Tong-QFT:2007,Padmanabhan-QFT:2016,Parker&Toms:2009,Wald-QFT:2004}\, and general relativity (GR) \cite{MTW:1973,Dirac-GR:1975,Wald-GR:1984,Lasenby:2006} -- the inertial effects that distinguish accelerated motion are explained by the difference between geodesic and non-geodesic motion in the background spacetime \cite{Frankel:1997ec,Nakahara:2003nw}. 
If the spacetime manifold no longer exists, how are we to account for the observed existence of {\bf pseudo-forces}, including the pseudo-acceleration induced by a gravitational field? 
  
The only plausible models of which we are aware, capable of implementing inertial effects in a purely relational way, involve a {\bf best-matching} procedure like that proposed by Barbour and Bertotti \cite{Barbour-Bertotti:1982,Mercati:2018}. 
These models are certainly interesting and it may, for all we know, be possible to embed the relational QRF models within the Barbour-Bertotti framework. 
However, even if this turns out to be the case, it is not relevant to our criticism. 
To the best of our knowledge, the proponents of the relational QRF framework -- as constructed in, for example, the large body of work \cite{Giacomini:2017zju,Castro-Ruiz:2019nnl,delaHamette:2020dyi,Giacomini:2021gei,Castro-Ruiz:2021vnq,Vanrietvelde:2018pgb,Hohn:2018iwn,Hohn:2018toe,delaHamette:2021oex,Krumm:2020fws,Hoehn:2019fsy,AliAhmad:2021adn,Vanrietvelde:2018dit,Hoehn:2023ehz,Hoehn:2023axh,DeVuyst:2025ezt} -- do not claim that it represents a `relational alternative' to canonical quantum theory. 
The claim is, instead, that the formalism is simply a part of canonical quantum mechanics -- which is {\bf inherently relational}, by its very nature \cite{Giacomini:2017zju,delaHamette:2020dyi,Hoehn:2019fsy,Vanrietvelde:2018pgb,Krumm:2020fws}.

This is, in fact, the main claim to fame of the relational formalisms. 
If true, it implies that previously unknown ``generalised symmetries'' have recently been discovered \cite{Giacomini:2017zju} -- that is, symmetries which existed at the heart of quantum theory, since it was first formulated in 1926, but which escaped the notice of Heisenberg, Schr{\" o}dinger, Weyl, Dirac, and others, and which also evaded the attention of later generations of physicists, until close to the quantum centenary \cite{Quantum-Centenary}. 
Big claims are made for these ``new'' symmetries -- including, but not limited to, that they allow us to generalise the concept of a classical diffeomorphism to the quantum regime, providing profound insights into the nature of quantum gravity \cite{delaHamette:2022cka,Kabel:2024lzr}, that they provide a quantum extension of the classical equivalence principle \cite{Giacomini:2020ahk,Giacomini:2021aof,Cepollaro:2021ccc}, that they explain the well-known contradictions between canonical quantisation, reduced quantisation, and Dirac quantisation for constrained Hamiltonian systems, incorporating all three into a universal framework \cite{Vanrietvelde:2018pgb,Hohn:2018iwn,Hohn:2018toe,delaHamette:2021oex}, and that they solve the well-known problem of how to define spin, operationally, within the rest frame of the particle, in canonical QFT \cite{Giacomini:2018gxh,Mikusch:2021kro}.

In our view, these claims make it essential to critically evaluate the founding assumptions on which these models are based. 
Given their radical nature, it is necessary to either verify, or disprove, the so-far unquestioned assumption that they are internally self-consistent. 
This leads us to the {\bf second problem}. 
In \cite{Vanrietvelde:2018pgb}, it is claimed that the relational framework emerges as a natural and unavoidable consequence of imposing Galilean invariance on a quantum system.
\footnote{Strictly, the ``toy model'' proposed in \cite{Vanrietvelde:2018pgb} considers only translational symmetry, but it is clear that a realistic completion of the model must encompass the whole Galilean group. Further steps towards this are taken in \cite{Vanrietvelde:2018dit}, in which it is claimed that rotational degrees of freedom can also be ``gauged'', in addition to translational degrees of freedom.} 
To us, this is a strange claim, since it is tantamount to the assertion that Galilean invariance implies Mach's Principle (i.e., the assertion that all physics is relational) \cite{Mach's_Principle:Book,Barbour:Mach's_Principle}. 

It is even stranger when one considers the fact that the Galilean group is the symmetry group of the Newtonian spacetime \cite{SEP:Newtonian_Space-time}, which has supposedly been done away with in the relational models -- its symmetries having been ``generalised'' to a larger group containing ``superpositions of Galilean transformations'' \cite{Giacomini:2017zju,Ballesteros:2020lgl,Ballesteros:2025ypr}.
According to the results of the Erlangen Program in classical geometry, any spacetime that possesses global symmetries is completely characterised by its symmetry group \cite{ErlangenProgram_Klein_1872,ErlangenProgram_EMS_2015,Kisil:2010,ErlangenProgram_Encycolpedia_Springer,Goenner:2015}. 
In other words, there is no such thing as a ``Euclidean space'' whose symmetries are described by a group other than the Euclidean group, and an analogous statement holds, also, for Galilean-invariant space. 
There is no geometry, other than the canonical Newtonian spacetime, which possesses global Galilean symmetries, and only global Galilean symmetries \cite{SEP:Newtonian_Space-time}. 
Within this spacetime, acceleration is absolute. 

How, then, can the relational models be, not only consistent with Galilean symmetries, but actually derived from them, as claimed in the current literature \cite{Vanrietvelde:2018pgb,Vanrietvelde:2018dit}? 
The short answer to this question is that they cannot. 
Contrary to these claims, we will show that the relational models \cite{Giacomini:2017zju,Vanrietvelde:2018pgb,Vanrietvelde:2018dit} actually violate the requirements of both Galilean symmetry and canonical quantum mechanics. 

The reasons for these violations are, however, quite subtle. 
The process of uncovering them yields insights into the nature of quantum mechanics, its physical degrees of freedom, and their classical limit. 
These insights lead us to define a `new' conceptual category of quantum mechanical variables, which are neither observable nor unobservable in the traditional sense \cite{Observable:QM-Compendium}. 
The term `new' is again placed in inverted commas because this category of variables was, in fact, always present in the canonical theory, since its inception in 1926. 
It is extremely surprising that it has not been identified, or characterised, until now.
\footnote{One may also ask how Heisenberg, Schr{\" o}dinger, Dirac and others could have `missed' its existence, but the pioneers of quantum theory were not interested in such topics. They did not ask, or attempt to answer in any systematic way, the question ``what does a quantum system look like, from the perspective of another quantum system?''. With the rise of the Copenhagen interpretation \cite{Heisenberg:1958}, ``observers'' were placed firmly on the other side of the ``Heisenberg cut''. (The founders did, however, care rather a lot about Galilean invariance, and the symmetries of the theory \cite{Weyl:1928}.) Although we are harsh on the relational QRF literature, we would be the first to admit the debt owed to all authors -- including those of a relational bent -- whose work has helped to raise awareness of this important and foundational problem.}   
We refer to these variables, whose existence cannot be verified by single measurements, but whose presence can be inferred from the statistics of repeated measurements on a quantum system, as ``detectables''. 
We show, in particular, that they generate non-Heisenberg terms in the uncertainty relations which characterise the statistics of the chosen QRF. 
Hence, far from being purely negative, we believe that these results -- including the rejection of the relational models -- form a basis for future progress in this field. 

Many of the results we utilise, to show where, how, and why the currently-in-vogue relational models go astray, are pedagogical in nature, and are available -- if one cares to look -- in well-known textbooks and review articles. 
Yet they are not so well known, and so well understood, as to make their understanding elementary and their acceptance unquestioned.
If they were, there would not now exist an alternative vision of how to create a Galilean-invariant quantum theory, in the form of the relational QRF literature \cite{Giacomini:2017zju,Ballesteros:2020lgl,Ballesteros:2025ypr,Vanrietvelde:2018pgb,Hohn:2018iwn,Hohn:2018toe,delaHamette:2021oex,Hoehn:2019fsy,AliAhmad:2021adn,Vanrietvelde:2018dit,Hoehn:2023ehz,Hoehn:2023axh,DeVuyst:2025ezt}.

On the one hand, it is important to include such gory detail, given the sheer volume of the literature that we wish to challenge. 
Were we not to include it, we would leave our conclusions open to question, to an unnecessary degree. 
On the other hand, there seems to be little value in repeating well-worn material from books and review papers in the main body of the text. 
For this reason, we have supplemented the main text with extensive appendices, which contain a somewhat unusual mix of near-pedagogical content, interspersed with far more advanced discussion and elaboration. 
In our view, both are absolutely necessary because -- as we aim to show -- it is the misunderstanding of such `basic' physics that has led to much of the confusion in the relational literature. 
And, therefore, to many erroneous claims.

Roughly speaking, the appendices represent the gauge-invariant description of the paper, and include all information which, though technically redundant, is nonetheless helpful for facilitating understanding. 
The main text represents the gauge-fixed description, in which all strictly non-necessary details have been omitted.

The structure of this paper is then as follows. 
In Sec. \ref{Sec.2}, we review the basic content of the perspective-neutral formalism \cite{Vanrietvelde:2018pgb} and related, relational, QRF theories \cite{Giacomini:2017zju}. 
This is kept deliberately brief, focussing only on the main elements of these models, and the reader is referred to the original works \cite{Giacomini:2017zju,Vanrietvelde:2018pgb} for full details. 
Our critique follows in Sec. \ref{Sec.3}. 
Because the studies in \cite{Giacomini:2017zju,Vanrietvelde:2018pgb} consider only three-particle systems, we do likewise in the main body of the text. 
The additional material required, in order to extend our results to a general, closed, and Galilean-invariant $N$-body system, is given in the appendices. 
For even greater simplicity, both these works treat only one-dimensional systems, but we do not go that far. 
In general, the results we critique represent three-dimensional generalisations of the results presented in \cite{Giacomini:2017zju,Vanrietvelde:2018pgb}, unless otherwise stated. 
Sections \ref{Sec.2} and \ref{Sec.3} are parallel in structure, dealing, in sequence, with the reduced quantisation, `canonical' quantisation, and ``Dirac quantisation'' procedures, considered in \cite{Vanrietvelde:2018pgb}. 
Although the authors of \cite{Vanrietvelde:2018pgb} use only the terms ``reduced quantisation'' and ``Dirac quantisation'', to refer to their methods, we show that their reduced quantisation scheme is really a misapplication of canonical quantisation, which is why the term `canonical', here, is in inverted commas. 

In Sec. \ref{Sec.X}, we consider how Galilean invariance is implemented in both classical Newtonian mechanics and canonical quantum theory. 
Specifically, we show how the implementation of Galilean invariance in the classical regime {\bf does not require} the imposition of the constraints $P_i \approx 0$. 
Hence, there is no reason to believe that setting $\widehat{P}_i \approx 0$ is required, in order to impose Galilean invariance in the quantum regime, although this forms the basic premise of the perspective-neutral framework \cite{Vanrietvelde:2018pgb,Vanrietvelde:2018dit}. 
Moreover, imposing these conditions requires us to impose a {\bf superselection rule} (SSR), for the net momentum of a closed quantum system, but no such rule exists in canonical quantum mechanics \cite{SSR:QM-Compendium,Giulini:2007fn}. 
Its imposition is therefore not justified, by the reasons given in \cite{Vanrietvelde:2018pgb,Vanrietvelde:2018dit}, and it must be regarded as an additional {\it ad hoc} assumption, designed to facilitate the development of a ``relational'' quantum theory that is compatible with Galilean symmetry \cite{Bargmann:1954gh,Levy-Leblond:1963qdx,Levy-Leblond:1967eic,Horzela:1991pa,Giulini:1995te,Csillag:2003}. 
Unfortunately, it fails in this task, since the resulting models \cite{Giacomini:2017zju,Vanrietvelde:2018pgb,Vanrietvelde:2018dit} actually violate Galilean invariance. 

In the final section of the main text, Sec. \ref{Sec.4}, we apply canonical, Galilean-invariant, quantum mechanics to a closed three-particle system and use this to show that QRFs imply GURs. 
This is the simplest, but perhaps most surprising result of our work. 
While we agree, with \cite{Vanrietvelde:2018pgb,Vanrietvelde:2018dit}, that the canonical Hilbert space contains both Galilean-invariant and non-Galilean-invariant subspaces, we disagree that the latter constitute an unphysical sector. 
The reasons are subtle, but, in short, we show how even non-Galilean-invariant variables give rise to Galilean-invariant variances, in both position and momentum space. 
These variances are responsible for the non-Heisenberg terms in the GUR, and can be measured indirectly by compiling statistics, about measurements of Galilean-invariant quantities. 
Thus, discarding the non-Galilean-invariant sector of the canonical Hilbert space discards {\bf physical information}. 
In the QRF picture, the information discarded is about the chosen ``reference system''.

Appendix \ref{App.A} covers the canonical treatment of an $N$-particle system in classical Newtonian mechanics, whereas Appendix \ref{App.B} considers its quantum counterpart. 
The general theory of constrained Hamiltonian systems, and the most popular strategies for their quantisation – canonical, reduced, and Dirac – are reviewed, extremely briefly, in Appendix C. 
The ultimate reasons for the failure of the ``gauge-fixing'' procedure, proposed in \cite{Vanrietvelde:2018pgb}, are also discussed at greater length in App. \ref{App.C.1}. 
The issues involved are again subtle and concern the distinctions -- both mathematical and physical -- between gauge symmetries and spacetime symmetries. 
However, the reader familiar with these topics may simply skip the Appendices, which account for almost a half of the total length of the manuscript, excluding the Bibliography.

Finally, before proceeding, a note on stylistics. 
We use double inverted commas for direct quotations and to refer to words and phrases with well-established meanings, in the current literature. 
Following semi-standardised conventions \cite{Teller:1994}, single inverted commas are used whenever a term is introduced without a strict definition, or when the meaning we ascribe to it may not
be universally accepted. 
In the latter case, the intended meaning of a particular term should be clear from the context.

\section{Review of the perspective-neutral and relational QRF formalisms} \label{Sec.2}

In the {\bf reduced quantisation} of a constrained Hamiltonian system the unphysical degrees of freedom in the canonical description are removed, before the classical observables are promoted to operators. 
This can be done in a rather {\it ad hoc} way, without any systematic treatment of the classical constraints, or by using the elegant treatment of constrained Hamiltonian systems developed by Dirac \cite{Dirac-CHS:1958,Dirac-CHS:1964,Henneaux-Teitelboim:1992,Rothe-RotheCHS:2010}. 
Confusingly, the latter is referred to as ``canonical quantisation'' \cite{Ita:2021cak,BarroseSa:2023rvz,Juhasz:2024twu} by some authors but as ``reduced quantisation'' by others \cite{Kunstatter:1991ds,Schleich:1990gd,Plyushchay:1994pk,Barvinsky:1996cg,Shimizu:1996vf}. 
In this work, we follow the former convention, since this allows us to distinguish clearly between different forms of ``reduction''. 

Hence, according to our meaning, {\bf canonical quantisation} is when the classical constraints are dealt with systematically, via the Dirac bracket, which is mapped to the quantum commutator in place of the classical Poisson bracket \cite{Dirac-CHS:1958,Dirac-CHS:1964,Henneaux-Teitelboim:1992,Rothe-RotheCHS:2010}. 
Note that this includes the case in which {\bf no classical constraints} are applied and no degrees of freedom are removed. 
In this case, the Dirac brackets equal the Poisson brackets, by construction.
Conversely, ``reduced quantisation'' is any form of reduction, prior to quantisation, in which the Dirac bracket is not utilised. 

In {\bf Dirac quantisation}, the constraints themselves are promoted to operators, before being solved in the quantum regime. 
Our usage of this term is consistent with that of all other authors. 
However, in Sec. \ref{Sec.3.3}, we challenge the idea that the {\it specific interpretation} of ``Dirac quantisation'', given in the relational QRF literature, is a valid implementation of this procedure.
   
These differing quantisation schemes often disagree with one another, for specific example systems, leading to inequivalent results \cite{Kunstatter:1991ds,Schleich:1990gd,Plyushchay:1994pk,Barvinsky:1996cg,Shimizu:1996vf,Ita:2021cak,BarroseSa:2023rvz,Juhasz:2024twu}. 
Recently a new formalism was proposed, which purports to be a universal framework, capable of encompassing all three methods and of explaining any differences between them, where such differences arise. 
This is known as the {\bf perspective-neutral formalism} \cite{Vanrietvelde:2018pgb}. 
It claims, in a nutshell, that the choice of QRF corresponds to a choice of ``gauge'', and that, using techniques developed in previous works \cite{Giacomini:2017zju}, all three quantisation schemes lead to the same ``relational'' model, for closed quantum systems \cite{Vanrietvelde:2018pgb,Hohn:2018iwn,Hohn:2018toe,delaHamette:2021oex}.

In this section, we briefly review how the reduced, `canonical', and Dirac quantisation schemes are realised within this framework, but our terminology differs from that used in \cite{Vanrietvelde:2018pgb,Hohn:2018iwn,Hohn:2018toe,delaHamette:2021oex}, which make no clear distinction between the first two schemes. 
Our reasons for drawing this distinction will become clear in Sec. \ref{Sec.3}; therein, we show that the ``reduced quantisation'', employed in the perspective-neutral model, is really a misapplication of canonical quantisation, in which the Dirac brackets are wrongly constructed. 

\subsection{Reduced quantisation in the perspective-neutral approach} \label{Sec.2.1}

The starting point for the analysis in \cite{Vanrietvelde:2018pgb}, in which the perspective-neutral formalism is introduced, is the canonical Hamiltonian for a system of three classical particles, 
\begin{eqnarray} \label{2.1}
H({\bf x},{\bf p}) &=& \frac{p_A^2}{2m_A} + \frac{p_B^2}{2m_B} + \frac{p_C^2}{2m_C} 
\nonumber\\
&+& V_{AB}(|{\bf x}_B-{\bf x}_A|) + V_{AC}(|{\bf x}_C-{\bf x}_A|) + V_{BC}(|{\bf x}_C-{\bf x}_B|) \, .
\end{eqnarray}
The authors then consider a frame in which the net momentum vanishes, 
\begin{eqnarray} \label{2.2}
{\bf P} := {\bf p}_A + {\bf p}_B+ {\bf p}_C = {\bf 0} \, , 
\end{eqnarray}
which can be used to rewrite (\ref{2.1}) as
\begin{eqnarray} \label{2.3}
H({\bf x},{\bf p}) &=& \frac{p_B^2}{2m_B} + \frac{p_C^2}{2m_C} + \frac{({\bf p}_B+{\bf p}_C)^2}{2m_A}  
\nonumber\\
&+& V_{AB}(|{\bf x}_B-{\bf x}_A|) + V_{AC}(|{\bf x}_C-{\bf x}_A|) + V_{BC}(|{\bf x}_C-{\bf x}_B|) \, . 
\end{eqnarray}
Next, they consider a frame whose coordinate origin traces out the trajectory of particle $A$, so that 
\begin{eqnarray} \label{2.4}
{\bf x}_A(t) = {\bf 0} \, , \, \, {\rm for \, \, all} \, \, t \, , 
\end{eqnarray}
and, hence, 
\begin{eqnarray} \label{2.5}
H({\bf x},{\bf p}) = \frac{p_B^2}{2m_B} + \frac{p_C^2}{2m_C} + \frac{({\bf p}_B+{\bf p}_C)^2}{2m_A}  + V_{AB}(|{\bf x}_B|) + V_{AC}(|{\bf x}_C|) + V_{BC}(|{\bf x}_C-{\bf x}_B|) \, .
\end{eqnarray}
In this frame, however, it is also necessary that
\begin{eqnarray} \label{2.6}
{\bf p}_A(t) = m_A\dot{{\bf x}}_A(t) = {\bf 0} \, , \, \, {\rm for \, \, all} \, \, t \, , 
\end{eqnarray}
yielding
\begin{eqnarray} \label{2.7}
{\bf p}_B + {\bf p}_C = {\bf 0} \, .
\end{eqnarray}
\footnote{In the original PN paper, it is claimed that there exists an alternative set of phase space coordinates in which $\dot{{\bf x}}_A = {\bf 0}$ and $\ddot{{\bf x}}_A = {\bf 0}$, for all $t$, by virtue of the condition ${\bf x}_A = {\bf 0}$ (\ref{2.4}), but in which $\dot{{\bf p}}_A = -\partial V/\partial {\bf x}_A \neq {\bf 0}$, in general. In other words, it is claimed that there exists a phase space coordinate system in which the coordinate acceleration of particle $A$ is zero, but in which its proper acceleration is correctly accounted for. In effect, this `decouples' $\ddot{{\bf x}}_A$ from $\dot{{\bf p}}_A$, and, hence, $\dot{{\bf x}}_A$ from ${\bf p}_A$. According to \cite{Vanrietvelde:2018pgb}, the transformation from the canonical phase space coordinates, used to construct the Hamiltonian (\ref{2.1}), to the new coordinate system in which ${\bf p}_A \neq m_A\dot{{\bf x}}_A$, is implemented by means of a Lagrange multiplier, plus what the authors refer to as a ``classical gauge fixing''. They start by constructing the total Hamiltonian, $H_{\rm tot} := H + \boldsymbol{\lambda}.{\bf P}$, in order to impose the constraint ${\bf P} \approx {\bf 0}$, and claim that this follows from the singular Lagrangian $L^{*} := L - E_{\rm CoM}$, where $L$ is the canonical Lagrangian, corresponding to $H$, and $E_{\rm CoM}$ is the centre-of-mass energy in $O$'s external frame. (See Eq. (57) in Appendix A of \cite{Vanrietvelde:2018pgb}.) We find this very strange, since $L^{*}$ does not contain Lagrange multipliers. They then claim that, because $L^{*}$ possesses a gauge symmetry, being invariant under the simultaneous transformations ${\bf x}_J \mapsto {\bf x}_J + {\bf f}(t)$, $\dot{{\bf x}}_J \mapsto \dot{{\bf x}}_J + \dot{{\bf f}}(t)$ for all $J \in \left\{A,B,C\right\}$, where ${\bf f}(t)$ is an arbitrary function of time, three degrees of freedom must be ``gauge fixed''. (Actually, \cite{Vanrietvelde:2018pgb} works in one spatial dimension, so considers ``gauge fixing'' only one degree of freedom, but the generalisation of their argument to three dimensions is straightforward.) They further claim that these degrees of freedom can be chosen at will, and, therefore, choose to ``gauge fix'' the configuration space variables of the chosen reference particle, $A$, setting ${\bf x}_A(t) = {\bf 0}$ for all $t$. Allegedly, this yields $\boldsymbol{\lambda} \approx -{\bf p}_A/m_A =: \boldsymbol{\lambda}_A$ as the  ``solution'' of the constraint equations, in $A$'s ``relational'' reference frame, where ${\bf p}_A \approx -{\bf p}_B-{\bf p}_C \neq {\bf 0}$ but $\dot{{\bf x}}_A = {\bf 0}$. However, it is not difficult to see that the zero net momentum constraint, at the level of the Hamiltonian, must be implemented as the zero net velocity constraint, at the level of the Lagrangian. The necessary total Lagrangian is then $L_{\rm tot} := L - \boldsymbol{\lambda}.M\dot{{\bf X}}$, where $M$ is the total mass of the system and ${\bf X}$ is the centre-of-mass coordinate, in $O$'s frame. This Lagrangian is regular and does not possess a gauge symmetry, so no gauge fixing is required. Furthermore, the Euler-Lagrange equations for $L_{\rm tot}$ yield the unique solution $\boldsymbol{\lambda} \approx \dot{{\bf X}} = {\bf P}/M \approx {\bf 0}$ (when ${\bf P} \approx {\bf 0}$). Finally, the total Lagrangian for $L^{*}$ is $L_{\rm tot}^{*} := L^{*} - \boldsymbol{\lambda}.M\dot{{\bf X}}$, yielding $H_{\rm tot}^{*} := H^{*} + \boldsymbol{\lambda}.{\bf P}$ in place of $H_{\rm tot} := H + \boldsymbol{\lambda}.{\bf P}$. We do not see why one would gauge the three centre-of-mass degrees of freedom, via the map $L \mapsto L^{*}$, then reintroduce them via a Lagrange multiplier term, $-\boldsymbol{\lambda}.M\dot{{\bf X}}$, as \cite{Vanrietvelde:2018pgb} appears to do, or why this term should be applied to the ungauged Lagrangian, when, allegedly, the constraint it imposes is needed to remove unwanted gauge freedom. In any case, the zero net momentum constraint is trivial, in the sense that it is perfectly compatible with the dynamics of the unconstrained system. Its only effect is to partially fix the abstract inertial frame $O$, forcing it comove with the centre of mass. The remaining ``residual gauge freedom'' (really, spacetime-frame freedom), which can then be fixed, corresponds to choosing the constant displacement between $O$'s coordinate origin the centre-of-mass coordinate, ${\bf X}$. In an upcoming work \cite{Lake-Miller-upcoming}, we will present the analysis summarised here in more detail. Suffice to say, we reject the claim that there exists a phase space coordinate system in which $\dot{{\bf x}}_A = {\bf 0}$, but ${\bf p}_A \neq m_A\dot{{\bf x}}_A$, and in which the zero net momentum condition also holds. It follows that setting ${\bf x}_A = {\bf 0}$ (\ref{2.4}), and, hence, ${\bf p}_A = m_A\dot{{\bf x}}_A = {\bf 0}$ (\ref{2.6}), yields ${\bf p}_B + {\bf p}_C = {\bf 0}$ (\ref{2.7}), as claimed above.}

The Hamiltonian (\ref{2.5}) then becomes
\begin{eqnarray} \label{2.8}
H({\bf x},{\bf p}) = \frac{p_B^2}{2m_B} + \frac{p_C^2}{2m_C} + V_{AB}(|{\bf x}_B|) + V_{AC}(|{\bf x}_C|) + V_{BC}(|{\bf x}_C-{\bf x}_B|) \, ,
\end{eqnarray}
by straightforward substitution of (\ref{2.7}), but, in this case, the two kinetic terms may also be combined into one, i.e., $p_B^2/(2\mu_{BC}) = p_C^2/(2\mu_{BC})$, where $\mu_{BC}$ is the reduced mass of the particle pair $BC$.

In \cite{Vanrietvelde:2018pgb}, it is claimed that the `derivation' of (\ref{2.5}), given above, is equivalent to imposing the classical constraints $P_i \approx 0$, using the procedure developed by Dirac \cite{Dirac-CHS:1958,Dirac-CHS:1964,Henneaux-Teitelboim:1992,Rothe-RotheCHS:2010}. 
We would agree with this, were it not for the necessity of also imposing (\ref{2.4}), in order to ``jump'' into $A$'s frame.
\footnote{The one-dimensional equivalents of the conditions (\ref{2.2}) and (\ref{2.4}) are given in Eqs. (2) and (6) of \cite{Vanrietvelde:2018pgb}, respectively.} 
As we will show, in Sec. \ref{Sec.3.1}, this condition is not compatible with the requirement that ${\bf P} \approx {\bf 0}$, if either $V_{AB} \neq 0$ or $V_{AC} \neq 0$. 

Hence, \cite{Vanrietvelde:2018pgb} does not apply the logic of its assumptions in a self-consistent way. 
The Hamiltonian (\ref{2.5}), with ${\bf p}_B \neq -{\bf p}_C$, is taken as its starting point, although this is not compatible with imposing {\it both} (\ref{2.2}) and (\ref{2.4}). 
But this does not much matter, as these two assumptions are also incompatible with each other; in the formalism presented, no restrictions are placed on the forms of the inter-particle potentials -- beyond the usual requirement that the total potential be translation-invariant -- and it is generally assumed that the reference particle, $A$, interacts with particles $B$ and/or $C$, i.e., that it does not form an isolated subsystem of the three-particle state \cite{Vanrietvelde:2018pgb,Vanrietvelde:2018dit}.

Quantising (\ref{2.5}), by promoting all classical quantities to operators, gives 
\begin{eqnarray} \label{2.9}
\widehat{H}({\bf x},{\bf p}) = \frac{\widehat{p}_B^2}{2m_B} + \frac{\widehat{p}_C^2}{2m_C} + \frac{(\widehat{{\bf p}}_B+\widehat{{\bf p}}_C)^2}{2m_A}  + \widehat{V}_{AB}(|{\bf x}_B|) + \widehat{V}_{AC}(|{\bf x}_C|) + \widehat{V}_{BC}(|{\bf x}_C-{\bf x}_B|) \, ,
\end{eqnarray}
whereas quantising (\ref{2.8}), in the same way, we obtain 
\begin{eqnarray} \label{2.10}
\widehat{H}({\bf x},{\bf p}) = \frac{\widehat{p}_B^2}{2m_B} + \frac{\widehat{p}_C^2}{2m_C} + \widehat{V}_{AB}(|{\bf x}_B|) + \widehat{V}_{AC}(|{\bf x}_C|) + \widehat{V}_{BC}(|{\bf x}_C-{\bf x}_B|) \, ,
\end{eqnarray}
with $\widehat{p}_B^2/(2m_B) + \widehat{p}_C^2/(2m_C) = \widehat{p}_B^2/(2\mu_{BC}) = \widehat{p}_C^2/(2\mu_{BC})$. 
Historically, (\ref{2.10}) was first posited, as the ``relational Hamiltonian'' of the {\bf closed three-particle system}, in \cite{Giacomini:2017zju}. 
Later, when the assumptions (\ref{2.2}) and (\ref{2.4}) were first written down explicitly, in \cite{Vanrietvelde:2018pgb}, it became clear how to `derive' this expression from the canonical Hamiltonian (\ref{2.1}). 

However, this `derivation' does not take account of the additional inference, that $\widehat{{\bf p}}_B = -\widehat{{\bf p}}_C$, which follows from combining (\ref{2.2}) with (\ref{2.6}) -- and, therefore, from combining (\ref{2.2}) with (\ref{2.4}).
Hence, in order to obtain the Hamiltonian for the ``relational'' three-particle state, used in the perspective-neutral formalism \cite{Vanrietvelde:2018pgb}, we must still assume that $\widehat{{\bf p}}_B \neq -\widehat{{\bf p}}_C$, despite the fact that both (\ref{2.2}) and (\ref{2.4}) are supposed to hold identically, in this model. 
In order to obtain the ``relational'' Hamiltonian used in \cite{Giacomini:2017zju}, we must set $\widehat{{\bf p}}_B = -\widehat{{\bf p}}_C$ in the third term of the perspective-neutral Hamiltonian, but then use $\widehat{{\bf p}}_B \neq -\widehat{{\bf p}}_C$ in the first two terms. 

When referring to (\ref{2.9}), it is stated that ``this reduction simply discards particle $A$’s position and momentum from among the physical degrees of freedom and we pick the remaining ones as coordinates of the reduced phase space. 
We thus end up with a theory for $N-1$ particles -- as seen by $A$'' \cite{Vanrietvelde:2018pgb}, but the same statement also holds for the Hamiltonian (\ref{2.10}). 
In this sense, the minor differences between the different relational models are mostly cosmetic. 
The key feature, which they {\bf all} share, is that the degrees of freedom corresponding to the chosen ``reference particle'' are simply excised from the theory.

From here on, we refer to the original model presented by Giacomini, Castro-Ruiz and Brukner \cite{Giacomini:2017zju} as either the ``GCB model'' or the ``relational model''. 
The term ``relational model'', however, also applies to all follow-up works, in which the extended formalisms recover the original GCB theory, in some limit. 
This includes all models developed in \cite{delaHamette:2020dyi,Krumm:2020fws,Vanrietvelde:2018pgb,Ballesteros:2020lgl,Ballesteros:2025ypr,Garmier:2025soc}, in which this is explicitly stated. 

Clearly, these `derivations' of the quantum Hamiltonians (\ref{2.9})-(\ref{2.10}) represent forms of {\bf reduced quantisation}, since the ``unphysical'' degrees of freedom in the canonical, classical, Hamiltonian are first removed, before the remaining phase space variables are promoted to operators. 
We stress that, according to the claims made in \cite{Giacomini:2017zju,Vanrietvelde:2018pgb,Hohn:2018iwn,Hohn:2018toe,delaHamette:2021oex}, these expressions describe an {\bf arbitrary} three-particle system, from the perspective of the QRF associated with particle $A$. 
That is, no restrictions are placed on the forms of the inter-particle potentials, $\widehat{V}_{AB}$, $\widehat{V}_{AC}$ and $\widehat{V}_{BC}$, beyond the requirement that the total potential obeys translational symmetry.

\subsection{`Canonical' quantisation in the perspective-neutral approach} \label{Sec.2.2}

We now review the arguments, also made in \cite{Vanrietvelde:2018pgb}, that the ``reduced quantisation'' of the closed three-particle system is equivalent to Dirac's method of {\bf canonical quantisation} \cite{Dirac-CHS:1958,Dirac-CHS:1964,Henneaux-Teitelboim:1992,Rothe-RotheCHS:2010}. 
The authors do not use this terminology, but do make use of the {\bf Dirac bracket} \cite{Giacomini:2017zju,Vanrietvelde:2018pgb,Hohn:2018iwn,Hohn:2018toe,delaHamette:2021oex}, which is the standard technique of canonical quantisation \cite{Ita:2021cak,BarroseSa:2023rvz,Juhasz:2024twu}. 
They begin by noting that a closed classical system, subject to Galilean symmetries, can be recast as a {\bf constrained Hamiltonian system}, since its canonical description contains unphysical degrees of freedom, associated with the inertial motion of its centre of mass \cite{Kibble&Berkshire:2004,McCall:2011,Arnold:1978,Marsden&Ratiu:1999}. 
They refer to the unphysical phase space coordinates as ``gauge'' degrees of freedom, and to relational, Galilean-invariant quantities as ``gauge-invariant Dirac observables'' \cite{Vanrietvelde:2018pgb}. 

Specifically, they claim that ``jumping'' to the frame of particle $A$ implies the existence of {\bf second-class constraints},
\footnote{The notation used in \cite{Vanrietvelde:2018pgb} is ${\bf x}_A \overset{!}{=} {\bf 0}$, not ${\bf x}_A \approx {\bf 0}$. However, on our reading of the text, both ${\bf x}_A \overset{!}{=} {\bf 0}$ and ${\bf P} \approx {\bf 0}$ define constraint surfaces in the canonical phase space. (See, for example, Fig. 1 in \cite{Vanrietvelde:2018pgb}.) The primary difference between ${\bf x}_A \overset{!}{=} {\bf 0}$ and ${\bf P} \approx {\bf 0}$, therefore, is that the relevant constraints are not implemented in the same way. Specifically, whereas ${\bf P} \approx {\bf 0}$ is imposed by means of a Lagrange multiplier -- which, if implemented correctly, is a valid and well-defined mathematical procedure -- the notation ${\bf x}_A \overset{!}{=} {\bf 0}$, used in the perspective-neutral formalism, indicates an ad hoc and ill-defined mathematical step, in the which the ${\bf x}_A$ coordinate is set equal to zero `by hand'. (See the upcoming work \cite{Lake-Miller-upcoming} for further details.) In this work, we use more standard notation, in which the weak equality denotes the restriction to a constraint surface, regardless of how, exactly, the constraint is imposed \cite{Henneaux-Teitelboim:1992}.} 
\begin{eqnarray} \label{2.11}
{\bf x}_A \approx {\bf 0} \, , \quad {\bf P} \approx {\bf 0} \, , 
\end{eqnarray}
and that these, in turn, yield the {\bf Dirac bracket} 
\begin{eqnarray} \label{2.12}
\left\{F,G\right\}_{\rm DB} = \left\{F,G\right\}_{\rm PB} - \left\{F,P_i\right\}_{\rm PB}\left\{x_A^i,G \right\}_{\rm PB} + \left\{F,x_A^i\right\}_{\rm PB}\left\{P_i,G\right\}_{\rm PB} \, , 
\end{eqnarray}
for any two functions of the canonical phase space coordinates, $F({\bf x},{\bf p})$ and $G({\bf x},{\bf p})$ \cite{Vanrietvelde:2018pgb}.
\footnote{Equation (\ref{2.11}) is the three-dimensional equivalent of Eq. (6) in \cite{Vanrietvelde:2018pgb}. Below this, it is explicitly stated that ``this implies that the constraints
become second class''.} 
(The reader unfamiliar with constrained Hamiltonian systems, or desirous of a quick refresher, is referred to Appendix \ref{App.C}.) 
Setting $F =x_A^i$ and $G = P_j$ in (\ref{2.12}) then yields 
\begin{eqnarray} \label{2.13}
\left\{x_A^i,P_j\right\}_{\rm DB} = \left\{x_A^i,P_j\right\}_{\rm PB} - \left\{x_A^i,P_k\right\}_{\rm PB}\left\{x_A^k,P_j\right\}_{\rm PB}  +  \left\{x_A^i,x_A^k\right\}_{\rm PB}\left\{P_k,P_j\right\}_{\rm PB} 
= 0 \, ,
\end{eqnarray}
which, upon quantisation, is mapped to 
\begin{eqnarray} \label{2.14}
\frac{1}{i\hbar}[\widehat{x}_A^i,\widehat{P}_j] = 0 \, . 
\end{eqnarray}
On the basis of this result, it is also claimed that setting
\begin{eqnarray} \label{2.15}
\widehat{{\bf x}}_A \approx {\bf 0} \, , \quad \widehat{{\bf P}} \approx {\bf 0} \, , 
\end{eqnarray}
in canonical quantum theory, again yields (\ref{2.9}) \cite{Vanrietvelde:2018pgb}. 
The precise mechanism by which the canonical quantum Hamiltonian,
\begin{eqnarray} \label{2.16}
\widehat{H}({\bf x},{\bf p}) &=& \frac{\widehat{p}_A^2}{2m_A} + \frac{\widehat{p}_B^2}{2m_B} + \frac{\widehat{p}_C^2}{2m_C} 
\nonumber\\
&+& \widehat{V}_{AB}(|{\bf x}_B-{\bf x}_A|) + \widehat{V}_{AC}(|{\bf x}_C-{\bf x}_A|) + \widehat{V}_{BC}(|{\bf x}_C-{\bf x}_B|) \, ,
\end{eqnarray}
is mapped to (\ref{2.9}), is outlined in Sec. \ref{Sec.3.3}. 
Ultimately, however, the Dirac bracket (\ref{2.13}) stems from the individual relations
\begin{eqnarray} \label{2.17}
\left\{x_A^i,p_{Aj}\right\}_{\rm DB} = 0 \, , \quad \left\{x_B^i,p_{Bj}\right\}_{\rm DB} = \delta^{i}{}_{j} \, , \quad \left\{x_C^i,p_{Cj}\right\}_{\rm DB} = \delta^{i}{}_{j} \, ,
\end{eqnarray}
where
\begin{eqnarray} \label{2.18}
{\bf x}_A \approx {\bf 0} \, , \quad {\bf p}_A \approx {\bf 0} \, , 
\end{eqnarray}
which yield
\begin{eqnarray} \label{2.19}
[\widehat{x}_A^i,\widehat{p}_{Aj}] = 0 \, , \quad [\widehat{x}_B^i,\widehat{p}_{Bj}] = i\hbar\delta^{i}{}_{j}\widehat{\mathbb{I}} \, , \quad [\widehat{x}_C^i,\widehat{p}_{Cj}] = i\hbar\delta^{i}{}_{j}\widehat{\mathbb{I}} \, ,
\end{eqnarray}
and
\begin{eqnarray} \label{2.20}
\widehat{{\bf x}}_A \approx {\bf 0} \, , \quad \widehat{{\bf p}}_A \approx {\bf 0} \, ,
\end{eqnarray} 
upon quantisation.
\footnote{Strictly, the canonical operator $\widehat{{\bf x}}_A$ is ill-defined in the perspective-neutral formalism, at least in the ``relational'' QRF of particle $A$, since the canonical degrees of freedom associated with the chosen ``observer'' are simply excised from the theory \cite{Vanrietvelde:2018pgb,Giacomini:2017zju}. However, on our reading, this means that it can be set equal to zero, without loss of generality. This is the concrete meaning of the expression $\widehat{{\bf x}}_A \approx {\bf 0}$, in (\ref{2.20}).}
Simply substituting these conditions into (\ref{2.16}) gives the Hamiltonian (\ref{2.10}), originally proposed in \cite{Giacomini:2017zju}, but the method proposed in \cite{Vanrietvelde:2018pgb} is said to yield (\ref{2.9}) instead.

It is important to note that, according to both the perspective-neutral \cite{Vanrietvelde:2018pgb} and relational \cite{Giacomini:2017zju} QRF formalisms, the operators $\widehat{{\bf p}}_B$, $\widehat{{\bf p}}_C$ and $\widehat{{\bf x}}_B$, 
$\widehat{{\bf x}}_C$, appearing in the relevant Hamiltonians, represent {\bf complete sets of observables}. 
This claim is based on two supporting arguments. 
The first is that the {\bf relational} displacements and momenta,
\begin{eqnarray} \label{2.21}
\widehat{{\bf x}}_{B|A} = \widehat{{\bf x}}_B - \widehat{{\bf x}}_A \, , \quad \widehat{{\bf x}}_{C|A} = \widehat{{\bf x}}_C - \widehat{{\bf x}}_A \, , 
\nonumber\\ 
\widehat{{\bf p}}_{B:A} = \widehat{{\bf p}}_B - \frac{m_B}{m_A}\widehat{{\bf p}}_A \, , \quad \widehat{{\bf p}}_{C:A} = \widehat{{\bf p}}_C - \frac{m_C}{m_A}\widehat{{\bf p}}_A \, , 
\end{eqnarray}
are observables, because they are Galilean (or ``gauge'') invariant \cite{Vanrietvelde:2018pgb,Hohn:2018iwn,Hohn:2018toe,delaHamette:2021oex}. 
Here, we use the symbol $\widehat{{\bf p}}_{J:I} := {\bf p}_J - (m_J/m_I){\bf p}_I$ to denote the momentum of particle $J$ `seen' by particle $I$, in order to distinguish this from the canonical ``relational momentum'' 
$\widehat{{\bf p}}_{J|I} := \mu_{IJ}({\bf p}_J/m_J - {\bf p}_I/m_I)$. 
(See Appendix \ref{App.A} for details.) 
The second is that we may identify 
\begin{eqnarray} \label{2.22}
\widehat{{\bf x}}_{B|A} = \widehat{{\bf x}}_B \, , \quad \widehat{{\bf x}}_{C|A} = \widehat{{\bf x}}_C \, , 
\nonumber\\ 
\widehat{{\bf p}}_{B:A} = \widehat{{\bf p}}_B \, , \quad \widehat{{\bf p}}_{C:A} = \widehat{{\bf p}}_C \, , 
\end{eqnarray}
in particle $A$'s QRF. 
In the language of \cite{Vanrietvelde:2018pgb}, the variables (\ref{2.21}) are ``gauge-invariant'', whereas those in (\ref{2.22}) are ``gauge-fixed''. 
This means that, in the QRF of particle $A$, all {\bf physical degrees of freedom} can be defined {\bf relative} to $A$, so that choosing a QRF is equivalent to fixing a gauge \cite{Vanrietvelde:2018pgb,Hohn:2018iwn,Hohn:2018toe,delaHamette:2021oex}. 
Hence, 
\begin{eqnarray} \label{2.23}
\widehat{{\bf x}}_A|\Psi\rangle \approx {\bf 0} \, , \quad \widehat{{\bf p}}_A|\Psi\rangle \approx {\bf 0} \, ,
\end{eqnarray}
for all {\bf physical states} $|\Psi\rangle$, in $A$'s frame, according to these assumptions. 
This is why these operators do not appear in either the relational (\ref{2.10}) or the perspective-neutral (\ref{2.9}) Hamiltonians. 
Using $\widehat{{\bf P}} \approx {\bf 0}$, also, gives
\begin{eqnarray} \label{2.24}
\widehat{{\bf P}}|\Psi\rangle \approx {\bf 0} \, , \quad (\widehat{{\bf p}}_B + \widehat{{\bf p}}_C)|\Psi\rangle \approx {\bf 0} \, ,
\end{eqnarray}
but this conclusion appears not to have been drawn in \cite{Vanrietvelde:2018pgb}.

In most models which are ``relational by construction'' \cite{Giacomini:2017zju}, the conditions (\ref{2.23}) are satisfied by expanding an arbitrary state of the {\bf three-particle system} as
\begin{eqnarray} \label{2.25}
|\Psi\rangle = \int \Psi({\bf x}_B,{\bf x}_C) |{\bf x}_B\rangle_B |{\bf x}_C\rangle_C {\rm d}^3x_B{\rm d}^3x_C  
= \int \tilde{\Psi}({\bf p}_B,{\bf p}_C) |{\bf p}_B\rangle_B |{\bf p}_C\rangle_C {\rm d}^3p_B{\rm d}^3p_C \, , 
\end{eqnarray}
in the QRF of particle $A$, where $\Psi({\bf x}_B,{\bf x}_C)$ and $\tilde{\Psi}({\bf p}_B,{\bf p}_C)$ are normalised over a six-dimensional volume,
\begin{eqnarray} \label{2.26}
\langle \Psi | \Psi \rangle = \int |\Psi({\bf x}_B,{\bf x}_C)|^2 {\rm d}^3x_B{\rm d}^3x_C = \int |\tilde{\Psi}({\bf p}_B,{\bf p}_C)|^2 {\rm d}^3p_B{\rm d}^3p_C = 1 \, .
\end{eqnarray}
However, the same number of {\bf independent} quantum mechanical degrees of freedom are required, in order to expand this state in the basis associated with any other frame \cite{Giacomini:2017zju}. 
Generalising these results to $N$ particles, we see that the number of physical degrees of freedom is $3(N-1)$, exactly as in the classical case \cite{Kibble&Berkshire:2004,McCall:2011,Arnold:1978,Marsden&Ratiu:1999}.

An exception is the model proposed in a follow-up work on ``switching perspectives'' in the perspective-neutral formalism \cite{Vanrietvelde:2018dit}. 
In this paper it is claimed that, because the reduction from $3N$ degrees of freedom, to $3(N-1)$, is due to the ``gauge modes'' associated with translational invariance, the number of physical degrees of freedom can be reduced still further, to $3(N-2)$, by imposing rotational invariance. 

In \cite{Vanrietvelde:2018pgb,Hohn:2018iwn,Hohn:2018toe,delaHamette:2021oex}, the method summarised in this subsection is also referred to as a ``reduced quantisation'' scheme. 
As already stated, we disagree with this classification: when Dirac brackets are introduced, in order to systematically deal with the classical constraints, the quantisation scheme is -- or at least should be -- {\bf canonical} \cite{Ita:2021cak,BarroseSa:2023rvz,Juhasz:2024twu}. 
In Sec. \ref{Sec.3}, we show that the canonical quantisation of a closed, classical, Galilean-invariant system -- when properly applied -- is {\bf not} equivalent to the process of reduced quantisation, as claimed in the perspective-neutral approach \cite{Vanrietvelde:2018pgb}. 

\subsection{Dirac quantisation in the perspective-neutral approach} \label{Sec.2.3}

Finally, it is also claimed that (\ref{2.9}) can be derived by a process of {\bf Dirac quantisation}, using the methods developed in \cite{Vanrietvelde:2018pgb}. 
The specification of the Dirac quantisation formalism represents the bulk of this text and we review it here only briefly. 
The interested reader is referred to the original works \cite{Vanrietvelde:2018pgb,Hohn:2018iwn,Hohn:2018toe,delaHamette:2021oex}, for further details. 

The basic idea behind this type of quantisation is that the classical constraints should be ``promoted'' to quantum operators then ``solved'' directly in the quantum regime 
\cite{Kunstatter:1991ds,Schleich:1990gd,Plyushchay:1994pk,Barvinsky:1996cg,Shimizu:1996vf,Ita:2021cak,BarroseSa:2023rvz,Juhasz:2024twu}. 
As it is interpreted in \cite{Vanrietvelde:2018pgb}, it represents a previously unknown type of quantisation, which has not appeared before in the literature. 
Therein, it is claimed that, instead of imposing $P_i \approx 0$ at the classical level then performing a {\bf reduced quantisation} to give the ``relational three-particle state'' (\ref{2.25}), the conditions $\widehat{P}_i \approx 0$ should be imposed by acting on the ``kinematical three-particle state'', which is described by nine quantum mechanical degrees of freedom, with a series of well-defined operators. 
This is, according to \cite{Vanrietvelde:2018pgb}, the correct way to construct a Galilean-invariant quantum theory, containing only ``gauge-invariant Dirac observables''. 

The problem, of course, is that it is by no means clear how the ``kinematical states'' should be defined, or what imposing the conditions $\widehat{P}_i \approx 0$ on them really {\it means}, concretely. 
In the classical theory of constrained Hamiltonian systems, imposing the constraints $P_i \approx 0$ means that the components of the net momentum are set equal to zero only after all the relevant Poisson brackets have been defined \cite{Dirac-CHS:1958,Dirac-CHS:1964,Henneaux-Teitelboim:1992,Rothe-RotheCHS:2010}. 
The {\bf literal} quantum analogue of this would be to interpret $\widehat{P}_i \approx 0$ as follows: 
First, we define $\widehat{P}_i$ in the usual (canonical) way, as $\widehat{P}_i := \widehat{p}_{Ai} + \widehat{p}_{Bi} + \widehat{p}_{Ci}$, where each individual operator, $\widehat{p}_{Ai}$, $\widehat{p}_{Bi}$ and $\widehat{p}_{Ci}$ has a non-trivial eigenspectrum \cite{Rae:2002,Isham:1995,Dirac:1958,vonNeumann:1955}. 
Hence, $\widehat{P}_i \neq 0$. 
Second, we construct all the relevant commutators involving $\widehat{P}_i$. 
Finally, we set $\widehat{P}_i = 0$ at the level of the Hamiltonian and the quantum equations of motion. 

But there is an obvious problem with this approach, namely, that -- unlike in the classical case, where there is no contradiction between a variable possessing a fixed value, in the solutions to the equation of motion, yet still acting as a dynamical quantity in the Poisson bracket-formulation of the equations of motion \cite{Dirac-CHS:1958,Dirac-CHS:1964,Henneaux-Teitelboim:1992,Rothe-RotheCHS:2010} -- this is not the case for quantum operators. 
Does it make sense to set $\widehat{P}_i \neq 0$, in the commutator $[\widehat{O},\widehat{H}] = [\widehat{O},(\widehat{P}^2/(2M) + \widehat{U})]$, but to then set $\widehat{P}_i = 0$ in the Heisenberg and Schr{\" o}dinger equations? 

The issue is that, unlike classical variables, quantum operators can be time-independent, without possessing fixed constant values (such as zero). 
That is, they can be {\bf stationary} while still possessing standard spectral representations \cite{Rae:2002,Isham:1995,Dirac:1958,vonNeumann:1955}. 
This is why the concrete implementation of the innocent-looking expression, $\widehat{P}_i \approx 0$, is by no means obvious -- even in the case of reduced quantisation (see Secs. \ref{Sec.2.1}-\ref{Sec.2.2}) -- and, in fact, why it is by no means clear whether it actually has any {\bf physical meaning}. 

The perspective-neutral formalism answers these doubts resolutely, boldly claiming that (a) the expression $\widehat{P}_i \approx 0$ {\bf does} have physical meaning, and (b) that it should be interpreted as the restriction which determines the ``gauge-fixed'' {\bf physical states} of Galilean-invariant quantum mechanics \cite{Bargmann:1954gh,Levy-Leblond:1963qdx,Levy-Leblond:1967eic,Horzela:1991pa,Giulini:1995te,Csillag:2003}. 
It is therefore claimed that imposing $\widehat{P}_i \approx 0$ is {\bf equivalent} to imposing the condition
\begin{eqnarray} \label{2.27}
\widehat{P}_i |\Psi\rangle^{\rm phys} :=
\left(\widehat{p}_{Ai} + \widehat{p}_{Bi} + \widehat{p}_{Ci}\right) |\Psi\rangle^{\rm phys} \overset{!}{=} 0 \, .
\end{eqnarray}
The notation ``$\overset{!}{=}$'', used here, is not ours, but is taken from \cite{Vanrietvelde:2018pgb}. 
It is intended to indicate that, {\it if} the state $|\Psi\rangle$ satisfies the relevant equations -- indicated by the presence of the exclamation mark -- then it belongs to the {\bf physical Hilbert space} of ``gauge-invariant'' (that is, Galilean-invariant) states, whereas, if it does not satisfy these equations, it is an {\bf unphysical state} in the {\bf kinematical Hilbert space} of the theory \cite{Vanrietvelde:2018pgb}. 
(It then acquires the relevant superscript, either ``${\rm phys}$'' or ``${\rm kin}$'', accordingly.)
The kinematical Hilbert space is defined as
\begin{eqnarray} \label{2.28}
\overline{\mathcal{H}}_{ABC}^{\rm kin} := \overline{\mathcal{H}}_A \otimes \overline{\mathcal{H}}_B \otimes \overline{\mathcal{H}}_C \, , 
\end{eqnarray}
where the overline indicates the rigged basis completion of the relevant canonical Hilbert space \cite{Isham:1995,vonNeumann:1955}. 
That is, the space involving unphysical distributions, as well as physical, normalisable states \cite{Rae:2002,Dirac:1958}.

Concretely, the kinematical states are defined as
\begin{eqnarray} \label{2.29}
|\Psi\rangle^{\rm kin} :=
\int \Psi^{\rm kin}({\bf p}_A,{\bf p}_B,{\bf p}_C)|{\bf p}_A\rangle_A|{\bf p}_B\rangle_B|{\bf p}_C\rangle_C {\rm d}^3p_A{\rm d}^3p_B{\rm d}^3p_C \, , 
\end{eqnarray}
where $\Psi^{\rm kin}({\bf p}_A,{\bf p}_B,{\bf p}_C)$ is {\bf not normalisable}, with respect to the integration measure ${\rm d}^3p_A{\rm d}^3p_B$ ${\rm d}^3p_C$. 
This should be compared with the physical state of the three-particle system in canonical quantum theory, in which the wave function {\it is} normalised with respect this integration measure \cite{Rae:2002,Isham:1995,Dirac:1958,vonNeumann:1955}. 
The ``physical states'' $|\Psi\rangle^{\rm phys}$ are then selected by applying the operator $\delta(\widehat{{\bf P}})$ to the kinematical states $|\Psi\rangle^{\rm kin}$,
\begin{eqnarray} \label{2.30}
\delta(\widehat{{\bf P}}): |\Psi\rangle^{\rm kin} \mapsto |\Psi\rangle^{\rm phys} \, .
\end{eqnarray}
The reader is referred to \cite{Vanrietvelde:2018pgb} for details. 

For our purposes, what is important is that the $|\Psi\rangle^{\rm phys}$, thus obtained, can be transformed via a {\bf passive unitary transformation}, 
\begin{eqnarray} \label{2.31}
|\Psi\rangle^{\rm phys} \mapsto \widehat{\mathcal{T}}_{A,BC}|\Psi\rangle^{\rm phys} =: |\Psi\rangle_{A,BC} \, , \quad
{}^{\rm phys}\langle\Psi | \mapsto {}^{\rm phys}\langle\Psi |\widehat{\mathcal{T}}_{A,BC}^{\dagger} =: {}_{A,BC}\langle\Psi | \, , 
\end{eqnarray}
\begin{eqnarray} \label{2.32}
\widehat{O}^{\rm phys} \mapsto \widehat{\mathcal{T}}_{A,BC}\widehat{O}^{\rm phys}\widehat{\mathcal{T}}_{A,BC}^{\dagger}
=: \widehat{O}_{A,BC} \, , 
\end{eqnarray}
where
\begin{eqnarray} \label{2.33}
\widehat{\mathcal{T}}_{A,BC} := \exp\left[\frac{i}{\hbar}(\widehat{{\bf p}}_B + \widehat{{\bf p}}_C).\widehat{{\bf x}}_A \right] \, , 
\end{eqnarray}
to give
\begin{eqnarray} \label{2.34}
|\Psi\rangle_{A,BC} := \widehat{\mathcal{T}}_{A,BC}|\Psi\rangle^{\rm phys} = |p=0\rangle_A \otimes |\Psi\rangle_{BC|A} \, ,
\end{eqnarray}
with
\begin{eqnarray} \label{2.35}
|\Psi\rangle_{BC|A} := \int \Psi_{BC|A}({\bf p}_B,{\bf p}_C)|{\bf p}_B\rangle_B|{\bf p}_C\rangle_C {\rm d}^3p_B{\rm d}^3p_C \, , 
\end{eqnarray}
and
\begin{eqnarray} \label{2.36}
\Psi_{BC|A}({\bf p}_B,{\bf p}_C) := \Psi^{\rm kin}(-{\bf p}_B-{\bf p}_C,{\bf p}_B,{\bf p}_C) \, , 
\end{eqnarray}
using the notation of \cite{Vanrietvelde:2018pgb}. 
The state $|\Psi\rangle_{BC|A} $ is supposed to represent the ``view of the system from the perspective of particle $A$'' and is obtained from $|\Psi\rangle_{A,BC}$ (\ref{2.34}) by acting with the ``quantum reduction map''
\begin{eqnarray} \label{2.37}
\widehat{\Pi}_A := (\sqrt{2\pi\hbar})^3\langle {\bf q}_A = {\bf 0}|_A \otimes \widehat{\mathbb{I}}_B \otimes \widehat{\mathbb{I}}_C 
= \int {\rm d}^3p' \langle {\bf p}'|_A \otimes \widehat{\mathbb{I}}_B \otimes \widehat{\mathbb{I}}_C \, .
\end{eqnarray}
The last operation is referred to as ``constraint trivialisation'' in \cite{Vanrietvelde:2018pgb}.

This procedure yields a reduced state space, $\mathcal{H}_{BC|A} $, which describes only {\bf six quantum mechanical degrees of freedom}. 
This should be contrasted with the corresponding state space in standard quantum mechanics, $\mathcal{H}_{ABC}$, which encodes the {\bf nine degrees of freedom} used to describe the canonical three-particle state \cite{Rae:2002,Isham:1995,Dirac:1958,vonNeumann:1955}. 
Like all other relational models \cite{Giacomini:2017zju,Castro-Ruiz:2019nnl,delaHamette:2020dyi,Krumm:2020fws,Giacomini:2021gei,Castro-Ruiz:2021vnq,delaHamette:2022cka,Kabel:2024lzr,Giacomini:2020ahk,Giacomini:2021aof,Cepollaro:2021ccc,Giacomini:2018gxh,Mikusch:2021kro,Ballesteros:2020lgl,Ballesteros:2025ypr}, the perspective-neutral framework \cite{Vanrietvelde:2018pgb,Hohn:2018iwn,Hohn:2018toe,delaHamette:2021oex,Hoehn:2019fsy,AliAhmad:2021adn,Vanrietvelde:2018dit,Hoehn:2023ehz,Hoehn:2023axh,DeVuyst:2025ezt} is not grounded in canonical quantum theory \cite{Rae:2002,Isham:1995,Dirac:1958,vonNeumann:1955}. 
Given that its rationale is, supposedly, the imposition of Galilean invariance, we find this doubly odd, since canonical quantum mechanics is already Galilean-invariant, so long as one chooses to work with Galilean-invariant operators as observables. 
There is no need to impose a state space reduction of any kind, in order to achieve this \cite{Bargmann:1954gh,Levy-Leblond:1963qdx,Levy-Leblond:1967eic,Horzela:1991pa,Giulini:1995te,Csillag:2003}.

Similar comments also hold for the special-relativistic and gravitational extensions of the non-relativistic relational models \cite{delaHamette:2022cka,Kabel:2024lzr,Giacomini:2020ahk,Giacomini:2021aof,Cepollaro:2021ccc,Giacomini:2018gxh,Mikusch:2021kro,Hoehn:2020epv,Apadula:2022pxk,Kabel:2022cje,Carrozza:2021gju,Goeller:2022rsx,Kabel:2023jve,Chen:2026kui,Ballesteros:2020lgl,Ballesteros:2025ypr}. 
In accordance with the relational philosophy, as originally espoused in works such as \cite{Giacomini:2017zju,Vanrietvelde:2018pgb}, these follow-up studies also excise the degrees of freedom of the chosen reference system -- which is, in general, non-inertial -- and/or make no distinction between inertial and accelerating frames, whether classical or quantum \cite{Hoehn-Presentation:CERN-2019,Hoehn-Presentation:HK-2020,Giacomini-Presentation:QGAAT-2023,Overstreet:2022zgq,Giacomini:2017zju:PeerReview}.

\section{Critique of the perspective-neutral and relational formalisms} \label{Sec.3}

In this section, we provide details of the technical errors, as well as the deeper conceptual misunderstandings, that lead to the erroneous claims in \cite{Giacomini:2017zju,Vanrietvelde:2018pgb} and in a large body of related work. 
(See, for example, \cite{Giacomini:2017zju,Castro-Ruiz:2019nnl,delaHamette:2020dyi,Krumm:2020fws,Giacomini:2021gei,Castro-Ruiz:2021vnq,delaHamette:2022cka,Kabel:2024lzr,Giacomini:2020ahk,Giacomini:2021aof,Cepollaro:2021ccc,Giacomini:2018gxh,Mikusch:2021kro,Ballesteros:2020lgl,Ballesteros:2025ypr,Vanrietvelde:2018pgb,Hohn:2018iwn,Hohn:2018toe,delaHamette:2021oex,Hoehn:2019fsy,AliAhmad:2021adn,Vanrietvelde:2018dit,Hoehn:2023ehz,Hoehn:2023axh,DeVuyst:2025ezt,Hoehn:2020epv,Apadula:2022pxk,Kabel:2022cje,Carrozza:2021gju,Goeller:2022rsx,Kabel:2023jve,Chen:2026kui} and references therein.)
This means, unfortunately, that we must be highly critical of some authors. 
But this criticism is necessary, as these misconceptions have gone unchallenged in the literature for far too long -- so long, that they appear to be accepted as ``facts'' by a large section of the research community.
\footnote{By this, we mean that the assumptions of a handful of related models are now widely accepted as the legitimate `starting point' for further work \cite{Giacomini:2017zju,Castro-Ruiz:2019nnl,delaHamette:2020dyi,Krumm:2020fws,Giacomini:2021gei,Castro-Ruiz:2021vnq,delaHamette:2022cka,Kabel:2024lzr,Giacomini:2020ahk,Giacomini:2021aof,Cepollaro:2021ccc,Giacomini:2018gxh,Mikusch:2021kro,Ballesteros:2020lgl,Ballesteros:2025ypr,Vanrietvelde:2018pgb,Hohn:2018iwn,Hohn:2018toe,delaHamette:2021oex,Hoehn:2019fsy,AliAhmad:2021adn,Vanrietvelde:2018dit,Hoehn:2023ehz,Hoehn:2023axh,DeVuyst:2025ezt,Hoehn:2020epv,Apadula:2022pxk,Kabel:2022cje,Carrozza:2021gju,Goeller:2022rsx,Kabel:2023jve,Chen:2026kui}. These assumptions are regarded as ``facts'' in the sense that their validity is not critically analysed, but is taken as given, at the beginning of all follow-up studies. The follow-up studies that make use of the formalisms presented in \cite{Giacomini:2017zju,Vanrietvelde:2018pgb}, without challenging the basic assumptions of the ``relational'' viewpoint, on which these models are based -- such as the equivalence of inertial and non-inertial frames, or the idea that the degrees of freedom of the chosen ``reference particle'' can be safely excised from the description of a closed system -- are a case in point.}

We proceed systematically, dealing first with the errors implicit in the reduced quantisation scheme, summarised in Sec. \ref{Sec.2.1}. 
We then show that these errors carry forward, to the `canonical' quantisation summarised in Sec. \ref{Sec.2.2}. 
Here, new errors are also introduced, due to a misunderstanding of the relevant constraints. 
Finally, we show that the Dirac quantisation scheme, summarised in Sec. \ref{Sec.2.3}, cumulatively inherits the contradictions of the previous two schemes, and introduces new ones. 
We conclude that the perspective-neutral formalism \cite{Vanrietvelde:2018pgb}, and all related relational models, such as \cite{Giacomini:2017zju}, are highly non-canonical -- and, furthermore, that they are not internally self-consistent. 

\subsection{Critique of the reduced quantisation method, used in relational models} \label{Sec.3.1}

A {\bf literal} jump from the external inertial frame $O$, to the frame centred on the trajectory of particle $A$, corresponds to the following transformation of the canonical Hamiltonian (\ref{2.1}):
\begin{eqnarray} \label{3.1}
H \mapsto H_{BC:A} &:=& \frac{({\bf p}_O + m_A\dot{{\bf X}})^2}{2m_A} + \frac{({\bf p}_{B} + m_B\dot{{\bf x}}_O)^2}{2m_B} + \frac{({\bf p}_{C} + m_C\dot{{\bf x}}_O)^2}{2m_C} 
\nonumber\\
&+& V_{AB}(|{\bf x}_B|) + V_{AC}(|{\bf x}_C|) + V_{BC}(|{\bf x}_C-{\bf x}_B|) \, ,
\end{eqnarray}
where ${\bf x}_O := -{\bf x}_A$ denotes the displacement of $O$'s (abstract) coordinate origin, from the physical particle $A$, with ${\bf p}_O := -{\bf p}_A$. 
The first term in $H_{BC:A}$ represents the kinetic energy of particle $A$, relative to the common centre of mass of the $ABC$ system, which is equal to the apparent work done by the pseudo-force experienced in $A$'s accelerated frame of reference \cite{Kibble&Berkshire:2004,McCall:2011,Arnold:1978,Marsden&Ratiu:1999}. 
The second and third terms represent the kinetic energies of particles $B$ and $C$, respectively, relative to $A$. 
The non-relativistic inter-particle potentials retain the same form as in the canonical Hamiltonian, $H$ (\ref{2.1}), but are now expressed in terms of the coordinate system centred on particle $A$

It is important to understand that, here, the variable ${\bf p}_O = m_A\dot{{\bf x}}_{O}$ must be regarded as generating an {\bf effective external potential}, which couples non-minimally to the momenta of particles $B$ and $C$. 
Its presence in the modified Hamiltonian indicates the existence of a {\bf pseudo-force}, which is peculiar to $A$'s accelerated frame, and which is not a dynamical variable of the system {\bf in that frame} \cite{Kibble&Berkshire:2004,McCall:2011,Arnold:1978,Marsden&Ratiu:1999}. 
Likewise, the canonical centre-of-mass velocity, $\dot{{\bf X}}$, must also be treated as an external constant, since this is not necessarily conserved from $A$'s perspective.
The usual relations
\begin{eqnarray} \label{3.2}
\left\{x_B^i,p_{Bj}\right\}_{\rm PB} = \delta^{i}{}_{j} \, , \quad \left\{x_B^i,p_{Cj}\right\}_{\rm PB} = 0 \, ,
\nonumber\\
\left\{x_C^i,p_{Bj}\right\}_{\rm PB} = 0 \, , \quad  \left\{x_C^i,p_{Cj}\right\}_{\rm PB} = \delta^{i}{}_{j} \, , 
\end{eqnarray}
allow us to use Hamilton's equations for the remaining variables, yielding
\begin{eqnarray} \label{3.3}
m_B\ddot{{\bf x}}_B = -\frac{\partial}{\partial {\bf x}_B}(V_{AB} + V_{BC}) + m_B\ddot{{\bf x}}_O \, , \, \, \,  
m_C\ddot{{\bf x}}_C = -\frac{\partial}{\partial {\bf x}_C}(V_{AC} + V_{BC}) + m_C\ddot{{\bf x}}_O \, .
\end{eqnarray}
In this construction, it is clear that the fictitious forces
\begin{eqnarray} \label{3.4}
{\bf F}_{O:A} := m_A\ddot{{\bf x}}_O \, \quad 
{\bf F}_{B:A} := m_B\ddot{{\bf x}}_O \, \quad 
{\bf F}_{C:A} := m_C\ddot{{\bf x}}_O \, ,
\end{eqnarray}
appear to act on particles $A$, $B$ and $C$, respectively, in addition to the real forces sourced by the inter-particle potentials. 
We are, however, unable to solve for ${\bf x}_O(t)$, since the `equation of motion' for particle $A$ is simply $-m_A\ddot{{\bf x}}_O + {\bf F}_{O:A} = {\bf 0}$, by construction.
\footnote{See, also, \cite{R.R.Rafokov:2026} for an alternative approach.} 
It cannot be anything else, except this {\bf identity}, {\it if} we wish our frame of reference to literally inhabit $A$'s trajectory. 
Instead, we must solve for ${\bf x}_A(t)$ in the {\bf external inertial frame}, in which Newton's laws are canonically formulated \cite{Kibble&Berkshire:2004,McCall:2011,Arnold:1978,Marsden&Ratiu:1999}. 
We may then use $\ddot{{\bf x}}_A = -m_A^{-1}\partial V/\partial{\bf x}_A$, as usual, and, from this, determine the pseudo-forces (\ref{3.4}). 
The relevant expressions are then substituted into Eqs. (\ref{3.3}), and the motions of particles $B$ and $C$, as seen from $A$'s accelerated frame, are determined by solving all three equations as a closed system.
\footnote{These results also have important implications for the quantum regime. It may be shown that the quantum analogue of (\ref{3.1}), $\widehat{H}_{BC:A}$ is {\bf not} unitarily equivalent to the canonical quantum three-particle Hamiltonian (\ref{2.16}). Similarly, the Hamiltonian associated with $B$'s accelerated QRF, $\widehat{H}_{AC:B}$, is not unitarily equivalent to either the canonical Hamiltonian, or to $\widehat{H}_{BC:A}$, and analogous comments also hold for $\widehat{H}_{AB:C}$. This means that the operators representing ``jumps'' from $O$'s classical frame, to the QRFs associated with the individual particles, {\bf cannot be unitary}. Perhaps more importantly, it also means that QRF-to-QRF transitions -- where the latter represent ``jumps'' between the perspectives of different material ``reference systems'', as in most of the existing relational literature \cite{Giacomini:2017zju,Castro-Ruiz:2019nnl,delaHamette:2020dyi,Krumm:2020fws,Giacomini:2021gei,delaHamette:2022cka,Kabel:2024lzr,Giacomini:2020ahk,Giacomini:2021aof,Cepollaro:2021ccc,Giacomini:2018gxh,Mikusch:2021kro,Ballesteros:2020lgl,Ballesteros:2025ypr,Vanrietvelde:2018pgb,Hohn:2018iwn,Hohn:2018toe,delaHamette:2021oex,Hoehn:2019fsy,AliAhmad:2021adn,Vanrietvelde:2018dit,Hoehn:2023ehz,Hoehn:2023axh,DeVuyst:2025ezt,Hoehn:2020epv,Apadula:2022pxk,Kabel:2022cje,Carrozza:2021gju,Goeller:2022rsx,Kabel:2023jve,Chen:2026kui} -- {\bf cannot be unitary} either. 
To be clear, the QRF transition operators formulated in \cite{Giacomini:2017zju}, and related works, are {\bf unitary by construction}. 
But this means that they {\it cannot} represent transitions between the perspectives of differently accelerated particles, in either the position space or momentum space representations. 
Such perspectives cannot be unitarily related, because they are not equivalent descriptions of equivalent physics: the pseudo-forces associated with particle $A$'s frame are unique to that frame, and other embodied frames will `see things differently'. 
Ultimately, this is why the canonical degrees of freedom associated with the frame of the embodied observer are physically relevant: one cannot account for such `perspectival differences', if these are simply excised from the model. 
We will have more to say about this issue, in an upcoming work. 
(See also \cite{Busch:2016rui,Loveridge:2016tnh,Loveridge:2017pcv,Carette:2023wpz, Castro-Ruiz:2021vnq,Lake:2023nua,Knopki:2025bab,Garmier:2025soc,Lake:2026zrk}, for alternative models of non-unitary QRF-to-QRF transitions.) } 

Equations (\ref{3.3}) reduce to the equations of motion (EOM) for ${\bf x}_B$ and ${\bf x}_C$, derived from the canonical Hamiltonian (\ref{2.8}), if and only if we impose the condition $\ddot{{\bf x}}_O(t) = {\bf 0}$, so that all three pseudo-forces (\ref{3.4}) vanish. 
This is, in general, impossible, if either $V_{AB} \neq 0$ or $V_{AC} \neq 0$. 
Under no circumstances is the Hamiltonian `seen' in $A$'s accelerated frame, $H_{BC:A}$ (\ref{3.1}), equivalent to the ``reduced'' Hamiltonian of the perspective-neutral formalism \cite{Vanrietvelde:2018pgb}, (\ref{2.3}), with ${\bf p}_B + {\bf p}_C \neq {\bf 0}$. 

However, we may obtain a Hamiltonian that is very similar to (\ref{2.3}) by instead performing the coordinate transformation
\begin{eqnarray} \label{3.5}
\left\{{\bf x}_A,{\bf x}_B,{\bf x}_C;{\bf p}_A,{\bf p}_B,{\bf p}_C\right\} \mapsto \left\{{\bf X},{\bf x}_{B|A},{\bf x}_{C|A};{\bf P},\boldsymbol{\pi}_{B|A},\boldsymbol{\pi}_{C|A}\right\} \, ,
\end{eqnarray}
where
\begin{eqnarray} \label{3.6}
\boldsymbol{\pi}_{B|A} := {\bf p}_B - \frac{m_B}{M}{\bf P} \, , \quad \boldsymbol{\pi}_{C|A} := {\bf p}_C - \frac{m_C}{M}{\bf P} \, ,
\end{eqnarray}
and 
\begin{eqnarray} \label{3.7}
{\bf X} := \frac{m_A{\bf x}_A + m_B{\bf x}_B + m_C{\bf x}_C}{M} \, .
\end{eqnarray}
In the new phase space coordinates, $\left\{{\bf X},{\bf x}_{B|A},{\bf x}_{C|A};{\bf P},\boldsymbol{\pi}_{B|A},\boldsymbol{\pi}_{C|A}\right\}$, which also form a canonically conjugate set, the canonical Hamiltonian (\ref{2.1}) can be rewritten as
\begin{eqnarray} \label{3.8}
H({\bf x}_{\rm rel},\boldsymbol{\pi}_{\rm rel}) &=& \frac{P^2}{2M} + \frac{\pi_{B|A}^2}{2m_B} + \frac{\pi_{C|A}^2}{2m_C} + \frac{(\boldsymbol{\pi}_{B|C}+\boldsymbol{\pi}_{C|A})^2}{2m_A} 
\nonumber\\
&+&  V_{AB}(|{\bf x}_{B|A}|) + V_{AC}(|{\bf x}_{C|A}|) + V_{BC}(|{\bf x}_{C|A}-{\bf x}_{B|A}|) \, .
\end{eqnarray}
We are then free to boost to a frame in which the net momentum is zero. 

This is equivalent to imposing the {\bf primary first-class constraints} $P_i \approx 0$ \cite{Dirac-CHS:1958,Dirac-CHS:1964,Henneaux-Teitelboim:1992,Rothe-RotheCHS:2010}, but it is important to understand that such a transformation is possible {\bf only} because (\ref{3.8}) still represents the {\bf external} view of the three-particle system, from the perspective of an {\bf arbitrary, abstract, inertial frame}. 
The change of phase space coordinates, (\ref{3.5}), which re-expresses (\ref{2.1}) in terms of the relational variables $\left\{{\bf x}_{B|A},{\bf x}_{C|A};\boldsymbol{\pi}_{B|A},\boldsymbol{\pi}_{C|A}\right\}$ and the centre-of-mass variables $\left\{{\bf X},{\bf P}\right\}$, does not take us literally into the frame of $A$'s trajectory. 
We cannot, therefore, set ${\bf x}_A \approx 0$ in this frame. 

Hence, far from being the correct ``relational'' implementation of Galilean symmetry, the classical model on which the perspective-neutral formalism is based actually violates the requirements of Galilean invariance, by insisting that this invariance be maintained even in an accelerating frame. 
This erroneous view seems to be linked to the claim -- made, for example, in \cite{Vanrietvelde:2018pgb,Hoehn-Presentation:CERN-2019,Hoehn-Presentation:HK-2020,Giacomini-Presentation:QGAAT-2023,Overstreet:2022zgq,Giacomini:2017zju:PeerReview} -- that ``all the laws of physics are the same in every reference frame''. 

So far as we can gather, from a review of the now rather vast literature on relational QRFs (see \cite{Giacomini:2017zju,Castro-Ruiz:2019nnl,delaHamette:2020dyi,Krumm:2020fws,Giacomini:2021gei,Castro-Ruiz:2021vnq,delaHamette:2022cka,Kabel:2024lzr,Giacomini:2020ahk,Giacomini:2021aof,Cepollaro:2021ccc,Giacomini:2018gxh,Mikusch:2021kro,Ballesteros:2020lgl,Ballesteros:2025ypr,Vanrietvelde:2018pgb,Hohn:2018iwn,Hohn:2018toe,delaHamette:2021oex,Hoehn:2019fsy,AliAhmad:2021adn,Vanrietvelde:2018dit,Hoehn:2023ehz,Hoehn:2023axh,DeVuyst:2025ezt,Hoehn:2020epv,Apadula:2022pxk,Kabel:2022cje,Carrozza:2021gju,Goeller:2022rsx,Kabel:2023jve,Chen:2026kui} and references therein), the logic of this view is as follows:
(i) The Weak Equivalence Principle (WEP) states the equivalence between uniform acceleration and motion in a constant and uniform gravitational field.
This view can be attributed to \cite{Giacomini:2017zju}, and related works, but is not correct. 
(See App. \ref{App.A.1}.) 
(ii) If the WEP asserts the equivalence between uniform acceleration, and motion in a constant and uniform gravitational field, then the general EP must assert the equivalence between non-uniform acceleration, and motion in a non-uniform gravitational field. 
This is a logical, and, indeed, inescapable conclusion, if one believes statement (i), but is incorrect because statement (i) is incorrect. 
(iii) The general EP asserts the equivalence of all Locally Inertial Frames (LIFs). 
This means that, for motion in an arbitrary gravitational field, all non-gravitational laws of physics take their special-relativistic forms. 
This statement is correct \cite{MTW:1973,Dirac-GR:1975,Wald-GR:1984,Lasenby:2006}. 
It leads to incorrect conclusions only when it is married to statements (i) and (ii); 
{\it if} one puts all three statements together, then one is led, inexorably, to believe in the equivalence of accelerated and inertial motion, because the former is, supposedly, equivalent to motion in a gravitational field -- in which all non-gravitational laws of physics take their standard `inertial' forms.

Thus, in a large chunk of the relational QRF literature -- including, but not limited to, the references cited above -- the supposed equivalence between accelerated and inertial motion is believed to be a consequence of the theory of classical general relativity \cite{MTW:1973,Dirac-GR:1975,Wald-GR:1984,Lasenby:2006}. 
In our view, the relational literature also falsely conflates the EP, which has physical significance, with the principle of general (passive) covariance, which does not \cite{Norton:1993eqc}. 
This leads to talk about a theory of ``quantum general covariance'', which asserts the equivalence of all quantum frames of reference, regardless of whether they are associated with accelerated or non-accelerated particles \cite{Hoehn-Presentation-fxqi-2021}. 
When the authors of \cite{Vanrietvelde:2018pgb} refer to a ``gravity-inspired symmetry principle'', which they do not explicitly define, but which is said to motivate their model, they appear to be claiming that this ``Machian'' view of general relativity should be maintained in the quantum regime, by extending the equivalence of all classical frames to include QRFs.

We categorically disagree with this. 
In classical Newtonian mechanics, a pseudo-force, 
\begin{eqnarray} \label{3.9}
{\bf F}_{{\rm CoM}:A} := -M\ddot{{\bf x}}_A \, , 
\end{eqnarray}
also acts on the apparent centre of mass, from $A$'s perspective, giving
\begin{eqnarray} \label{3.10}
\dot{{\bf P}} = M\ddot{{\bf X}} = {\bf 0} \mapsto \dot{{\bf P}}_{{\rm CoM}:A} = M\ddot{{\bf X}}_{{\rm CoM}|A} = M(\ddot{{\bf X}} - \ddot{{\bf x}}_A ) = -M\ddot{{\bf x}}_A \, . 
\end{eqnarray}
Therefore, it is simply not possible to impose {\bf both} the conditions ${\bf x}_A \approx {\bf 0}$ and ${\bf P} \approx {\bf 0}$ (\ref{2.11}), if particle $A$ is accelerated. 
In the general case, with arbitrary inter-particle potentials, there is {\bf no frame} that satisfies the requirements of the reduced quantisation procedure, used in the perspective-neutral formalism \cite{Vanrietvelde:2018pgb}.

Another way to see this is to note that the quantum version of (\ref{3.8}), in which {\bf all} classical variables are promoted to operators, is unitarily equivalent to the quantum version of (\ref{2.1}) (i.e, (\ref{2.16})), so that the two contain equivalent physical information, expressed with respect to different Hilbert space bases \cite{Rae:2002,Isham:1995,Dirac:1958,vonNeumann:1955}. 
(See Appendix \ref{App.B} for details.) 
If the net momentum term and location information for particle $A$ are simply discarded -- setting $\widehat{{\bf P}} = 0$ and $\widehat{{\bf x}}_A = 0$ -- one obtains the quantum Hamiltonian (\ref{2.9}) used in the perspective-neutral formalism \cite{Vanrietvelde:2018pgb} (in a slightly different notation) but {\bf physical information has been lost}. 
The expression (\ref{2.9}) is {\bf not} the Hamiltonian for the three-particle system, as `seen' by particle $A$, as the authors of \cite{Vanrietvelde:2018pgb} claim, but the Hamiltonian of a three-particle system `seen' from the perspective of an {\bf external inertial frame}, from which an arbitrary chunk of physical information has been removed. 
Similar remarks hold for the Hamiltonian (\ref{2.10}), used in \cite{Giacomini:2017zju}, only the chunk is slightly different. 

\subsection{Critique of the `canonical' quantisation method, used in relational models} \label{Sec.3.2}

Our comments regarding the classical models (\ref{2.8}) and (\ref{2.3}), which underpin the relational \cite{Giacomini:2017zju} and perspective-neutral \cite{Vanrietvelde:2018pgb} frameworks, respectively, also hold for the quantum models proposed in these formalisms, (\ref{2.10}) and (\ref{2.9}). 
Yet, in \cite{Vanrietvelde:2018pgb}, it is claimed that (\ref{2.3})/(\ref{2.9}) can be derived from a rigorous implementation of translation symmetry, using the Dirac bracket to provide a systematic treatment of the relevant constraints. 
Clearly, something has gone wrong with this procedure and we can see two errors in the `canonical' quantisation method, proposed in these works. 

The first error is a belief that the conditions ${\bf x}_A \approx {\bf 0}$ and ${\bf P} \approx {\bf 0}$ (\ref{2.11}) represent a system of second-class constraints, on the closed, classical, three-particle system. 
As shown in Sec. \ref{Sec.3.1}, there is no classical frame of reference in which both these conditions can be satisfied simultaneously, in the presence of arbitrary particle interactions. 
The proposed mapping, $\left\{x_A^i,P_j\right\}_{\rm PB} = \delta^{i}{}_{j} \mapsto \left\{x_A^i,P_j\right\}_{\rm DB} = 0 \mapsto (i\hbar)^{-1}[\widehat{x}_A^i,\widehat{P}_j] = 0$ \cite{Vanrietvelde:2018pgb}, is therefore invalid, since the Dirac bracket used in the intermediate step is not properly defined. 
This is consistent with the fact that, in canonical classical mechanics, the $P_i \approx 0$ represent primary first-class constraints, so that there cannot exist any other constraints which fail to commute with them \cite{Dirac-CHS:1958,Dirac-CHS:1964,Henneaux-Teitelboim:1992,Rothe-RotheCHS:2010}.
\footnote{Imposing these constraints simply fixes the values of the momenta, $P_i$, that are canonically conjugate to the classically ``ignorable coordinates'' $X^i$ \cite{Kibble&Berkshire:2004,McCall:2011,Arnold:1978,Marsden&Ratiu:1999}.}

The second error is the belief that all {\bf classically ignorable} coordinates remain unphysical in the quantum regime. 
This is what an {\it ad hoc} process of {\bf reduced quantisation} tells us, but it is {\bf not} consistent with a {\bf canonical quantisation} procedure, when the latter is properly applied. 

There are many ways to demonstrate the fact that a closed system of $N$ quantum particles requires a total of $3N$ quantum mechanical degrees of freedom -- not $3(N-1)$ or less \cite{Vanrietvelde:2018pgb,Vanrietvelde:2018dit} -- for its full description, and that, conversely, using fewer degrees of freedom results in a loss of {\bf physical information}. 
On an intuitive level, the physical relevance of the centre-of-mass operators, $\widehat{{\bf X}}$ and $\widehat{{\bf P}}$, can be explicitly demonstrated. 
More generally, this statement can be proven by comparing the precise mechanisms by which (a) the equations of motion for a closed classical system are made invariant under the Poisson bracket representation of the centrally-extended Galilean group, and (b) the Schr{\" o}dinger equation for a closed quantum system is made invariant under its commutator representation. 
The rigorous argument is provided in Sec. \ref{Sec.X}, referring to the standard references \cite{Bargmann:1954gh,Levy-Leblond:1963qdx,Levy-Leblond:1967eic,Horzela:1991pa,Giulini:1995te,Csillag:2003}. 
Here, we first give an outline of the intuitive reasoning.

To see the physical relevance of the centre-of-mass operators, $\widehat{{\bf X}}$ and $\widehat{{\bf P}}$, it is sufficient to note that the conditions
\begin{eqnarray} \label{3.11}
\widehat{{P}}_i|\Psi\rangle \overset{!}{=} 0 \, , 
\end{eqnarray} 
are {\bf not} the correct quantum analogues of the classical constraints,
\begin{eqnarray} \label{3.12}
P_i \approx 0 \, . 
\end{eqnarray}
The problem with the reasoning used in (\ref{3.11}) is that (\ref{3.12}) are not the classical conditions that need to be quantised. 
The {\bf classical} physics remains unaffected, regardless of whether the external inertial frame coincides with the trajectory of centre of mass of the system, or moves with a constant nonzero velocity with respect to it. 
Hence, for a general inertial frame $O$, the relevant conditions are
\begin{eqnarray} \label{3.13}
P_i \approx Mv_i  \, ,
\end{eqnarray}
where ${\bf v} = [v_x,v_y,v_z]^{\rm T}$ is the {\bf constant velocity} of the classical centre of mass, relative to the frame $O$. 
Quantising these conditions gives
\begin{eqnarray} \label{3.14}
\widehat{P}_i = M\widehat{v}_i  \, ,
\end{eqnarray}
where the $\widehat{v}_i$'s are {\bf time-independent} operators. 
The requirement that these operators be time-independent places no constraints on the number of degrees of freedom, used to expand the physical states of the theory \cite{Rae:2002,Isham:1995,Dirac:1958,vonNeumann:1955}. 

Conversely, imposing the conditions
\begin{eqnarray} \label{3.15}
\widehat{{P}}_i|\Psi\rangle \overset{!}{=} Mv_i|\Psi\rangle 
\end{eqnarray} 
defines $|\Psi\rangle$ as an {\bf eigenstate} of the operator $\widehat{\bf{P}}$, which effectively `kills' three degrees of freedom \cite{Vanrietvelde:2018pgb}. 
The states satisfying (\ref{3.11}) form a subset of this subset, corresponding to the specific eigenvalues $v_i = 0$. 
More importantly, these states are actually {\bf unphysical}, in the canonical theory \cite{Rae:2002,Isham:1995,Dirac:1958,vonNeumann:1955}, since the continuous eigenspectrum of $\widehat{{P}}_i$ implies that they are non-normalisable. 
In this sense, canonical Galilean-invariant quantum mechanics predicts the very opposite of the relational models -- which, we again see, are highly non-canonical in nature. 

Another way of looking at this result is to say that, unlike the classical equation ${\bf P} := {\bf p}_A + {\bf p}_B + {\bf p}_C$ (\ref{2.2}), which represents both a {\bf definition} of the net momentum and a {\bf specification} of its particular value, in a given inertial frame $O$, the corresponding equation in the quantum regime, $\widehat{{\bf P}} := \widehat{{\bf p}}_A + \widehat{{\bf p}}_B + \widehat{{\bf p}}_C$, is just the definition of the operator $\widehat{{\bf P}}$. 
It is defined in terms of three other {\bf independent} operators, corresponding to a total of nine independent quantum mechanical degrees of freedom, and has no fixed value for a {\bf physical}, normalisable, state. 

One may, of course, wonder what happened to the three ``gauge'' degrees of freedom, which can be safely removed in the classical limit. 
How did they suddenly {\it become} physical, in the quantum world? 
The answer to this question is subtle, but important, and is not much explored in the existing literature on the foundations of quantum theory \cite{Landsman:2017}. 
To answer it, it is better to first flip it around, to ask why there are exactly three {\bf unphysical} degrees of freedom, in the canonical description of the {\bf classical} system? 

Ultimately, this is because the classical description contains no analogue of the delocalised quantum wave function, whereas the mathematical description of quantum systems requires us to treat both operators {\it and} states. 
The hermitian operators have classical counterparts, in the mathematical description of classical systems, but the states do not. 
Practically, this means that there is no classical counterpart of the canonical Schr{\" o}dinger equation \cite{Rae:2002,Isham:1995,Dirac:1958,vonNeumann:1955}. 
It is the invariance of this equation -- under the centrally-extended Galilean group \cite{Bargmann:1954gh,Levy-Leblond:1963qdx,Levy-Leblond:1967eic,Horzela:1991pa,Giulini:1995te,Csillag:2003} 
-- which proves that {\it none} of the $3N$ canonical degrees of freedom can be safely dispensed with, without erasing physical information about the system. 
(See Sec. \ref{Sec.X}.)
 
Hence, in canonical quantum mechanics, an arbitrary physical state of the closed three-particle system is expanded as
\begin{eqnarray} \label{3.16}
|\Psi\rangle &=& \int \Psi({\bf x}_A,{\bf x}_B,{\bf x}_C) |{\bf x}_A\rangle_A|{\bf x}_B\rangle_B |{\bf x}_C\rangle_C {\rm d}^3x_A{\rm d}^3x_B{\rm d}^3x_C  
\nonumber\\
&=& \int \tilde{\Psi}({\bf p}_A,{\bf p}_B,{\bf p}_C) |{\bf p}_A\rangle_A|{\bf p}_B\rangle_B |{\bf p}_C\rangle_C {\rm d}^3p_A{\rm d}^3p_B{\rm d}^3p_C \, , 
\end{eqnarray}
in the canonical basis, where $\Psi({\bf x}_A,{\bf x}_B,{\bf x}_C)$ and $\tilde{\Psi}({\bf p}_A,{\bf p}_B,{\bf p}_C)$ are each normalised over nine-dimensional volumes, 
\begin{eqnarray} \label{3.17}
\langle \Psi | \Psi \rangle = \int |\Psi({\bf x}_A,{\bf x}_B,{\bf x}_C)|^2 {\rm d}^3x_A{\rm d}^3x_B{\rm d}^3x_C 
= \int |\tilde{\Psi}({\bf p}_A,{\bf p}_B,{\bf p}_C)|^2{\rm d}^3p_A{\rm d}^3p_B{\rm d}^3p_C = 1 \, . 
\end{eqnarray}
In short, this means that {\bf all} relational models suffer from the same fatal flaw -- namely, that they do not contain a sufficient number of quantum mechanical degrees for freedom, to describe the physical systems they purport to describe \cite{Giacomini:2017zju,Vanrietvelde:2018pgb,Vanrietvelde:2018dit}.
\footnote{It is tempting to think that the so-called ``extra particle'' framework, introduced in \cite{Castro-Ruiz:2021vnq,Garmier:2025soc}, may have solved this problem. At first glance, it looks like the purpose of this framework is to `put back' the degrees of freedom of the embodied frame, that were excised from both the classical \cite{Kibble&Berkshire:2004,McCall:2011,Arnold:1978,Marsden&Ratiu:1999} and canonical quantum \cite{Rae:2002,Isham:1995,Dirac:1958,vonNeumann:1955} theories, in previous relational models. 
But, logically, this should just lead {\it back} to canonical, Galilean-invariant, quantum mechanics \cite{Bargmann:1954gh,Levy-Leblond:1963qdx,Levy-Leblond:1967eic,Horzela:1991pa,Giulini:1995te,Csillag:2003} -- and, hence, to exactly the conclusions drawn here, in Sec. \ref{Sec.4} and Appendices \ref{App.A}-\ref{App.C}. 
Yet this is not the case as the ``extra particle'' models \cite{Castro-Ruiz:2021vnq,Garmier:2025soc} remain consistent  with previous relational theories, such as those presented in \cite{Giacomini:2017zju,Vanrietvelde:2018pgb}. In \cite{Castro-Ruiz:2021vnq}, it is stated that ``our framework restricted to the zero-charge sector for a general group reduces to the QRF transformation found in Ref. \cite{delaHamette:2020dyi}. For the case of translations, this recovers formally the perspective-neutral computation of the QRF transformation developed in Ref. \cite{Giacomini:2017zju}'' and in \cite{Garmier:2025soc} that ``It has been shown \cite{delaHamette:2021oex,Vanrietvelde:2018pgb} that the QRF transformations of \cite{Giacomini:2017zju} and \cite{delaHamette:2020dyi} are mathematically equivalent to those of the perspective-neutral framework in the case of regular QRFs. Therefore, they are also obtained mathematically from our framework in the regular case by restricting to the trivial representation $r = 0$''. (In these quotations, the relevant references have been renumbered to match our bibliography.) Here, $r$ denotes the ``charge associated to the global $G$-action'', which ``enumerates all irreducible representations of $G$'', where $G$ is the symmetry group of the theory \cite{Garmier:2025soc}. We find this latter claim strange because, according to Wigner's classification, it is only the invariants $C$, associated to Casimir operators of the relevant Lie algebra, that can be used to label irreps \cite{Wigner:1939cj,Bargmann:1948ck}, whereas ``charges'' \cite{Noether'sTheorems} cannot be used for this purpose. But this does not much matter, for the point we wish to make here. We stress only that, {\bf under no possible circumstances} is canonical quantum mechanics equivalent to the relational models presented in refs. \cite{Giacomini:2017zju,Vanrietvelde:2018pgb}. These theories contain ``unmoved movers'' \cite{Aristotle,Aristotle:SEP}, violate Galilean symmetry \cite{Bargmann:1954gh,Levy-Leblond:1963qdx,Levy-Leblond:1967eic,Horzela:1991pa,Giulini:1995te,Csillag:2003}, and are not compatible with the law of action and reaction, at either the classical \cite{Kibble&Berkshire:2004,McCall:2011,Arnold:1978,Marsden&Ratiu:1999} or the quantum level \cite{Rae:2002,Isham:1995,Dirac:1958,vonNeumann:1955}. 
(See Secs. \ref{Sec.X} and Appendices \ref{App.A}-\ref{App.C}.) 
The issue of whether the ``global charge'' is experimentally accessible, to embodied observers, is also addressed in \cite{Doat:2025klp}. However, this study makes no mention of pseudo-forces, or of the critical distinction between accelerated and inertial frames. In other words, it engages the question within the context of existing ``relational'' theories, in which no such distinction is made \cite{Giacomini:2017zju,Vanrietvelde:2018pgb}. For Galilean symmetries, the relevant charge is the net momentum of the closed system, and, as we stressed in Sec. \ref{Sec.2.1}, this is not conserved in the frame associated with a general, interacting, subsystem. In our view, this renders the question moot: the global charge cannot be measured, {\it as a conserved global charge}, in the QRF of an accelerated particle (contrary to recent claims in \cite{delaHamette:2025hpd}). What {\it can} be determined, from within such a frame, is the statistical spread $\Delta_{\Psi}P_i$, and its position-space counterpart, $\Delta_{\Psi}X^i$. (See Sec. \ref{Sec.4}.)}

In Sec. \ref{Sec.4}, we elucidate some of the {\bf physical} consequences of the degrees of freedom that are routinely discarded in the relational QRF literature. 
Before then, we complete this section by analysing the Dirac quantisation procedure, also proposed in \cite{Vanrietvelde:2018pgb}.

\subsection{Critique of the Dirac quantisation method, used in relational models} \label{Sec.3.3}

Our previous criticisms, already voiced in Secs. \ref{Sec.3.1}-\ref{Sec.3.2}, are also applicable to the ``Dirac quantisation'' scheme, proposed in \cite{Vanrietvelde:2018pgb} and summarised in Sec. \ref{Sec.2.3}. 
In this subsection, we need only add that the so-called ``Dirac constraint'' (\ref{2.27}) is not implemented by applying the canonical operators $\widehat{P}_i := \widehat{p}_{Ai} + \widehat{p}_{Bi} + \widehat{p}_{Ci}$ to states in the Hilbert space of canonical quantum theory. 
Doing this yields three distinct eigenstates with zero net momentum, i.e., those states satisfying the equation
\begin{eqnarray} \label{3.18}
\widehat{{\bf P}}|\phi_{{\bf P}}\rangle = {\bf P} |\phi_{{\bf P}}\rangle \, , 
\end{eqnarray}
with ${\bf P} = {\bf 0}$. 
In the position space representation of the wave mechanics picture, these states take the form
\begin{eqnarray} \label{3.19}
\phi_{{\bf 0}}^{(1)}({\bf x}) &\propto& e^{\frac{i}{\hbar} \left({\bf p}_A.{\bf x}_A + {\bf p}_B.{\bf x}_B - ({\bf p}_A + {\bf p}_B).{\bf x}_C \right)} \, ,
\nonumber\\
\phi_{{\bf 0}}^{(2)}({\bf x}) &\propto& e^{\frac{i}{\hbar}\left({\bf p}_A.{\bf x}_A - ({\bf p}_A + {\bf p}_C).{\bf x}_B + {\bf p}_C.{\bf x}_C \right)} \, , 
\nonumber\\
\phi_{{\bf 0}}^{(3)}({\bf x}) &\propto& e^{\frac{i}{\hbar} \left(-({\bf p}_B + {\bf p}_C).{\bf x}_A + {\bf p}_B.{\bf x}_B + {\bf p}_C.{\bf x}_C \right)} \, . 
\end{eqnarray}

Clearly, the condition (\ref{3.18}) has nothing to do with imposing translational symmetry. 
This may be imposed, as usual, by constructing a quantum equation of motion that is invariant under a general {\bf active transformation} of the centrally-extended Galilean group \cite{Bargmann:1954gh,Levy-Leblond:1963qdx,Levy-Leblond:1967eic,Horzela:1991pa,Giulini:1995te,Csillag:2003}. 
Hence, contrary to the claims made in \cite{Vanrietvelde:2018pgb,Hohn:2018iwn,Hohn:2018toe,delaHamette:2021oex,Hoehn:2019fsy,AliAhmad:2021adn,Vanrietvelde:2018dit,Hoehn:2023ehz,Hoehn:2023axh,DeVuyst:2025ezt}, the condition (\ref{2.27}) -- which is implemented by applying the operator $\delta(\widehat{{\bf P}})$ (\ref{2.30}) to the distributional states (\ref{2.29}) -- is {\bf not required}, in order to implement the translation-invariance of non-relativistic quantum theory. 
In fact, it is no more required than (\ref{3.19}) is required for this purpose. 
For this reason, and the others mentioned in Secs. \ref{Sec.3.1}-\ref{Sec.3.2}, it is simply not correct to say that the imposition of Galilean symmetries restricts the number of physical degrees of freedom in canonical quantum theory, or that it leads, necessarily, to the relational \cite{Giacomini:2017zju} or perspective neutral \cite{Vanrietvelde:2018pgb} models, advocated in the majority of the existing QRF literature \cite{Giacomini:2017zju,Castro-Ruiz:2019nnl,delaHamette:2020dyi,Krumm:2020fws,Giacomini:2021gei,Castro-Ruiz:2021vnq,delaHamette:2022cka,Kabel:2024lzr,Giacomini:2020ahk,Giacomini:2021aof,Cepollaro:2021ccc,Giacomini:2018gxh,Mikusch:2021kro,Ballesteros:2020lgl,Ballesteros:2025ypr,Vanrietvelde:2018pgb,Hohn:2018iwn,Hohn:2018toe,delaHamette:2021oex,Hoehn:2019fsy,AliAhmad:2021adn,Vanrietvelde:2018dit,Hoehn:2023ehz,Hoehn:2023axh,DeVuyst:2025ezt,Hoehn:2020epv,Apadula:2022pxk,Kabel:2022cje,Carrozza:2021gju,Goeller:2022rsx,Kabel:2023jve,Chen:2026kui}. 

The physical nature of the centre-of-mass degrees of freedom is manifested in many ways, but perhaps the most obvious is in the Ehrenfest Theorems \cite{Ehrenfest:QM-Compendium}. 
When applied to a closed quantum system, these theorems state that the expectation value of the operator $\widehat{{\bf X}}$, which is again defined from the perspective of an external inertial frame $O$, follows a classical inertial trajectory: $d\langle \widehat{{\bf X}}\rangle_{\Psi}/dt = \langle \widehat{{\bf v}}\rangle_{\Psi}$. 
(See, also, App \ref{App.B.1}.) 
For canonical quantum states, $\langle\widehat{{\bf X}}\rangle_{\Psi}$ is not well-defined if $\widehat{{\bf P}} \approx {\bf 0}$ ($\widehat{{\bf P}}|\Psi\rangle \overset{!}{=} {\bf 0}$). 
Hence, these constraints are not compatible with the theorems \cite{Ehrenfest:QM-Compendium}, or with the standard uncertainty principle for the centre-of-mass phase space coordinates, $\Delta_{\Psi}X^i\Delta_{\Psi}P_j \geq (\hbar/2)\delta^{i}{}_{j}$. 

In the perspective-neutral formalism \cite{Vanrietvelde:2018pgb}, this problem is `solved' by constructing what the authors refer to as a ``physical inner product'' 
(Eq. (69) in Appendix C of that text). 
For the three-dimensional extrapolation of their model, this  is defined as $(\psi^{\rm phys},\phi^{\rm phys}) := {}^{\rm kin}\langle \psi | \delta(\widehat{{\bf P}}) |\phi \rangle^{\rm kin}$, where $\langle \, . \, | \, . \, \rangle$ denotes the inner product on the ``kinematical Hilbert space'' $\overline{\mathcal{H}}_{ABC}^{\rm kin}$ (\ref{2.28}). 
This ``kinematical inner product'', $\langle \, . \, | \, . \, \rangle$, is equivalent to the usual inner product, for {\bf physical three-particle states} in canonical quantum mechanics \cite{Rae:2002,Isham:1995,Dirac:1958,vonNeumann:1955}, and, once again, we see that the relational models \cite{Giacomini:2017zju,Vanrietvelde:2018pgb} are not compatible with the canonical, Galilean-invariant, theory \cite{Bargmann:1954gh,Levy-Leblond:1963qdx,Levy-Leblond:1967eic,Horzela:1991pa,Giulini:1995te,Csillag:2003}. 
Specifically, while a maximally-dispersed centre-of-mass is impossible in the latter \cite{Ehrenfest:QM-Compendium}, it is considered physical in the former, in which the conditions $\Delta_{\Psi}X \rightarrow \infty$, $\Delta_{\Psi}P_i = 0$ are referred to as a manifestation of ``gauge invariance''.
\footnote{The second paragraph in Sec. 4.3 of \cite{Vanrietvelde:2018pgb} reads ``Hence, physical states as zero-eigenstates of $\widehat{P}$ must be maximally spread out over $q_{\rm cm}$. But this is gauge invariance: to smear/average over the gauge orbit. Indeed, this is precisely what the improper projector (23) does.''. The improper projection (23) is reproduced as (\ref{2.30}), in Sec. \ref{Sec.2.3} of the present work.} 

To be clear, our objection is not that the construction of the alternative norm fails, but that the theory it produces discards the centre-of-mass spreads, without adequate physical justification. 
Moreover, in the resulting QRF model \cite{Vanrietvelde:2018pgb}, eliminating the centre-of-mass subspace is equivalent to eliminating the subspace of the chosen reference particle, $A$, which is typically {\bf accelerated} by the presence of inter-particle potentials. 
In any viable formalism, these must be {\bf inequivalent deletions}, unless particle $A$ forms an isolated subsystem. 
(Since only then is its motion inertial.) 
Hence, we disagree with the claim that {\it any} subspace can be safely excised from the canonical theory. 
In Sec \ref{Sec.4}, we examine some of the {\bf measurable physical consequences} of these discarded degrees of freedom. 
Before then, in Sec. \ref{Sec.X}, we give detailed arguments for why the centre-of-mass degrees of freedom must be considered as physical, in canonical quantum mechanics, meaning that they cannot be ``gauge-fixed''. 
These arguments are based on the standard account of how Galilean invariance is implemented, in this theory \cite{Bargmann:1954gh,Levy-Leblond:1963qdx,Levy-Leblond:1967eic,Horzela:1991pa,Giulini:1995te,Csillag:2003}, and in canonical classical mechanics \cite{Kibble&Berkshire:2004,McCall:2011,Arnold:1978,Marsden&Ratiu:1999}. 

\section{Galilean invariance in classical and quantum mechanics} \label{Sec.X}

In \cite{Vanrietvelde:2018pgb}, it is claimed that the correct way to impose three primary first-class constraints, $P_i \approx 0$, on a closed, classical, $N$-particle system, is to subtract the centre-of-mass energy from the canonical Lagrangian \cite{Dirac-CHS:1958,Dirac-CHS:1964,Henneaux-Teitelboim:1992,Rothe-RotheCHS:2010}. 
In this section, we again deal with a three-particle system, as a simple example, but our results generalise straightforwardly to arbitrary $N$.
Hence, while the canonical Lagrangian,
\begin{eqnarray} \label{X.1}
L &:=& \frac{1}{2}m_A\dot{{\bf x}}_A^2 + \frac{1}{2}m_B\dot{{\bf x}}_B^2 + \frac{1}{2}m_C\dot{{\bf x}}_C^2 
\nonumber\\
&-& V_{AB}(|{\bf x}_B-{\bf x}_A|) - V_{AC}(|{\bf x}_C-{\bf x}_A|) - V_{BC}(|{\bf x}_C-{\bf x}_B|) \, , 
\end{eqnarray}
is {\bf regular}, but not Galilean-invariant, the modified Lagrangian,
\begin{eqnarray} \label{X.2}
L^{*} := L - \frac{1}{2}M\dot{{\bf X}}^2 \, , 
\end{eqnarray}
is Galilean-invariant, but {\bf singular}. 
Here, these terms refer to the Hessian \cite{Hessian}; a Lagrangian is said to be regular (singular) if its Hessian matrix is regular (singular), in the usual sense \cite{Boas}. 

There are various ways to show that, although the canonical Lagrangian (\ref{X.1}) is invariant under rotations, as well as both spatial and temporal translations, it is not invariant under Galilean boosts. 
Perhaps the easiest, however, is to first rewrite it as 
\begin{eqnarray} \label{X.2a}
L &:=& \frac{1}{2}M\dot{{\bf X}}^2 + \frac{1}{2}m_B(\dot{{\bf x}}_B - \dot{{\bf X}})^2 + \frac{1}{2}m_C(\dot{{\bf x}}_C - \dot{{\bf X}})^2 
+ \frac{[m_B(\dot{{\bf x}}_B - \dot{{\bf X}}) + m_C(\dot{{\bf x}}_C - \dot{{\bf X}})]^2}{2m_A}
\nonumber\\
&-& V_{AB}(|{\bf x}_B-{\bf x}_A|) - V_{AC}(|{\bf x}_C-{\bf x}_A|) - V_{BC}(|{\bf x}_C-{\bf x}_B|) \, .
\end{eqnarray}
It is then clear that under a Galilean boost,
\begin{eqnarray} \label{X.2b}
{\bf x}_A \mapsto {\bf x}_A - {\bf v}t \, , \quad 
{\bf x}_B \mapsto {\bf x}_B - {\bf v}t \, , \quad 
{\bf x}_C \mapsto {\bf x}_C - {\bf v}t \, , \quad 
({\bf X} \mapsto {\bf X} - {\bf v}t) \, , \quad 
\end{eqnarray}
$L$ transforms as
\begin{eqnarray} \label{X.2c}
L \mapsto L' := L - M\dot{{\bf X}}.{\bf v} + \frac{1}{2}Mv^2 \, . 
\end{eqnarray}
It is then also clear why $L^*$ (\ref{X.2}) is boost-invariant, and, therefore, exactly Galilean-invariant.

The Euler-Lagrange equations obtained from $L^{*}$ (\ref{X.2}), i.e., $m_J(\ddot{x}_J - \ddot{X}) = -\partial V/\partial {\bf x}_J$,  where $J \in \left\{A,B,C\right\}$ and $V := V_{AB} + V_{AC} + V_{BC}$, are of course different from their canonical counterparts, $m_J\ddot{x}_J = -\partial V/\partial {\bf x}_J$. 
But they become equivalent, {\it after} one imposes the conditions $P_i \approx 0$, which also require that $\dot{P}_i = M\ddot{X}_i \approx 0$, for all $t$. 
Hence, both formulations yield $\ddot{x}_J = -m_J^{-1}\partial V/\partial {\bf x}_J$, with $\sum_{J=1}^{N}\partial V/\partial {\bf x}_J = 0$, and, therefore, $\ddot{{\bf X}} = 0$. 
This is, of course, exactly the point: one would not want the imposition of the constraints $P_i \approx 0$ to change the equations of motion for the individual particles, since this would result in a distinct physical theory. 

The picture that emerges is as follows: starting with the canonical description of the system, (\ref{X.1})/(\ref{X.2a}), we perform the usual analysis to find that {\bf the canonical Lagrangian contains exactly three 
independent, non-Galilean invariant, configuration space variables}: $X^i(t) = v^{i}t + x_0^{i}$. 
With this knowledge, we see that it is `safe' to impose the constraints $P_i \approx 0$, first subtracting the term $(1/2)M\dot{{\bf X}}^2$ from $L$. 
This procedure is completely equivalent to keeping the $(1/2)M\dot{{\bf X}}^2$ term, using (\ref{X.1})/(\ref{X.2a}) to determine the Euler-Lagrange equations, then setting $v^i = 0$ (with a strong equality) in the {\it solutions} to the equations of motion.
 
Performing the latter gives $X^i(t) = x_0^{i}$, for all $t$, which ensures that the external inertial frame comoves with the classical centre of mass. 
But this does not completely fix the frame $O$, because the constant displacement ${\bf x}_0$ remains undetermined. 
Setting $P_i \approx 0$ also `fixes the external frame', to exactly the same degree.

{\bf This ``gauge fixing'', or rather ``frame fixing'', is therefore optional}. 
If one chooses not to perform it, one is led to the conclusion that the inertial motion of the centre of mass of a closed, classical, and Galilean-invariant system is {\it not} Galilean-invariant. 
It follows that it is unmeasurable, and therefore unphysical. 
If one chooses, instead, to impose the constraints $P_i \approx 0$, then one effectively assumes this conclusion, from the beginning. 
The only rationale for doing so is to invoke the results of the canonical analysis, which uses the regular Lagrangian $L$ (\ref{X.1}). 
All other equations of motion remain unaffected by this choice and the final result is the well-known one: generalising to the $N$-particle case, we see that this is described by $3N$ canonical degrees of freedom, but that only $3(N-1)$ of them are physical \cite{Kibble&Berkshire:2004,McCall:2011,Arnold:1978,Marsden&Ratiu:1999}. 

{\bf It is important to understand that this result does not follow from the imposition of constraints}. 
To say that it does would be a logical error, putting the inference before the predicate \cite{Logic}. 
Hence, while it is correct to say that ``we impose three first-class constraints because the system contains three unphysical degrees of freedom'', it is {\bf incorrect} to say that ``the system contains three unphysical degrees of freedom because we impose three first-class constraints''. 
This may be true mathematically, if one defines ``the system'' as $L^{*}$ (\ref{X.2}), but it is not true physically.
Indeed, this would be the wrong thing to do, if we did not know that it was `safe' to impose {\it exactly} these constraints, based on the canonical analysis.
\footnote{Another way to say this is that three canonical degrees of freedom can be gauged, because they are unphysical; they are not unphysical because they are gauge degrees of freedom.}

We believe that the perspective-neutral formalism gets the statements above back-to-front \cite{Vanrietvelde:2018pgb,Vanrietvelde:2018dit}. 
If one believes the second assertion, rather than the first, then one is led to believe that the three constraints {\bf must} be imposed, before the classical system can be quantised. 
(This is then just a part of {\it defining} what we mean by ``the classical system''; but its physical rationale is unclear.)
If one believes the first assertion, rather than the second, one is instead led to the opposite conclusion -- namely, that all $3N$ canonical degrees of freedom {\bf must} be quantised, before one decides which (if any) of these are unphysical in the quantum regime. 
This is not something that can be decided, {\it a priori}, from the within the {\it classical} regime. 

The relevant analysis must, therefore, be carried out at the quantum level. 
However, it would again be a logical error to equate this with the necessity of imposing a ``Dirac quantisation'' scheme \cite{Ita:2021cak,BarroseSa:2023rvz,Juhasz:2024twu,Kunstatter:1991ds,Schleich:1990gd,Plyushchay:1994pk,Barvinsky:1996cg,Shimizu:1996vf}. 
Hence, for completeness, we now consider the effects of imposing the constraints $P_i \approx 0$, prior to quantisation, before considering their Dirac-quantisation counterparts, $\widehat{P}_i \approx 0$. 
Along the way, we also compare both options with the canonical theory, in which the operators $\widehat{P}_i$ are time-independent, but not ``gauge-fixed''. 

The Lagrangian $L$ (\ref{X.1}) corresponds to the canonical Hamiltonian (\ref{2.1}), but $L^{*}$ (\ref{X.2}) corresponds to the modified Hamiltonian
\begin{eqnarray} \label{X.3}
H^{*} := H - \frac{P^2}{2M} \, .
\end{eqnarray}
This is equivalent to the {\bf internal energy} of the system, $U$, and its Galilean-invariance is easy to see if one performs the coordinate transformation (\ref{3.5}), 
giving
\begin{eqnarray} \label{X.4}
U({\bf x}_{\rm rel},\boldsymbol{\pi}_{\rm rel}) &=& \frac{\pi_{B|A}^2}{2m_B} + \frac{\pi_{C|A}^2}{2m_C} + \frac{(\boldsymbol{\pi}_{B|A} + \boldsymbol{\pi}_{C|A})^2}{2m_A}
\nonumber\\ 
&+&  V_{AB}(|{\bf x}_{B|A}|) + V_{AC}(|{\bf x}_{C|A}|) + V_{BC}(|{\bf x}_{C|A}-{\bf x}_{B|A}|) \, . 
\end{eqnarray}
Hamilton's equations for (\ref{X.4}) are the same as those for $H$ itself, but minus the equations for $\dot{{\bf X}}$ and $\dot{{\bf P}}$. 
Thus, the phase space variables ${\bf X}$ and ${\bf P}$, which are the only non-Galilean-invariant variables in the canonically-conjugate set $\left\{{\bf X},{\bf x}_{B|A},{\bf x}_{C|A};{\bf P},\boldsymbol{\pi}_{B|A},\boldsymbol{\pi}_{C|A}\right\}$, are also ``gauge fixed'' in the classical Hamiltonian. 
In the classical theory, removing the term $P^2/(2M)$ from $H$ (\ref{2.1}) makes sense -- or, at least, poses no fundamental problems -- because the $X^i$ coordinates are {\bf ignorable} \cite{Kibble&Berkshire:2004,McCall:2011,Arnold:1978,Marsden&Ratiu:1999}.

But is this still the case in quantum mechanics? 
We now show that it is not. 
The reason is that, in quantum theory, we have both hermitian operators, representing observables, {\it and} states on which these operators act. 
While the former must satisfy the Heisenberg equation -- which, in general, gives rise to operator-analogues of the classical equations of motion -- the latter must satisfy the Schr{\" o}dinger equation, which has no classical analogue  \cite{Rae:2002,Dirac:1958,Isham:1995,vonNeumann:1955}. 
These two pictures must be consistent with one another.

Yet it is straightforward to show that, in order for the Schr{\" o}dinger equation to be Galilean invariant, the quantum Hamiltonian {\bf cannot} be Galilean invariant. 
Specifically, under an {\bf active} boost belonging to the {\bf centrally-extended Galilean group}, given by the operator
\begin{eqnarray} \label{X.5}
\widehat{S}_{b}({\bf v}) := \exp\left[\frac{i}{\hbar}{\bf v}.(M\widehat{{\bf X}} - \widehat{{\bf P}}t)\right] \, , 
\end{eqnarray}
the canonical quantum Hamiltonian, $\widehat{H} = \widehat{P}^2/(2M) + \widehat{U}$, transforms as 
\begin{eqnarray} \label{X.6}
\widehat{H} \mapsto \widehat{H}' := \widehat{S}_{b}\widehat{H}\widehat{S}_{b}^{\dagger} = \widehat{H} - \widehat{{\bf P}}.{\bf v} + \frac{1}{2}Mv^2 \, \widehat{\mathbb{I}} \, . 
\end{eqnarray}
The time derivative term also transforms as 
\begin{eqnarray} \label{X.7}
i\hbar\frac{{\rm d}}{{\rm d}t} \mapsto  
i\hbar\frac{{\rm d}}{{\rm d}t} - \widehat{{\bf P}}.{\bf v} + \frac{1}{2}Mv^2 \, \widehat{\mathbb{I}} \, ,
\end{eqnarray}
so that the canonical Schr{\" o}dinger equation is {\bf exactly Galilean-invariant} \cite{Bargmann:1954gh,Levy-Leblond:1963qdx,Levy-Leblond:1967eic,Horzela:1991pa,Giulini:1995te,Csillag:2003}.
\footnote{Note that, although the $\frac{1}{2}Mv^2$ term in both (\ref{X.6}) and (\ref{X.7}) is the classical Bargmann cocycle \cite{Bargmann:1954gh}, the $\widehat{{\bf P}}.{\bf v}$ term in these equations differs from the classical generator, $M\dot{{\bf X}}.{\bf v} = {\bf P}.{\bf v}$ in (\ref{X.2c}). The two would be equivalent if and only if the relation $\widehat{{\bf P}} = {\bf P} \, \widehat{\mathbb{I}}$ held: this does not hold, however, because the quantum operator $\widehat{{\bf P}}$ admits superpositions of values, whereas the classical variable ${\bf P}$ is single-valued, for a given system. This is what separates the quantum from the classical case.} 
Only $\widehat{U}$ (not $\widehat{H}$) represents the {\bf observable energy} of the system, as this must be given by a Galilean-invariant operator. 
But this does not change the fact that we must use the {\it full} canonical Hamiltonian, which includes the {\bf quantised centre-of-mass energy}, in order to obtain Galilean-invariant time-evolution; the state $|\Psi_t\rangle := \exp[-(i/\hbar)(\widehat{P}^2/(2M) + \widehat{U})t]|\Psi_0\rangle$ transforms in the correct way, in order to ensure this, whereas $|\Psi_t\rangle := \exp[-(i/\hbar)\widehat{U}t]|\Psi_0\rangle$ does not.

We now compare this standard analysis with the ``gauge fixed'' quantum theory, proposed in the perspective-neutral framework \cite{Vanrietvelde:2018pgb}. 
In this formalism, the ``physical Schr{\" o}dinger equation'',
\begin{eqnarray} \label{X.8}
i\hbar\frac{{\rm d}}{{\rm d}t}|\Psi\rangle^{\rm phys} = \widehat{U}|\Psi\rangle^{\rm phys} \, , 
\end{eqnarray}
must be derived from the ``kinematical Schr{\" o}dinger equation'', 
\begin{eqnarray} \label{X.9}
i\hbar\frac{{\rm d}}{{\rm d}t}|\Psi\rangle^{\rm kin} = \widehat{H}|\Psi\rangle^{\rm kin} \, , 
\end{eqnarray}
by imposing the conditions $\widehat{P}_i|\Psi\rangle^{\rm phys} \overset{!}{=} 0$, to define the ``physical states'' of the system \cite{Vanrietvelde:2018pgb}. 

However, as shown in Sec. \ref{Sec.3}, the $\widehat{U}$ used in (\ref{X.8}) does {\it not} represent the internal energy of the system `seen' in particle $A$'s QRF, as the authors of \cite{Vanrietvelde:2018pgb} claim, but the internal energy of the system in the external, {\bf classical}, inertial frame $O$.
\footnote{The exact notation used in \cite{Vanrietvelde:2018pgb}, for the ``physical Schr{\" o}dinger equation'', is $i\partial_{t}|\psi\rangle_{BC|A} = \widehat{H}_{BC|A}|\psi\rangle_{BC|A}$, where $|\psi\rangle_{BC|A}$ is the one-dimensional counterpart of the state defined in (\ref{2.35}) and $\widehat{H}_{BC|A} = \widehat{p}_B^2 + \widehat{p}_C^2 + \widehat{p}_B\widehat{p}_C + \widehat{V}(q_B,q_C)$. These equations are numbered as (38) and (37), respectively, in the published version of that text. The Hamiltonian $\widehat{H}_{BC|A}$ is structurally equivalent to the one-dimensional restriction of $\widehat{U}$, obtained by quantising (\ref{X.4}), with unit particle masses $m_A = m_B = m_C = 1$. (Only the notation is different.)}  
(And, therefore, in any other classical inertial frame, $O'$, which is related to $O$ by a transformation of the centrally-extended Galilean group; see Appendices \ref{App.A} and \ref{App.B} for the generalisation of this result to the $N$-particle case.) 
This depends on only the translation and Galilean boost-invariant displacements and momenta, and, in our notation, the latter are written as $\widehat{\boldsymbol{\pi}}_{B|A}$ and $\widehat{\boldsymbol{\pi}}_{C|A}$. 
In the Hilbert space basis $|{\bf P}\rangle_{\rm CoM}|\boldsymbol{\pi}_{B|A}\rangle_{B|A}|\boldsymbol{\pi}_{C|A}\rangle_{C|A}$, these operators act trivially on the $|{\bf P}\rangle_{\rm CoM}$ subspace, but does this entitle us to impose $\widehat{{\bf P}} \approx {\bf 0}$ and to then perform the ``constraint trivialisation'' procedure (\ref{2.37})?

We see no physical or mathematical reason to impose this condition, since, in the canonical basis $|{\bf p}_A\rangle_{A}|{\bf p}_{B}\rangle_{B}|{\bf p}_{C}\rangle_{C}$, $\widehat{\boldsymbol{\pi}}_{B|A} := \widehat{{\bf p}}_B - (m_B/M)\widehat{{\bf P}}$ and $\widehat{\boldsymbol{\pi}}_{C|A} := \widehat{{\bf p}}_C - (m_C/M)\widehat{{\bf P}}$ act nontrivially across all three subspaces. 
If we insist on ``trivialising'' one of them, we are forced to {\it forbid} the net momentum of the system from existing in a superposition of states, relative to $O$'s sharp classical frame. 
In other words, we are forced to impose a {\bf superselection rule} (SSR) on the net momentum of a closed quantum system \cite{SSR:QM-Compendium,Giulini:2007fn}.

We have three comments to make, regarding this possibility: 
Firstly, we note that, as shown in \cite{Bargmann:1954gh,Levy-Leblond:1963qdx,Levy-Leblond:1967eic,Horzela:1991pa,Giulini:1995te,Csillag:2003}, such a restriction is certainly not {\it necessary}, in order to impose exact Galilean invariance in canonical quantum mechanics. 
Given two inequivalent physical theories, one of which is gauge-invariant and one of which is gauge-fixed, we consider it far safer to choose the {\bf manifestly gauge-invariant} option \cite{GaugeTheory:Healy2007}. 
Second, it is in fact well-known that there is no superselection rule for the net momentum of a closed system, in canonical quantum theory \cite{SSR:QM-Compendium,Giulini:2007fn}.
\footnote{See, for example, the review by Giulini, in which it is stated that ``there are many additive conserved quantities, like momentum and angular momentum, for which certainly no SSRs is at work'' \cite{SSR:QM-Compendium,Giulini:2007fn}.} 
The results of \cite{Vanrietvelde:2018pgb}, and the relational program in general, therefore conflict with the consensus view of this topic, prior to the original GCB paper \cite{Giacomini:2017zju}. 
(Likewise, there is no superselection rule for the net angular momentum \cite{SSR:QM-Compendium,Giulini:2007fn}, which conflicts with the claims made in \cite{Vanrietvelde:2018dit}.) 
Third, if one accepts that the operators $\widehat{\boldsymbol{\pi}}_{B|A}$ and $\widehat{\boldsymbol{\pi}}_{C|A}$ -- or, equivalently, $\widehat{{\bf p}}_B$ and $\widehat{{\bf p}}_C$, using the notation in \cite{Vanrietvelde:2018pgb,Vanrietvelde:2018dit} -- must be defined as projections over nine independent quantum mechanical degrees of freedom, rather than six (in order to avoid the SSR), then it is straightforward to see that (\ref{X.8}) is {\bf not Galilean invariant}. 
Specifically, the right-hand side of this equation is manifestly invariant under Galilean boosts, but the left-hand side transforms as shown in (\ref{X.7}). 
Therefore, the ``physical Schr{\" o}dinger equation'' of \cite{Vanrietvelde:2018pgb} is not boost-invariant, and cannot represent the time-evolution of {\bf physical states}.  

It is helpful to recap the logical steps of our argument, point-by-point. 
These are as follows:
\begin{enumerate}

\item If we insist that $\widehat{U}$ in (\ref{X.8}) is the Hamiltonian in particle $A$’s frame, and that all variables therein are physical, and therefore Galilean-invariant -- as asserted in \cite{Vanrietvelde:2018pgb} -- then canonical Galilean invariance is actually violated, since this is tantamount to asserting that it holds in an accelerated frame. 
Hence, the state denoted $|\Psi\rangle^{\rm phys}$ cannot represent a {\bf physical state} of a viable quantum theory. 

\item We must, therefore, abandon the idea that $\widehat{U}$ represents the Hamiltonian in $A$’s QRF; we must recognise, instead, that it represents the internal energy of the three-particle system, as seen from the external inertial frame $O$. 
This is a {\bf classical} frame of reference.

\item Given that $\widehat{U}$ is the internal energy of the three-particle system, in the classical frame $O$, it must act on the canonical, non-reduced, Hilbert space. 
This corresponds to nine independent quantum mechanical degrees of freedom \cite{Rae:2002,Dirac:1958,Isham:1995,vonNeumann:1955}. 

\item $\widehat{U}$ is, therefore, a Galilean-invariant quantity in canonical quantum mechanics \cite{Bargmann:1954gh,Levy-Leblond:1963qdx,Levy-Leblond:1967eic,Horzela:1991pa,Giulini:1995te,Csillag:2003}.

\item Because $\widehat{U}$ is a Galilean-invariant quantity in canonical quantum mechanics, (\ref{X.8}) is not a Galilean invariant equation. 
Hence, the state denoted $|\Psi\rangle^{\rm phys}$ cannot represent a {\bf physical state} of a viable quantum theory. 

\end{enumerate}
We are thus led into a catch-22 situation: 
If $\widehat{U}$ is the Hamiltonian in $A$'s QRF, Galilean symmetry is violated, and $|\Psi\rangle^{\rm phys}$ is not a physical state. 
If, alternatively, $\widehat{U}$ is the Hamiltonian in the CRF $O$, then Galilean symmetry is violated, and $|\Psi\rangle^{\rm phys}$ is not a physical state. 
The only solution, that we can see, is to replace $\widehat{U}$ with the full Hamiltonian $\widehat{H} = \widehat{P}^2/(2M) + \widehat{U}$, in (\ref{X.8}); i.e., to restore canonical, Galilean-invariant, quantum theory, in place of the proposed ``relational'' model \cite{Vanrietvelde:2018pgb,Vanrietvelde:2018dit}.

In summary, we have shown that in classical mechanics one may subtract the centre-of-mass energy from the canonical Lagrangian, reformulating the canonical theory as a gauge theory in which the centre-of-mass variables represent gauge degrees of freedom. 
We may therefore perform a gauge-fixing procedure, which sets $P_i \approx 0$ and $X^i(0) = X_0^i \approx 0$, and the resulting reduced Hamiltonian is Galilean-invariant. 
(Note that this is different to the ``gauge-fixing'' performed in\cite{Vanrietvelde:2018pgb}, which sets the three canonical degrees of freedom, associated an arbitrary interacting particle, $I$, equal to zero `by hand', by imposing both $P_i \approx 0$ {\it and} $x_I^i \approx 0$.) 
However, in the quantum realm, the time-evolution of quantum states is given by the Schr{\" o}dinger equation, which has no classical counterpart \cite{Rae:2002,Dirac:1958,Isham:1995,vonNeumann:1955}. 
In this equation, enforcing Galilean invariance in the Hamiltonian actually {\bf violates} Galilean invariance in the equation of motion. 
This is because the $\widehat{P}^2/(2M)$ term in the (quantised) canonical Hamiltonian, $\widehat{H} = \widehat{P}^2/(2M) + \widehat{U}$, is needed, in order to counter-balance the transformation of the time-derivative term under a general Galilean boost. 
It is, therefore, not possible to `kill' the {\bf quantised} centre-of-mass degrees of freedom, by imposing the conditions $\widehat{P}_i|\Psi\rangle \approx 0$ ($\widehat{P}_i|\Psi\rangle \overset{!}{=} 0$), without violating the Galilean invariance of the theory. 
This is why there is no superselection rule for the net momentum of a closed quantum system \cite{SSR:QM-Compendium,Giulini:2007fn}.

\section{Detectables versus observables: the relation between QRFs and GURs} \label{Sec.4}

It is straightforward to show that certain operators, which are themselves {\bf not} Galilean-invariant, nevertheless possess Galilean-invariant variances, in both position space and momentum space. 
These include the canonical phase space coordinates of an $N$-particle system, $\left\{\widehat{{\bf x}}_I,\widehat{{\bf p}}_I\right\}_{I=1}^{N}$, and the centre of mass coordinates $\left\{\widehat{{\bf X}},\widehat{{\bf P}}\right\}$.
\footnote{For rotations, the meaning of this invariance is that $\Delta_{\Psi}x_{J}^{i'} = \Delta_{\Psi}R^{i'}{}_{i}x_{J}^{i} = \Delta_{\Psi}x_{J}^{i}$ and $\Delta_{\Psi}p_{Ji'} = \Delta_{\Psi}R^{i}{}_{i'}p_{Ji} = \Delta_{\Psi}p_{Ji}$, i.e., after the rotation, the wave packet spread in the $x^{i'}$-direction is equal to its spread in the $i$-direction of the unrotated coordinate system. Clearly, this means that the wave packet maintains both its size and its shape, in both position and momentum space. (For translations and Galilean boosts, the coordinate axes still `point in the same direction', both before and after the transformation.)} 
According to the standard logic of canonical, Galilean-invariant, quantum mechanics, these variances should be {\bf measurable} \cite{Bargmann:1954gh,Levy-Leblond:1963qdx,Levy-Leblond:1967eic,Horzela:1991pa,Giulini:1995te,Csillag:2003}. 
But how can this be? 
How can we compile statistics, about repeated measurements of $\widehat{{\bf X}}$, if $\widehat{{\bf X}}$ itself cannot be measured?

We now show that there is a way to measure the variances $\left\{(\Delta_{\Psi}x_I^i)^2,(\Delta_{\Psi}p_{Ii})^2\right\}_{I=1}^{N}$ and $\left\{(\Delta_{\Psi}X^i)^2,(\Delta_{\Psi}P_{i})^2\right\}$, for all $i \in \left\{x,y,z\right\}$, without measuring any of the individual operators, $\left\{\widehat{x}_I^i,\widehat{p}_{Ii}\right\}_{I=1}^{N}$ or $\left\{\widehat{X}^i,\widehat{P}_{i}\right\}$. 
For simplicity, we return again to an example three-particle system, but our results hold more generally, for arbitrary $N$. 
For even greater simplicity, we consider a state that is separable, from the perspective of the external inertial frame $O$,
\begin{eqnarray} \label{4.1}
|\Psi\rangle &:=& \int \psi({\bf x}_A)\phi({\bf x}_B)\xi({\bf x}_C) |{\bf x}_A\rangle_A|{\bf x}_B\rangle_B|{\bf x}_C\rangle_C {\rm d}^3x_A{\rm d}^3x_B{\rm d}^3x_C
\nonumber\\
&:=& \int \tilde{\psi}({\bf p}_A)\tilde{\phi}({\bf p}_B)\tilde{\xi}({\bf p}_C)|{\bf p}_A\rangle_A|{\bf p}_B\rangle_B|{\bf p}_C\rangle_C {\rm d}^3p_A{\rm d}^3p_B{\rm d}^3p_C \, . 
\end{eqnarray}
This means that it will also be separable, from the perspective of any other inertial frame, $O'$, which is related to $O$ by a transformation of the centrally-extended Galilean group \cite{Bargmann:1954gh,Levy-Leblond:1963qdx,Levy-Leblond:1967eic,Horzela:1991pa,Giulini:1995te,Csillag:2003}. 
In this scenario, it is physically meaningful to associate the individual wave functions $\psi({\bf x}_A)$, $\phi({\bf x}_B)$, and $\xi({\bf x}_C)$, with the particles $A$, $B$ and $C$, respectively. 
For non-separable states, this will no longer be the case, but it is straightforward to generalise our results for arbitrary wave functions. 
For now, however, we consider only functions of the form $\Psi({\bf x}_A,{\bf x}_B,{\bf x}_C) := \psi({\bf x}_A)\phi({\bf x}_B)\xi({\bf x}_C)$ 
and $\tilde{\Psi}({\bf p}_A,{\bf p}_B,{\bf p}_C) := \tilde{\psi}({\bf p}_A)\tilde{\phi}({\bf p}_B)\tilde{\xi}({\bf p}_C)$.

Using standard manipulations, we then obtain 
\begin{eqnarray} \label{4.2}
(\Delta_{\Psi}x^i_{B|A})^2 = (\Delta_{\phi}x^i_{B})^2 + (\Delta_{\psi}x^i_{A})^2 \, , 
\end{eqnarray}
for the variance of the observable, Galilean-invariant, and relational displacement $\widehat{x}^i_{B|A} := \widehat{x}^i_{B} - \widehat{x}^i_{A}$. 
Similarly, the variance of particle $B$'s momentum, as `seen' by particle $A$, $\widehat{p}_{B:Aj} := \widehat{p}_{Bj} - (m_B/m_A)\widehat{p}_{Aj}$, is given by
\begin{eqnarray} \label{4.3}
(\Delta_{\Psi}p_{B:Aj})^2 = (\Delta_{\phi}p_{Bj})^2 + \left(\frac{m_B}{m_A}\right)^2(\Delta_{\psi}p_{Aj})^2 \, .
\end{eqnarray}
Using the canonical uncertainty relations for particle $B$’s wave function, 
\begin{eqnarray} \label{4.4}
\Delta_{\phi}x^i_{B}\Delta_{\phi}p_{Bj} \geq \frac{\hbar}{2}\delta^{i}{}_{j} \, , 
\end{eqnarray}
and relabelling the uncertainties for particle $A$ such that 
\begin{eqnarray} \label{4.5}
\Delta_{\psi}x^i_{A} =: \sigma_A^i \, , \quad \Delta_{\psi}p_{Aj} =: \tilde{\sigma}_{Aj} \, , 
\end{eqnarray}
we obtain the GUR
\begin{eqnarray} \label{4.6}
(\Delta_{\Psi}x^i_{B|A})^2(\Delta_{\Psi}p_{B:Aj})^2 &\geq& \left(\frac{\hbar}{2}\right)^2(\delta^{i}{}_{j})^2 + \left(\frac{m_B}{m_A}\right)^2(\tilde{\sigma}_{Aj})^2(\Delta_{\Psi}x^i_{B|A})^2 
\nonumber\\
&+& (\sigma_A^i)^2(\Delta_{\Psi}p_{B:Aj})^2  - \left(\frac{m_B}{m_A}\right)^2(\sigma_A^i)^2(\tilde{\sigma}_{Aj})^2 \, . 
\end{eqnarray}

This expression contains two uncertainties, $\Delta_{\Psi}x^i_{B|A}$ and $\Delta_{\Psi}p_{B:Aj}$, which are {\bf directly measurable}. 
That is, we can perform repeated measurements, on an ensemble of identically prepared systems, of the Galilean-invariant observables $\widehat{x}^i_{B|A}$ and $\widehat{p}_{B:Aj}$, then compute $\Delta_{\Psi}x^i_{B|A}$ and $\Delta_{\Psi}p_{B:Aj}$ in the usual way. 
It also contains two uncertainties, $\Delta_{\psi}x^i_{A}$ and $\Delta_{\psi}p_{Aj}$, that are {\bf not directly measurable}, in general. 
(See the comments below (\ref{4.7}), regarding the restriction to the two-particle case, and/or non-separable states with $N \geq 2$.) 
These quantify the widths of particle $A$'s wave function, in real space and momentum space. 
Because neither $\widehat{x}^i_{A}$ nor $\widehat{p}_{Aj}$ is invariant under translations, or Galilean boosts, we cannot perform single-shot experiments to determine the eigenvalues $x^i_{A}$ and $p_{Aj}$. 
This means that we cannot compile statistics to determine $\Delta_{\psi}x^i_{A}$ and $\Delta_{\psi}p_{Aj}$. 
These quantities have been relabelled as $\sigma_A^i $ and $\tilde{\sigma}_{Aj}$, respectively, in order to emphasise this distinction.

But we do not need to compile statistics, about repeated measurements of $\widehat{x}^i_{A}$ nor $\widehat{p}_{Aj}$, in order to determine the spreads (\ref{4.5}). 
We need only compile statistics for repeated measurements of the translation- and boost-invariant observables, $\widehat{x}^i_{B|A}$ and $\widehat{p}_{B:Aj}$, then plot the variance $(\Delta_{\Psi}x^i_{B|A})^2$ as a function of $(\Delta_{\Psi}p_{B:Aj})^2$, or vice versa. 
If our measurements are sufficiently precise, the quantum fluctuations of the chosen reference particle -- in this case, particle $A$ -- will manifest in the presence of additional, non-Heisenberg, terms in the GUR (\ref{4.6}). 
In this relation, the quantities $(\sigma_A^i)^2$ and $(\tilde{\sigma}_{Aj})^2$ act as (possibly time-dependent) multiplicative factors, which determine the magnitudes of the non-Heisenberg contributions. 
Practically, both can be extracted from the Galilean-invariant data by a curve-fitting procedure, using an ansatz of the form (\ref{4.6}). 
Performing a similar analysis for all three particles, the sets $\left\{\sigma_A^i,\sigma_B^i,\sigma_C^i \right\}_{i=1}^{3}$ and $\left\{\tilde{\sigma}_{Ai},\tilde{\sigma}_{Bi},\tilde{\sigma}_{Ci} \right\}_{i=1}^{3}$ can then be used to determine $\Delta_{\Psi}X^i$ and $\Delta_{\Psi}P_i$. 

In fact, for separable states with $N \geq 3$, both the position and momentum space spreads of $A$'s QRF can be computed directly, from the Galilean-invariant data, using 
\begin{eqnarray} \label{4.7}
(\sigma_A^i)^2 = \frac{1}{2}[(\Delta_{\Psi}x_{B|A}^{i})^2 + (\Delta_{\Psi}x_{C|A}^{i})^2 - (\Delta_{\Psi}x_{C|B}^{i})^2] \, , 
\nonumber\\
(\tilde{\sigma}_{Aj})^2 = \frac{(\Delta_{\Psi}p_{C:Aj})^2 - (\Delta_{\Psi}p_{C:Bj})^2 + (m_C/m_B)^2(\Delta_{\Psi}p_{B:Aj})^2}{2(m_C/m_A)^2} \, ,
\end{eqnarray}
although this is not possible when $N = 2$, or when the state is non-separable, for any $N$. 
However, the important point is that the standard Heisenberg relations, which hold for the non-Galilean-invariant position and momentum operators, differ from that for the Galilean-invariant pair, $\widehat{x}^i_{B|A}$ and $\widehat{p}_{B:Aj}$ (\ref{4.6}). 
(Here, we again use the term `Galilean-invariant' rather loosely, subject to the caveat about rotations in footnote 21, and regard all translation- and boost-invariant quantities as bona fide observables, measurable in single-shot experiments.)

The non-Heisenberg terms in the GUR (\ref{4.6}) are due to the quantum fuzziness of the frame itself, which prevents particle $A$ from `seeing' the remaining quantum subsystems in an undistorted, `objective', way. 
The objective viewpoint, in which the Heisenberg bound is visible, is accessible only to the abstract spacetime frame $O$ -- which, we have emphasised, cannot correspond to any embodied, physical, and {\it quantum mechanical} observer. 
Thus, QRFs imply GURs, and the presence of the latter is the `smoking gun', which indicates the existence of the former. 
Near-future experiments may, in principle, confirm these predictions. 

While it is tempting to think that the Heisenberg bound may also be accessible, in particle $A$'s QRF, if the non-Heisenberg terms in (\ref{4.6}) sum to zero, it is straightforward to show -- using the commutation relation between $\widehat{x}_{B|A}^{i}$ and $\widehat{p}_{B:Aj}$, plus the standard Schr{\" o}dinger-Robertson relation \cite{Isham:1995} -- that these terms must sum to a minimum positive value of $(\hbar/2)^2[2(m_B/m_A) + (m_B/m_A)^2]$. 
This is what distinguishes the quantum scenario from the classical case: although the classical ensemble variances for measurements of $x_{B|A}^i$ and $p_{B:Aj}$ are also Galilean-invariant, both can be driven to zero, for perfect measurements on ensembles of identically-prepared systems. 
Hence, the classical theory supports no irreducible detectable structure, related to the choice of ``reference particle''. 

This structure gives a physical role to the quantised centre-of-mass coordinates, since the canonical Hilbert space basis $|{\bf x}_A\rangle_{A}|{\bf x}_{B}\rangle_{B}|{\bf x}_{C}\rangle_{C}$ is unitarily equivalent to the basis $|{\bf X}\rangle_{\rm CoM}|{\bf x}_{B|A}\rangle_{B|A}|{\bf x}_{C|A}\rangle_{C|A}$. 
(See. App. \ref{App.B.2}.) 
It follows, automatically, that the number of {\bf physical degrees of freedom}, required to characterise the {\bf quantum three-particle state}, is $9$ 
(not $6$ \cite{Giacomini:2017zju,Vanrietvelde:2018pgb}, or fewer \cite{Vanrietvelde:2018dit}). 
Generalising to the $N$-particle case, we see that the relevant distinction is between $3(N-1)$ physical degrees of freedom, in the classical limit, versus $3N$ physical degrees of freedom, in the quantum regime, as announced in the Abstract, and Introduction.

We call operators that are not Galilean-invariant, and therefore not measurable, but whose quantum mechanical spreads {\it are} Galilean-invariant, and therefore measurable, {\bf detectables} as opposed to {\bf observables}. 
It is important to understand that the quantum mechanical degrees of freedom, to which these operators correspond, are {\bf physical}, because they have measurable physical consequences. 
These consequences arise from the fact that the Galilean-invariant quantum mechanical spreads obey GURs carrying an irreducible non-classical floor; a structure that classical ensemble variances, which can be jointly driven to zero, do not possess. 
Despite this, {\bf all} relational QRF models discard these physical (but non-relational) degrees of freedom, as the very starting point of their respective constructions \cite{Giacomini:2017zju,Castro-Ruiz:2019nnl,delaHamette:2020dyi,Krumm:2020fws,Giacomini:2021gei,Castro-Ruiz:2021vnq,delaHamette:2022cka,Kabel:2024lzr,Giacomini:2020ahk,Giacomini:2021aof,Cepollaro:2021ccc,Giacomini:2018gxh,Mikusch:2021kro,Ballesteros:2020lgl,Ballesteros:2025ypr,Vanrietvelde:2018pgb,Hohn:2018iwn,Hohn:2018toe,delaHamette:2021oex,Hoehn:2019fsy,AliAhmad:2021adn,Vanrietvelde:2018dit,Hoehn:2023ehz,Hoehn:2023axh,DeVuyst:2025ezt,Hoehn:2020epv,Apadula:2022pxk,Kabel:2022cje,Carrozza:2021gju,Goeller:2022rsx,Kabel:2023jve,Chen:2026kui}.

Finally, we note that the GUR (\ref{4.6}) holds, regardless of whether we interpret $\widehat{\dot{{\bf p}}}_A = m_A\widehat{\ddot{{\bf x}}}_A$ as a genuine (quantised) force -- as it would be reckoned in $O$'s inertial frame -- or as the quantum analogue of the classical pseudo-force, which acts in the accelerated frame defined by $A$'s trajectory, in the classical limit of the quantum theory \cite{Kibble&Berkshire:2004,McCall:2011,Arnold:1978,Marsden&Ratiu:1999}. 
In the quantum regime, the position and momentum space widths of particle $A$'s wave function are the same, irrespective of whether they are determined by measurements made in a classical inertial frame, or directly in $A$'s quantum frame of reference. 
Although the QRF $A$ and the CRF $O$ will obtain different outcomes, for the results of many physical measurements, they will not disagree about the GUR (\ref{4.6}). 
In this sense, these results really do allow us to answer the question ``what does the Universe look like, from particle $A$'s perspective?'', and to ``jump'', {\bf literally}, into $A$'s QRF. 

This is impossible, in the relational models, in which $A$'s degrees of freedom simply don't exist \cite{Giacomini:2017zju,Vanrietvelde:2018pgb}. 
In our model, the Heisenberg limit -- in which the three-particle system reduces, {\it effectively}, to a two-particle system -- is recovered only as $m_A \rightarrow \infty$. 
For arbitrarily large values of $m_A$, the width of particle $A$'s wave packet can be made arbitrarily small, in both position space and velocity space, so that we may impose the simultaneous conditions 
\begin{eqnarray} \label{4.8}
\sigma_A^i = \Delta_{\psi}x^i_{A} \rightarrow 0 \, , \quad  \frac{\tilde{\sigma}_{Ai}}{m_A} = \frac{\Delta_{\psi}p_{Ai}}{m_A} = \Delta_{\psi}v_{Ai} \rightarrow 0 \, , 
\end{eqnarray}
for all $i \in \left\{x,y,z\right\}$. 
The GUR (\ref{4.6}) then reduces to the standard Heisenberg uncertainty relation, $\Delta_{\Psi}x^i_{B|A}\Delta_{\Psi}p_{B:Aj} \rightarrow \Delta_{\phi}x^i_{B}\Delta_{\phi}p_{Bj} \geq (\hbar/2)\delta^{i}{}_{j}$, and $A$'s {\bf accelerated QRF} reduces to a {\bf classical inertial frame}.

In fact, it is not necessary to consider the quantum theory of the perspective-neutral formalism, and related QRF models, in order to show that the ``relational'' approach contradicts observed experimental data. 
To illustrate this point, a good example system in the classical regime is comprised of three equal masses, two of which, $A$ and $B$, interact and rotate in circles about their common centre of mass. 
The third particle, $C$, does not interact with either $A$ or $B$ and is at rest with respect to the centre of mass of the $AB$ subsystem, maintaining a fixed displacement from it, perpendicular to the plane in which $A$ and $B$ rotate. 
(The shape traced out by the motion of the inter-particle displacements is then a cone, in which the diameter of the circle is $|{\bf x}_B-{\bf x}_A|$ and the side-length from the apex to a point on the circumference of the circle is $|{\bf x}_C-{\bf x}_A| = |{\bf x}_C-{\bf x}_B|$.) 
In the limit that the very weak gravitational attraction between the particles can be neglected, such a system could be comprised of three classical microspheres: one with a small positive charge, $+\delta$ ($A$), one with a small negative charge $-\delta$ ($B$), and one which is charge-neutral ($C$). 

In the ``relational'' models, the coordinate system could be chosen such that ${\bf x}_A = {\bf 0}$ and ${\bf x}_C-{\bf x}_A = {\bf x}_C$ is the $z$-axis of the coordinates whose origin coincides with the trajectory of $A$. 
In this picture, both particles $B$ and $C$ remain at fixed coordinate values, in $A$’s ``relational'' reference frame, just as they would if all three particles were at rest, with respect to one another (and with respect to the external reference frame O), with their relative positions forming a right-angled triangle.

In short, there is no way for the relational models to account for the rotational motion of particles $A$ and $B$, because they make no distinction between inertial and accelerated motion. 
Practically, this means that A’s proper acceleration, which can be expressed as $-\partial V({\bf x}_A,{\bf x}_B,{\bf x}_C)/\partial {\bf x}_A$ in the coordinate system of the external frame $O$, must be replaced by either $-\partial V({\bf x}_A,{\bf x}_B,{\bf x}_C)/\partial {\bf x}_A|_{{\bf x}_A={\bf 0}}$, or, worse still, $-\partial V({\bf x}_A={\bf 0},{\bf x}_B,{\bf x}_C)/\partial {\bf x}_A ={\bf 0}$, in the ``reduced'' picture of the relational framework. 
In either case, physical information, regarding this proper acceleration, is in general lost.

Considerations of this kind become especially acute if one considers the even more extreme “relational” model, proposed in \cite{Vanrietvelde:2018dit}, in which it is claimed that rotational invariance necessitates the imposition of the constraint ${\bf L} \approx {\bf 0}$, for the net angular momentum of a closed system, about its centre of mass. 
With this constraint in place, the first system described is considered unphysical, and its existence is proscribed by construction. 

\section{Discussion} \label{Sec.5}

Our criticisms of the relational QRF literature can be summarised, not-so-concisely, as follows:

There is confusion regarding the difference between an accelerated frame of reference, defined by the trajectory of a particle, and an external inertial frame, in which the Hamiltonian and equations of motion are expressed in terms of relational variables. 
In the relational models, an attempt is made to implement the former, at the classical level, before promoting the relevant variables to operators \cite{Giacomini:2017zju,Vanrietvelde:2018pgb}. 
But this step is implemented incorrectly, since no account is taken of the relevant {\bf pseudo-forces}, introduced by inertial effects. 
The result is a formalism that contains ``unmoved movers'' \cite{Aristotle,Aristotle:SEP}, which violate the law of action and reaction -- and, hence, all conservation laws due to Galilean symmetry \cite{Bargmann:1954gh,Levy-Leblond:1963qdx,Levy-Leblond:1967eic,Horzela:1991pa,Giulini:1995te,Csillag:2003}. 
Mathematically, this formalism is equivalent to taking the Hamiltonian for an open system of $N-1$ particles, in the presence of {\bf external potential} that is not back-reacted upon, even while sourcing the particles' accelerations, and reinterpreting it as the ``relational'' description of an $N$-particle state \cite{Giacomini:2017zju}. 
The same basic error is implemented, by manipulations at the quantum level, in the perspective neutral formalism \cite{Vanrietvelde:2018pgb}.

There is confusion regarding the nature and structure of symmetry transformations. 
For example, in \cite{Giacomini:2017zju}, operators representing so-called ``quantum superpositions of Galilean symmetries'' -- at least some of which, according to our reading of this work, are intended to be applied {\bf passively}
\footnote{These are the operators corresponding to ``superpositions of Galilean translations'' and ``superpositions of Galilean boosts'' \cite{Giacomini:2017zju}.} 
-- are listed in a table alongside their ``classical'' counterparts. 
But this is misleading, because Galilean symmetries -- like {\bf all} symmetry transformations -- are meaningful only when applied {\bf actively} \cite{Bargmann:1954gh,Levy-Leblond:1963qdx,Levy-Leblond:1967eic,Horzela:1991pa,Giulini:1995te,Csillag:2003}. 
In quantum mechanics, all operator expectation values are invariant under passive transformations, {\bf by construction}, since these transform both states and operators such that 
$\langle \Psi | \widehat{O} |\Psi\rangle \mapsto \langle \Psi | \widehat{S}^{-1}(\widehat{S}\widehat{O}\widehat{S}^{-1})\widehat{S} |\Psi\rangle$. 
This form of ``invariance'' proves nothing and does not, in any way, indicate that the operator $\widehat{S}$ represents a symmetry of the theory. 
Only when $\widehat{S}$ is applied {\bf actively}, yielding $\langle \Psi | \widehat{O} |\Psi\rangle \mapsto \langle \Psi | \widehat{S}^{-1}\widehat{O}\widehat{S} |\Psi\rangle =\langle \Psi | \widehat{O} |\Psi\rangle$, where $\widehat{O}$ is an observable, is the existence of a symmetry demonstrated. 
Conversely, when $\langle \Psi | \widehat{O} |\Psi\rangle \mapsto \langle \Psi | \widehat{S}^{-1}\widehat{O}\widehat{S} |\Psi\rangle \neq \langle \Psi | \widehat{O} |\Psi\rangle$, $\widehat{S}$ is {\bf not} a symmetry of the theory.

In general, the relational QRF literature makes no clear distinction between active and passive transformations \cite{Apadula:2022pxk,delaHamette:2020dyi,delaHamette:2021oex,delaHamette:2022cka,Kabel:2024lzr,Ballesteros:2020lgl,Ballesteros:2025ypr,Vanrietvelde:2018pgb,Vanrietvelde:2018dit} -- or, worse still, claims that invariance under passive transformations demonstrates the existence of ``new'' symmetries \cite{Giacomini:2017zju,Krumm:2020fws,Ballesteros:2020lgl,Ballesteros:2025ypr}. 
In addition, it often conflates two distinct concepts, namely, the {\bf covariance group} of a physical theory and its {\bf symmetry group} \cite{Norton:1993eqc}. 
We will have more to say about these issues, in an upcoming work. 

A related point concerns the failure to distinguish between the symmetries of a physical system and the symmetries of its particular configuration \cite{Vanrietvelde:2018pgb,Vanrietvelde:2018dit}. 
In a popular textbook on group theory \cite{Jones:1998}, the author H. F. Jones urges students, in the opening chapter, not to fall into this trap. 
He cites the example of the solar system, whose planar symmetry is a result of the conservation of angular momentum -- which, in turn, results from the full three-dimensional rotation-invariance of the Newtonian spacetime \cite{SEP:Newtonian_Space-time}. 
Yet, in work on the perspective-neutral formalism, in particular, there is much talk about ``colineations'' and ``total collisions'', etc., and about how the ``gauge-invariant phase space'' must exclude such ``pathological configurations'' \cite{Vanrietvelde:2018pgb,Vanrietvelde:2018dit}.

To us, such considerations are irrelevant to the question ``what are the symmetries of a theory?''. 
These are the symmetries of the equations of motion and the chosen set of observables \cite{Noether'sTheorems,Weyl:1928,Bargmann:1948ck,Bargmann:1954gh,Levy-Leblond:1963qdx,Levy-Leblond:1967eic,Horzela:1991pa,Giulini:1995te,Csillag:2003}. 
The symmetries of individual configurations are the symmetries of particular {\bf solutions} to the equations of motion, at particular moments in time \cite{Jones:1998}. 
As Jones points out, using the solar system example, these are two very different things. 
In canonical theories, only the latter have fundamental significance, in the sense that they are capable generating ``gauge transformations'', which restrict the number of physical degrees of freedom \cite{Dirac-CHS:1958,Dirac-CHS:1964,Henneaux-Teitelboim:1992,Rothe-RotheCHS:2010}. 

But this is no longer the case in relational models, where it is claimed, for example, that a translation- and rotation-invariant system of $N$ classical particles possesses only $3N-6$ physical degrees of freedom, because it possesses only $3N-6$ geometrically-independent relative distances, $|{\bf x}_J - {\bf x}_I|$, $I,J \in \left\{1,2, \dots N\right\}$ \cite{Vanrietvelde:2018dit}. 
An additional three degrees of freedom, beyond those associated with the inertial motion of the centre of mass, are then removed from the already-reduced Lagrangian of \cite{Vanrietvelde:2018pgb}. 
Specifically, the net rotational energy of the system, about its centre of mass, is also subtracted from the classical Lagrangian, along with the centre-of-mass energy, in order to bring the total number of degrees of freedom `in line' with relational thinking about the possible configurations of the particles in three-dimensional space \cite{Vanrietvelde:2018dit}. 
Clearly, this approach is not compatible with canonical classical mechanics \cite{Kibble&Berkshire:2004,McCall:2011,Arnold:1978,Marsden&Ratiu:1999}.
\footnote{The relations specifying the conservation of the net linear momentum -- mentioned in the Abstract as the cause of three unphysical degrees of freedom, in the canonical description of the system -- are distinguished from those specifying the conservation of the net angular momentum by being constant-coefficient and integrable \cite{Arnold:1978,Marsden&Ratiu:1999}. (By contrast, ${\bf L} := \sum_{J=1}^{N}m_J({\bf x}_J-{\bf X}) \times (\dot{{\bf x}}_J - \dot{{\bf X}})$ has configuration-dependent coefficients and $L_i \approx 0$ is not reducible to holonomic form. See App. \ref{App.C.1}.) This is why the latter do not, and, in fact, {\it cannot} function in the same way as the former, by restricting the number of physical degrees still further. In other words, it is why the rotational degrees of freedom cannot be ``gauged'', as the theory proposed in \cite{Vanrietvelde:2018dit} requires, without destroying physical information about the {\it classical system}, as well as its quantum counterpart. Imposing $P_i \approx 0$ places no restrictions on the Galilean-invariant variables, but imposing $L_i \approx 0$ restricts the system to states with zero net angular momentum. Such a restriction is not required by rotational symmetry \cite{Kibble&Berkshire:2004,McCall:2011,Arnold:1978,Marsden&Ratiu:1999} and artificially forbids configurations with $L_i \neq 0$. In these states, at least some particles {\it must} be accelerating, whereas this may or may not be the case for systems with $L_i = 0$. Only a theory that treats accelerated and inertial motion as physically equivalent, like the relational models \cite{Giacomini:2017zju,Vanrietvelde:2018pgb,Vanrietvelde:2018dit}, could regard $L_i \approx 0$ as a necessary criterion, to be satisfied by ``physical states''.}

The assertion that the number of physical degrees of freedom equals the number of independent relational variables is not justified by a correct understanding of spacetime symmetries, but appears to be motivated, instead, by mistaken beliefs about the theory of general relativity \cite{Vanrietvelde:2018pgb,Hoehn-Presentation:CERN-2019,Hoehn-Presentation:HK-2020,Giacomini-Presentation:QGAAT-2023,Overstreet:2022zgq,Giacomini:2017zju:PeerReview}. 
In particular, that diffeomorphism invariance \cite{Gaul:1999ys} and/or the Hole Argument \cite{Gomes:2023ljj,Gomes:2023qmy} imply Mach's Principle (``all physics is relational'' \cite{Mach's_Principle:Book,Barbour:Mach's_Principle}) and that general covariance implies the equivalence of accelerated and inertial frames (``all the laws of physics are the same in every reference frame'' \cite{Vanrietvelde:2018pgb,Hoehn-Presentation:CERN-2019,Hoehn-Presentation:HK-2020,Giacomini-Presentation:QGAAT-2023,Overstreet:2022zgq,Giacomini:2017zju:PeerReview}). 
The latter belief is used as the basis of a theory of ``quantum general covariance'', which, in turn, is based on a misunderstanding of classical general covariance \cite{MTW:1973,Dirac-GR:1975,Wald-GR:1984,Lasenby:2006}. 
Coming full-circle, the relational viewpoint has also been used to define a ``quantum Hole Argument'' \cite{Kabel:2024lzr} and a ``quantum superposition of diffeomorphisms'' \cite{delaHamette:2022cka,Norton:1993eqc}. 
The structure of the latter, which also play a key role in the former, closely resembles the so-called ``quantum superposition of Galilean transformations'', proposed in \cite{Giacomini:2017zju,Ballesteros:2020lgl,Ballesteros:2025ypr}. 
In an upcoming work, we will also examine whether these operations -- when applied {\bf actively}, rather than passively -- represent genuine symmetries. 

In summary, the issues we have identified cast doubt on high-profile claims about the relevance of the wider relational program for quantum gravity research -- in particular, on its claims to extend key concepts in classical gravitational physics to the quantum regime. 
These include, but are not limited to, ``quantum general covariance'', the ``quantum equivalence principle'', ``quantum local inertial frames'', ``quantum diffeomorphisms'', and the ``quantum Hole Argument''. 
The fundamental problem is that the relational formalisms, on which these concepts are based, treat acceleration as a relative, rather than an absolute quantity -- and, therefore, that they treat accelerated and inertial frames of reference as physically equivalent. 
This is incompatible with the results of general relativity, and, in fact, with any theory in which the existence of a spacetime manifold permits us to make a meaningful physical distinction between geodesic motion and physical acceleration (i.e., non-geodesic motion).
Further work is required, to scrutinise the claims made by each of these `relational quantum gravity' models, on its individual merits, in light of the concerns we have raised.

Finally, there is confusion regarding the way in which even Galilean invariance is implemented, in both classical non-relativistic mechanics and canonical quantum theory. 
As a result, certain ideas are taken over too literally, from the classical to the quantum realm. 
For example, although a ``gauge-fixing'' procedure is {\bf possible} in the classical theory, it is by no means {\bf required} \cite{Arnold:1978,Marsden&Ratiu:1999}, in order to obtain Galilean-invariant equations of motion. 
Why, then, should we believe that it is required in the quantum realm, necessitating the imposition of a ``Dirac quantisation'' scheme? 
No such scheme is required in canonical, Galilean-invariant, quantum theory \cite{Bargmann:1954gh,Levy-Leblond:1963qdx,Levy-Leblond:1967eic,Horzela:1991pa,Giulini:1995te,Csillag:2003}. 
Moreover, we have shown that imposing one -- as the authors of \cite{Vanrietvelde:2018pgb} suggest --  leads to the violation of Galilean invariance, rather than its successful implementation. 
Ultimately, this is because the Schr{\" o}dinger equation has no classical analogue \cite{Rae:2002,Isham:1995,Dirac:1958,vonNeumann:1955}. 
(See Sec. \ref{Sec.X}.)

This, in turn, casts doubt on the idea that either reduced quantisation, or Dirac quantisation \cite{Kunstatter:1991ds,Schleich:1990gd,Plyushchay:1994pk,Barvinsky:1996cg,Shimizu:1996vf}, are appropriate schemes for the quantisation of closed Galilean-invariant systems. 
Our analysis suggests that only canonical quantisation \cite{Rae:2002,Isham:1995,Dirac:1958,vonNeumann:1955}, in which no constraints are implemented at the classical level \cite{Dirac-CHS:1958,Dirac-CHS:1964,Henneaux-Teitelboim:1992,Rothe-RotheCHS:2010}, is sufficient for this task. 

Ironically, one must keep the non-Galilean invariant equations for the classical centre-of-mass coordinates. 
Quantising these, by promoting variables to operators, is equivalent to constructing the Heisenberg equations for $\widehat{{\bf X}}$ and $\widehat{{\bf P}}$, in which the non-Galilean invariant Hamiltonian $\widehat{H} = \widehat{P}^2/(2M) + \widehat{U}$ must be used. 
In this sense, quantisation and {\bf non-reduction} also commute (see \cite{Vanrietvelde:2018pgb,Vanrietvelde:2018dit} for a similar statement about quantisation and reduction). 
What matters, however, is whether they commute to form a Galilean-invariant theory.

This picture is equivalent to constructing the Schr{\" o}dinger equation, for the evolution of {\bf quantum states} (not operators), with $\widehat{H} = \widehat{P}^2/(2M) + \widehat{U}$. 
The $\widehat{P}^2/(2M)$ term is necessary, in this equation, in order to cancel additional terms that arise from Galilean transformations of the time-derivative, $i\hbar {\rm d}/{\rm d}t$. 
Without it, the theory as a whole actually {\bf violates Galilean symmetry} \cite{Bargmann:1954gh,Levy-Leblond:1963qdx,Levy-Leblond:1967eic,Horzela:1991pa,Giulini:1995te,Csillag:2003}.

Unfortunately, canonical quantisation -- in which the classical constraints are {\bf not} applied, nor “promoted” {\it as constraints} to the quantum regime  --  is the only quantisation scheme not considered in the perspective-neutral framework \cite{Vanrietvelde:2018pgb,Vanrietvelde:2018dit}. 
This is intimately related to all other relational models \cite{Giacomini:2017zju,Castro-Ruiz:2019nnl,delaHamette:2020dyi,Krumm:2020fws,Giacomini:2021gei,Castro-Ruiz:2021vnq,delaHamette:2022cka,Kabel:2024lzr,Giacomini:2020ahk,Giacomini:2021aof,Cepollaro:2021ccc,Giacomini:2018gxh,Mikusch:2021kro,Ballesteros:2020lgl,Ballesteros:2025ypr,Vanrietvelde:2018pgb,Hohn:2018iwn,Hohn:2018toe,delaHamette:2021oex,Hoehn:2019fsy,AliAhmad:2021adn,Vanrietvelde:2018dit,Hoehn:2023ehz,Hoehn:2023axh,DeVuyst:2025ezt,Hoehn:2020epv,Apadula:2022pxk,Kabel:2022cje,Carrozza:2021gju,Goeller:2022rsx,Kabel:2023jve,Chen:2026kui}, which, for the same reasons, are incompatible with Galilean invariance.

In our view, the confusion about how to implement Galilean symmetries stems from a wider misunderstanding of the nature of gauge invariance, and, in particular, of the physical and mathematical properties that distinguish gauge groups from spacetime symmetry groups. 
Mathematically, invariance of the classical equations of motion, under a gauge group, {\it necessarily} requires a singular Lagrangian, whereas their invariance under a spacetime symmetry group does not. 
This may be seen as a defining feature of gauge theories proper, as opposed to theories with only spacetime symmetries; although the former {\it must} be represented as constrained Hamiltonian systems, the latter need not be. 
(See App. \ref{App.C.1}.) 
Hence, recasting a theory that is invariant under a given spacetime group as a constrained Hamiltonian system, with the formal structure of a gauge theory -- as proposed in \cite{Vanrietvelde:2018pgb,Vanrietvelde:2018dit} for Galilean-invariant classical mechanics -- may give rise to a theory that is physically equivalent (by construction) at the classical level, but {\bf physically inequivalent} to its canonical quantum counterpart. 
This is, ultimately, how the perspective-neutral formalism, and other relational theories, are constructed.  
\\ \\ \indent
On a more positive note, we have shown exactly why, how, and where the relational theories go astray, in their attempts to implement what they refer to as ``gauge invariance'' \cite{Vanrietvelde:2018pgb,Vanrietvelde:2018dit}. 
(To us, this is really invariance under a spacetime symmetry group, which has different mathematical properties and physical implications than invariance under a gauge group.)
For a closed system in classical mechanics this means invariance under the Poisson bracket representation of the centrally-extended Galilean group, whereas, for a closed quantum system, it means invariance under its commutator representation \cite{Bargmann:1954gh,Levy-Leblond:1963qdx,Levy-Leblond:1967eic,Horzela:1991pa,Giulini:1995te,Csillag:2003}. 
Hence, this knowledge can serve as a basis for future progress in research on {\bf embodied} frames of reference, whether classical or quantum. 

The simplest way to summarise our findings is to say that the relational approach is incompatible with Galilean symmetry because the set of all {\bf relational variables}, $\mathcal{R}$, forms only a subset of the set of all {\bf Galilean-invariant variables}, $\mathcal{R} \subset \mathcal{G}$. 
Relational variables are observable because they are Galilean-invariant, but the converse is not true; Galilean-invariant observables need not be relational. 
The recent focus on relational quantities, as the be-all-and-end-all of physical theories \cite{Giacomini:2017zju,Castro-Ruiz:2019nnl,delaHamette:2020dyi,Krumm:2020fws,Giacomini:2021gei,Castro-Ruiz:2021vnq,delaHamette:2022cka,Kabel:2024lzr,Giacomini:2020ahk,Giacomini:2021aof,Cepollaro:2021ccc,Giacomini:2018gxh,Mikusch:2021kro,Ballesteros:2020lgl,Ballesteros:2025ypr,Vanrietvelde:2018pgb,Hohn:2018iwn,Hohn:2018toe,delaHamette:2021oex,Hoehn:2019fsy,AliAhmad:2021adn,Vanrietvelde:2018dit,Hoehn:2023ehz,Hoehn:2023axh,DeVuyst:2025ezt,Hoehn:2020epv,Apadula:2022pxk,Kabel:2022cje,Carrozza:2021gju,Goeller:2022rsx,Kabel:2023jve,Chen:2026kui}, is therefore not justified, by considerations of symmetry \cite{Bargmann:1954gh,Levy-Leblond:1963qdx,Levy-Leblond:1967eic,Horzela:1991pa,Giulini:1995te,Csillag:2003}. 
In fact, by excising all non-relational but still Galilean-invariant quantities the relational models \cite{Giacomini:2017zju,Vanrietvelde:2018pgb} actually discard {\bf physical information}. 

This includes information about the absolute accelerations of individual particles, which are characterised by non-geodesic motion in the Newtonian spacetime manifold \cite{SEP:Newtonian_Space-time}. 
Thus, by `reducing' spacetime, {\` a} la Aristotle, Descartes, Leibniz and Mach \cite{Aristotle,Aristotle:SEP,Descartes,Descartes:SEP,Leibniz:Arthur,Leibniz:SEP,Mach's_Principle:Book,Barbour:Mach's_Principle,Absolute_vs_Relational_Space:Classical_SEP,Absolute_vs_Relational_Space:Relativistic_SEP}  , to a system of relations between material bodies, the canonical spacetime manifold is effectively dispensed with in these theories \cite{Giacomini:2017zju,Vanrietvelde:2018pgb}. 
It is no surprise, then, that the relational models actually violate Galilean symmetry, which is the symmetry group of the Newtonian geometry \cite{SEP:Newtonian_Space-time}.
\footnote{Strictly, the symmetry group of the Newtonian geometry \cite{SEP:Newtonian_Space-time} is the canonical (unextended) Galilean group, whereas the dynamics of physical systems embedded within that geometry -- both classical and quantum -- is invariant under the centrally-extended Galilean group \cite{Bargmann:1954gh,Levy-Leblond:1963qdx,Levy-Leblond:1967eic,Horzela:1991pa,Giulini:1995te,Csillag:2003}.} 
This violation is subtle, however, since it amounts to the claim -- which all recent relational theories share -- that the laws of physics are the same in both inertial and accelerated frames of reference \cite{Giacomini:2017zju,Vanrietvelde:2018pgb,Hoehn:2023axh,Vanrietvelde:2018pgb,Vanrietvelde:2018dit,Hoehn:2019fsy,delaHamette:2021oex,Krumm:2020fws,delaHamette:2020dyi,Hoehn-Presentation:CERN-2019,Hoehn-Presentation:HK-2020,Giacomini-Presentation:QGAAT-2023,Overstreet:2022zgq,Giacomini:2017zju:PeerReview}. 

While these remarks hold true for the ``relational'' analysis -- or, rather, relational {\it amendments} -- of both classical and quantum theories, it is in the quantum regime that our own analysis is most fruitful. 
In \cite{Vanrietvelde:2018pgb}, it was shown that a  closed, classical, and Galilean-invariant system may be viewed as {\it either} (a) an unconstrained system with a regular Lagrangian (the canonical viewpoint), or (b) a constrained system with gauge-type symmetries and a singular Lagrangian, and that the two theories are physically equivalent at the classical level. 
It was then {\it assumed} (implicitly) that the second, non-canonical, model is the correct classical model to ``promote'' to the quantum level. 
A rather elaborate formalism was then constructed to support this idea. 
Using this formalism, it was claimed that, for closed Galilean-invariant systems, in particular, both reduced and Dirac quantisation schemes lead to the same set of ``relational'' theories \cite{Giacomini:2017zju,Vanrietvelde:2018pgb}. 
In this way, the perspective-neutral framework hoped to show that the recent wave of ``relational'' QRF models are compatible with Galilean symmetries. 
 
However, in this work, we have shown that these models actually violate Galilean invariance, and that, in order to restore it, one must first `undo' the $P_i \approx 0$ constraints, imposed in the constrained-system view of the classical theory. 
(See Sec. \ref{Sec.X}.)
This means that, contrary to recent claims \cite{Giacomini:2017zju,Vanrietvelde:2018pgb,Vanrietvelde:2018dit}, a closed system of $N$ quantum particles possesses exactly $\boldsymbol{3N}$ {\bf physical degrees of freedom} (excluding the spin degrees of freedom) \cite{Rae:2002,Isham:1995,Dirac:1958,vonNeumann:1955,Bargmann:1954gh,Levy-Leblond:1963qdx,Levy-Leblond:1967eic,Horzela:1991pa,Giulini:1995te,Csillag:2003}. 

Lastly, we showed that the three `extra' degrees of freedom in the quantum theory, which, in an appropriate basis, correspond to the three ignorable degrees of freedom at the classical level \cite{Kibble&Berkshire:2004,McCall:2011,Arnold:1978,Marsden&Ratiu:1999}, also contribute to the position- and momentum-space spreads of a chosen QRF, associated with one of the particles in the closed system. 
These spreads are Galilean-invariant, and, therefore, {\bf physically-measurable}, even though the position and momentum vectors, to which they refer, are not. 
We call such variables, which are not {\bf quantum observables} in the traditional sense \cite{Observable:QM-Compendium}, {\bf detectables}. 
In short, the detectable variables give rise to GURs, whose non-Heisenberg terms characterise the quantum fluctuations of the chosen QRF, whereas the traditional observables characterise the quantum fluctuations of the `rest of the Universe'. 
\\ \\
{\bf Acknowledgments}
ML thanks Tomasz Paterek, for helpful discussions about quantum theory, Shi-Dong Liang, for helpful discussions about classical mechanics, and Bernard Carr, for helpful discussions about general relativity. 
This work is supported by the Universitatea Babe\c s-Bolyai Grants to Support Competitiveness: AGC30869/24.02.2025, AGC30870/24.02.2025, AGC30871/24.02.2025, AGC30872/24.02.2025,
AGC30873/24.02.2025 and AGC30874/24.02.2025. 
MM acknowledges support from the Novo Nordisk Foundation (Grant No. NNF20OC0059939 ``Quantum for Life'').
\\ \\
{\bf Data Availability Statement}: No new data. 
\\ \\
{\bf AI-use Disclosure}: AI review reports were engaged during revisions of this manuscript, prior to journal submission. (Source: claude-fable-5.) 


\appendix

\section{The closed N-particle system in classical mechanics} \label{App.A}

\subsection{Classical dynamics in an external inertial frame} \label{App.A.1}

An {\bf isolated} system is a subsystem that does not interact with the rest of the Universe. 
The Hamiltonian for an isolated system of $N$ classical non-relativistic particles, from the perspective of an arbitrary {\bf inertial frame} of reference $O$, {\bf external} to the system, is \cite{Kibble&Berkshire:2004,McCall:2011,Arnold:1978,Marsden&Ratiu:1999}
\begin{eqnarray} \label{A.1}
H({\bf x},{\bf p}) = \sum_{J=1}^{N}\frac{p_J^2}{2m_J} + \sum_{K < L}V_{KL}(|{\bf x}_L-{\bf x}_K|) \, . 
\end{eqnarray}
Here, ${\bf x}_J(t)$ represents the displacement of the $J^{\rm th}$ particle, from the origin of $O$'s coordinates, $m_J$ is its mass, and ${\bf p}_J(t) = m_J\dot{{\bf x}}_J(t)$, where $\dot{{\bf x}}_J$ is its velocity relative to $O$. 

If the system is {\bf open}, the origin of $O$'s frame may be coincident with the location of another particle, aka ``the observer'', but this is not a necessary condition. 
$O$ may equally be interpreted as an unphysical frame -- that is, as an abstract inertial coordinate system on the Newtonian spacetime manifold \cite{SEP:Newtonian_Space-time}, whose origin is not identified with the location of any physical body. 
The mathematical description of the system of interest is the same, in either case. 
In the open-system scenario, the physics described by (\ref{A.1}) is {\bf approximate} since it does not account for the motion of the $N$ particles, relative to the rest of the Universe, or even for $O$'s internal degrees of freedom, if they exist.

However, if the system is {\bf closed}, and not merely isolated, then it {\it is} the Universe. 
In this case the physics described by (\ref{A.1}) can be {\bf exact}, but only if we still take care to ensure that the inertial frame considered is not {\bf embodied}, as only then do we obtain a {\bf complete} description of the system. 
This means that $O$ must be chosen such that its coordinate origin does not coincide with the trajectory of any particular particle. 

On the other hand, since all physical observers must be embodied as material subsystems, we may accept the {\bf incomplete} description of the system, from the perspective of a given inertial ``observer'', and try to piece together the dynamics of the system as a whole by comparing different partial viewpoints (i.e., by switching between different, embodied, frames). 
In principle, this is permissible, but we note that it is only possible for a given particle, let us call it $I$, to coincide with $O$ for all $t$, if it has no interactions with any other subsystem. 
Only then will it undergo the necessary inertial motion. 

Alternatively, if there exists even a single nonzero potential of the form $V_{IJ}(|{\bf x}_J-{\bf x}_I|)$, for some $J \neq I$, then $I$ inhabits an accelerated frame of reference. 
In such a frame, we cannot describe the system using {\bf inertial phase space coordinates} $\left\{{\bf x},{\bf p}\right\}_{J=1}^{N}$, alone, and the Hamiltonian cannot be written in the form (\ref{A.1}) \cite{Kibble&Berkshire:2004,McCall:2011,Arnold:1978,Marsden&Ratiu:1999}. 

Unfortunately, the helpful clarifying prefix ``inertial'' is often dropped in pedagogical introductions to classical mechanics, but it is still used frequently in more specialist literature \cite{Baszo:2016}. 
We include it, explicitly, in order to emphasise the distinction between these coordinates and those corresponding to non-inertial, accelerated, frames. 
When the accelerated frame is {\bf external} to the system, additional terms corresponding to the centrifugal, Euler, coriolis, and/or gravitational {\bf pseudo-forces} must be included in the Hamiltonian \cite{Kibble&Berkshire:2004,McCall:2011,Arnold:1978,Marsden&Ratiu:1999}. 
These take the form of effective external potentials.

Similar statements hold for any {\bf internal} frame, associated with the trajectory of an interacting particle -- except that, in this case, the external potentials can couple non-minimally to the particle momenta. 
(See Sec. \ref{Sec.3.1}.)
In this scenario, the degrees of freedom associated with the frame are effectively removed from the dynamical description of the system, since ${\bf x}_I(t) = {\bf 0}$ and ${\bf p}_I(t) = m_I\dot{{\bf x}}_I(t) = {\bf 0}$, for all $t$. 
This loss, however, must be compensated for by the introduction of pseudo-forces, which describe inertial effects \cite{Kibble&Berkshire:2004,McCall:2011,Arnold:1978,Marsden&Ratiu:1999}. 

To avoid such complications, we restrict our attention to external inertial frames, for the remainder of this section. 
We stress that it does not matter how the individual particles move {\it relative} to a frame of this kind. 
In particular, it does not matter whether $\ddot{{\bf x}}_J(t) = {\bf 0}$ or $\ddot{{\bf x}}_J(t) \neq {\bf 0}$, for any $J$. 
The frame is still inertial, in either case. 
If $\ddot{{\bf x}}_J(t) \neq {\bf 0}$, this indicates only that the $J^{\rm th}$ particle is accelerating, while $O$ continues to undergo inertial motion. 
This is because acceleration is an {\bf absolute}, not a relational, quantity. 
Physically, it is characterised by the departure from geodesic propagation in the background spacetime \cite{SEP:Newtonian_Space-time,MTW:1973,Dirac-GR:1975,Wald-GR:1984,Lasenby:2006,Norton:1993eqc,Lusanna:2019,Schwartz:2026}.  

In theories with a well-defined metric structure on the whole spacetime manifold, such as general relativity (GR), the relevant geodesics are time-like \cite{MTW:1973,Dirac-GR:1975,Wald-GR:1984,Lasenby:2006}. 
However, in Newton-Cartan gravity \cite{Schwartz:2026}, which represents the non-relativistic limit of GR, the background spacetime is constructed as a foliation of the four-dimensional manifold, by equal-time spatial hypersurfaces $\Sigma^3_t$. 
In this limit, $t \in \mathbb{R}$ is the absolute, Universal, time, so that no valid spacetime metric exists. 
Instead, there are two separate metric structures -- a space-like metric, $g_{ij}(t)$, defined on each $\Sigma^3_t$, and a time-like metric, defined on time-like submanifolds $(t_1,t_2) \subset \mathbb{R}$, which allows us to define time intervals $\Delta t := t_2 - t_1$. 
Despite this, an affine connection is still well-defined globally \cite{Schwartz:2026}. 
This allows us to compare space-like tensors at different times, and, hence, to describe the time-evolution of systems `in space' in a generally covariant way. 
The relevant geodesics, for the propagation of massive test bodies, are again time-like, and motion along them is parameterised by the absolute Universal time $t$. 

Hence, even in non-relativistic theories, a {\bf clear physical distinction} can be drawn between the motion of a massive test body in a gravitational field -- which propagates along the relevant time-like geodesic -- and acceleration, which is characterised by non-geodesic motion. 
All that is required, in order to describe this, is for the gravitational field to be identified with the effects of spacetime curvature \cite{Schwartz:2026}.
\footnote{Strictly, there is one gravitational field configuration, in the non-relativistic limit, for which gravity is {\it not} identified with the effects of curvature. In the Newton-Cartan theory, a constant and uniform gravitational field has zero Newton-Cartan curvature, and the relevant connection can be transformed into the globally flat connection \cite{Schwartz:2026}. However, this does not change the fact that, in all theories possessing a spacetime manifold, physical acceleration is associated with non-geodesic motion whereas both inertial motion in flat space, and locally-inertial motion in a gravitational field, is characterised by the geodesic motion of test particles \cite{MTW:1973,Schwartz:2026}.} 

It is an unhappy fact -- which causes much confusion -- that in classical mechanics the spacetime has zero curvature. 
This is because the Newtonian manifold has global structure $\mathbb{E}^3 \times \mathbb{R}$, where $\mathbb{E}^3$ is the three-dimensional Euclidean geometry \cite{SEP:Newtonian_Space-time}. 
This, in turn, requires that gravitational effects be described by means of a scalar potential, with no geometric significance.
\footnote{Derivations of the Newtonian potential for a point-mass, from the low-velocity and static weak-field limit of GR, typically make use of a mathematical sleight of hand that obscures how this geometric information is `lost'. 
A typical derivation proceeds by first showing that, in the low-velocity limit $v \ll c$, the geodesic equation reduces to $\ddot{x}_i = -\partial_i \sqrt{g_{00}}c^2$ \cite{Dirac-GR:1975}. 
The relevant static field is provided by the Schwarzschild solution, giving $\sqrt{g_{00}}c^2 = c^2\sqrt{1 - 2GM/(c^2r)} \simeq c^2 - GM/r$. 
Plugging this in to the previous expression and multiplying by the mass of the test particle, $m$, yields an expression that is superficially similar to Newton's second law for a potential $V(r) = -GM/r$. 
But this $V$ is {\bf not} the Newtonian potential. 
Since $r$ is the radial coordinate in the curved Schwarzschild geometry, $V(r)$ still carries geometric information. 
Hence, gravity still has a geometric origin and can be physically distinguished from acceleration. 
The sleight of hand occurs when, without being explicitly stated, the curved-space $r$ is replaced by the usual Euclidean expression, $r = \sqrt{x^2 + y^2 + z^2}$. 
This is the point at which the physical distinction between motion in a gravitational field, and acceleration, is formally abolished.} 
It is therefore impossible, in the classical formulation of Newtonian mechanics \cite{Kibble&Berkshire:2004,McCall:2011,Arnold:1978,Marsden&Ratiu:1999}, to draw a meaningful distinction between gravity and acceleration. 
Therein, {\it all} interactions are mediated by potentials, giving rise to forces, which cause the interacting particles to undergo non-geodesic motion.
This is a severe limitation of the classical formalism which is also significant for the formulation of canonical, non-relativistic, quantum mechanics (see Appendix \ref{App.B} for details) and for any ``gravitationally-inspired'' extensions thereof \cite{Giacomini:2017zju,Castro-Ruiz:2019nnl,Giacomini:2021gei,Vanrietvelde:2018pgb,Hohn:2018iwn,Hohn:2018toe,delaHamette:2021oex,Hoehn:2019fsy,AliAhmad:2021adn,Vanrietvelde:2018dit,Hoehn:2023ehz,Hoehn:2023axh,DeVuyst:2025ezt,delaHamette:2022cka,Kabel:2024lzr,Giacomini:2020ahk,Giacomini:2021aof,Cepollaro:2021ccc,Kabel:2022cje,Carrozza:2021gju,Goeller:2022rsx,Kabel:2023jve,Chen:2026kui,Hoehn-Presentation:CERN-2019,Hoehn-Presentation:HK-2020}. 

We discuss these points at length, here, as the confusion between acceleration and motion in a gravitational field is especially acute within the relational QRF literature, even when purely classical systems are considered. 
For example, in \cite{Giacomini:2017zju}, it is stated that the Weak Equivalence Principle (WEP) asserts the equivalence between constant and uniform acceleration and motion in a constant and uniform gravitational field, whereas, in \cite{Giacomini:2020ahk,Giacomini:2021aof,Cepollaro:2021ccc}, it is stated that the general EP asserts the equivalence of all locally inertial frames. 
While the latter statement is true -- since a {\bf local inertial frame} (LIF) is defined as a local non-rotating frame, undergoing geodesic propagation \cite{MTW:1973,Wald-GR:1984,Lasenby:2006,Norton:1993eqc} -- the former is false.
\footnote{Unfortunately, some undergraduate texts give incorrect accounts of the WEP, for example, by stating that it implies the equivalence of constant and uniform acceleration with motion in a constant and uniform gravitational field, due to Einstein’s elevator argument. (See \cite{Elevator-Argument} for a `pedagogical' account of the latter.) The comments and references above show why this account is incorrect. However, on the whole, the better “introductions” to general relativity, such as \cite{Lasenby:2006}, as well as graduate-level texts like \cite{MTW:1973,Wald-GR:1984}, give correct accounts of both the WEP and the general EP. The former is clearly more limited in scope, but has its uses, and is best understood as a statement asserting the universality of free-fall, i.e., that the free-fall of a body in a uniform gravitational field does not depend on its mass, or internal composition. This statement fails for non-uniform fields, in which it is only applicable to test bodies \cite{Lasenby:2006,MTW:1973,Wald-GR:1984}.} 
Curiously, these definitions appear to be regarded as mutually compatible, rather than as mutually incompatible, statements \cite{Giacomini:2020ahk,Giacomini:2021aof,Cepollaro:2021ccc}. 

In \cite{Vanrietvelde:2018pgb,Hoehn-Presentation:CERN-2019,Hoehn-Presentation:HK-2020,Giacomini-Presentation:QGAAT-2023,Overstreet:2022zgq,Giacomini:2017zju:PeerReview}, it is even stated that ``all the laws of physics are the same in every reference frame'', this being an alleged consequence of the theory of classical general relativity. 
It is then argued that a ``gravity-inspired symmetry principle", which ensures that this equivalence also holds in the quantum regime, should form the basis of all QRF theories \cite{Vanrietvelde:2018pgb,Vanrietvelde:2018dit}. 
To be absolutely clear, general relativity asserts the complete opposite, since it provides a physical basis for the distinction between LIFs, in which all non-gravitational laws adopt their special-relativistic forms, and accelerated frames, in which fictitious forces must also be accounted for \cite{SEP:Newtonian_Space-time,MTW:1973,Dirac-GR:1975,Wald-GR:1984,Lasenby:2006,Norton:1993eqc,Lusanna:2019,Schwartz:2026}. 

In the non-relativistic limit, this implies that the so-called ``acceleration due to gravity'' is the biggest  misnomer in all of physics. 
This acceleration is the physical acceleration required, due to the action of non-gravitational forces, in order for one body to maintain its spatial separation from another, in the presence of mutual gravitational attraction. 
It must be distinguished from gravitationally-induced pseudo-acceleration that causes the relative displacement between two bodies to vary with time, such that $\ddot{\bf x}_{J|I}(t) = \ddot{\bf x}_J(t) - \ddot{\bf x}_I(t) \neq {\bf 0}$, but, during which, neither body `feels' any acceleration \cite{Kibble&Berkshire:2004,McCall:2011,Arnold:1978,Marsden&Ratiu:1999}. 
The latter is induced by genuine (non-gravitational) forces, of the form ${\bf F}_{J|I}(t) = -\partial V_{IJ}(|{\bf x}_J - {\bf x}_I|)/\partial {\bf x}_J = \partial V_{IJ}(|{\bf x}_J - {\bf x}_I|)/\partial {\bf x}_I = -{\bf F}_{I|J}(t)$, whereas the former is not. 
For these reasons, we restrict the inter-particle potentials such that $V_{IJ}(|{\bf x}_J-{\bf x}_I|) \neq -Gm_Im_J/|{\bf x}_J-{\bf x}_I|$, from here on, in recognition of the fact that an accurate description of the gravity--acceleration distinction requires us to go beyond the regime of classical Newtonian mechanics. 

With this caveat in place, we can now complete our description of the $N$-particle system, subject to arbitrary non-gravitational interactions, from the perspective of an external inertial frame. 
To this end, we note that the inertial phase space coordinates are canonically conjugate,
\begin{eqnarray} \label{A.2}
\left\{({\bf x}_I)^i,({\bf p}_J)_j\right\}_{\rm PB} = \delta_{IJ}\delta^{i}{}_{j} \, , 
\quad
\left\{({\bf x}_I)^i,({\bf x}_J)^j\right\}_{\rm PB} = 0 \, , \quad \left\{({\bf p}_I)_i,({\bf p}_J)_j\right\}_{\rm PB} = 0 \, . 
\end{eqnarray}
Here, $\left\{\, . \, , \, . \, \right\}_{\rm PB}$ denotes the classical Poisson bracket and $i,j \in \left\{x,y,z\right\}$ label the individual components of a vector. 
For simplicity, we assume that ${\bf x}_I(t)$ and ${\bf p}_J(t)$ are expressed in terms of a global Cartesian coordinate system. 
Hamilton’s equations for (\ref{A.1}) then read
\begin{eqnarray} \label{A.3}
\frac{d {\bf x}_J}{d t} = \frac{\partial H}{\partial {\bf p}_{J}} \, , \quad 
\frac{d {\bf p}_J}{d t} = -\frac{\partial H}{\partial {\bf x}_{J}} \, , 
\end{eqnarray}
or, equivalently,
\begin{eqnarray} \label{A.4}
{\bf p}_{J} = m_J\dot{{\bf x}}_J \, , \quad 
\dot{{\bf p}}_J = -\sum_{K < L}\frac{\partial V_{KL}}{\partial {\bf x}_{J}} \, , 
\end{eqnarray}
giving
\begin{eqnarray} \label{A.5}
\ddot{{\bf x}}_J = -\frac{1}{m_J}\sum_{K < L}\frac{\partial V_{KL}}{\partial {\bf x}_{J}} \, . 
\end{eqnarray}

It is convenient to define the net momentum 
\begin{eqnarray} \label{A.6}
{\bf P} := \sum_{J=1}^{N}{\bf p}_J \, , 
\end{eqnarray}
and centre-of-mass coordinate, 
\begin{eqnarray} \label{A.7}
{\bf X} := \frac{1}{M}\sum_{J=1}^{N}m_J{\bf x}_J \, , 
\end{eqnarray}
where
\begin{eqnarray} \label{A.8}
M := \sum_{J=1}^{N}m_J
\end{eqnarray}
is the total mass of the system, and it is straightforward to show that these variables are also canonically conjugate, 
\begin{eqnarray} \label{A.9}
\left\{X^i,P_j\right\}_{\rm PB} = \delta^{i}{}_{j} \, . 
\end{eqnarray}
From (\ref{A.4}), we then have
\begin{eqnarray} \label{A.10}
 {\bf P} = M\dot{{\bf X}} \, , \quad \dot{{\bf P}} = {\bf 0} \, , 
\end{eqnarray}
or, equivalently,
\begin{eqnarray} \label{A.11}
{\bf P} = M{\bf v} \, , \quad \ddot{\bf{X}} = {\bf 0} \, . 
\end{eqnarray}
and, hence, 
\begin{eqnarray} \label{A.12}
{\bf X}(t) = {\bf v}t + {\bf x}_0 \, ,
\end{eqnarray}
where ${\bf v}$ is a constant velocity and ${\bf x}_0$ is a constant displacement vector. 
This shows that the centre of mass of an isolated system undergoes inertial motion, even though the individual particles may be accelerated, due to the presence of inter-particle potentials. 
Finally, we note that 
\begin{eqnarray} \label{A.12*}
\left\{x_0^i,v_j\right\}_{\rm PB} = \frac{\delta^{i}{}_{j}}{M} \, .
\end{eqnarray}

While both the canonical coordinates $\left\{{\bf x}_J,{\bf p}_J\right\}_{J=1}^{N}$ and the centre-of-mass coordinates $\left\{{\bf X},{\bf P}\right\}$ are often extremely useful, it is important to remember their most severe limitation, namely, that they do not represent observable quantities, as they are not invariant under translations and Galilean boosts \cite{Bargmann:1954gh,Levy-Leblond:1963qdx,Levy-Leblond:1967eic,Horzela:1991pa,Giulini:1995te}.

\subsection{Classical dynamics in an `internal' frame} \label{App.A.2}

We now ask if it is possible to ``jump'' from the perspective of the external inertial frame, $O$, to the perspective of an internal frame associated with one of the particles? 
The answer to this question, however, is surprisingly subtle. 

On the one hand, we have already seen that if the chosen reference particle $I$ interacts with other subsystems then we must modify the Hamiltonian (\ref{A.1}) to include the pseudo-forces due to inertial effects \cite{Kibble&Berkshire:2004,McCall:2011,Arnold:1978,Marsden&Ratiu:1999}. 
On the other hand, we are free to simply rewrite (\ref{A.1}) in terms of {\bf relational} displacements and momenta. 
Specifically, we can define
\begin{eqnarray} \label{A.13}
{\bf x}_{J|I} := {\bf x}_{J} - {\bf x}_{I}
\end{eqnarray}
as the displacement of the $J^{\rm th}$ particle, relative to the $I^{\rm th}$, and 
\begin{eqnarray} \label{A.14}
{\bf p}_{J|I} := \mu_{IJ}\left(\frac{{\bf p}_J}{m_J}-\frac{{\bf p}_I}{m_I}\right)
= \mu_{IJ}(\dot{{\bf x}}_J-\dot{{\bf x}}_I)
= \mu_{IJ}\dot{{\bf x}}_{J|I}
\end{eqnarray}
as its ``relative momentum'', where
\begin{eqnarray} \label{A.15}
\mu_{IJ} := \frac{m_Im_J}{m_I+m_J} 
\end{eqnarray}
is the relevant reduced mass. 
The Hamiltonian (\ref{A.1}) can then be rewritten as
\begin{eqnarray} \label{A.16}
H({\bf x}_{\rm rel},{\bf p}_{\rm rel}) &=& 
\frac{P^2}{2M} + \sum_{J \neq I}^{N}\frac{m_J}{\mu_{IJ}}\frac{p_{J|I}^2}{2\mu_{IJ}} - \frac{1}{2M}\left(\sum_{J \neq I}^{N}\frac{m_J}{\mu_{IJ}}{\bf p}_{J|I}\right)^2 
\nonumber\\
&+& \sum_{J \neq I}V_{IJ}(|{\bf x}_{J|I}|) + \sum_{K < L} V_{KL}(|{\bf x}_{L|I} - {\bf x}_{K|I}|) \, ,  
\end{eqnarray}
with $K,L \neq I$.

It is also convenient to distinguish between the ``free particle'' Hamiltonian for the centre of mass,
\begin{eqnarray} \label{A.16*}
H_{\rm CoM}({\bf P}) := \frac{P^2}{2M} \, , 
\end{eqnarray}
and the {\bf internal energy} of the system,
\begin{eqnarray} \label{A.16**}
U({\bf x}_{\rm rel},{\bf p}_{\rm rel}) := H({\bf x}_{\rm rel},{\bf p}_{\rm rel}) - \frac{P^2}{2M} \, . 
\end{eqnarray}
In classical mechanics, $H_{\rm CoM}$ is the generator of global time translations whereas $H = H_{\rm CoM} + U$ generates the time-evolution of the system \cite{Arnold:1978,Marsden&Ratiu:1999}. 
Geometrically, $H$ gives rise to a Hamiltonian vector field on the symplectic phase space manifold, ${\rm X}_{H}( \, . \, ) := \left\{ \, . \, ,H\right\}_{\rm PB}$, and the flow along ${\rm X}_{H}(F)$ corresponds to the time-evolution of $F(t)$, according to Hamilton's equations \cite{Frankel:1997ec,Nakahara:2003nw}. 

The choice of reference particle is arbitrary but we must not run away with the idea that (\ref{A.16}) represents, in a {\bf literal} sense, the Hamiltonian of the system `seen' by $I$. 
Although ${\bf x}_{J|I}$ (\ref{A.13}) represents the position of particle $J$, from the perspective of particle $I$, we cannot make an analogous statement for ${\bf p}_{J|I}$ (\ref{A.14}). 
Instead, the momentum of $J$ `seen' by $I$ should be properly defined as
\begin{eqnarray} \label{A.17}
{\bf p}_{J:I} := m_J\dot{{\bf x}}_{J|I} = \frac{m_J}{\mu_{IJ}}{\bf p}_{J|I} \, . 
\end{eqnarray}
In fact, both (\ref{A.14}) and (\ref{A.17}) are proportional to the relative velocity of $J$, with respect to $I$, and differ only in their constants of proportionality, each of which has units of mass. 

In the analysis that follows, which is canonical, we will have no need of ${\bf p}_{J:I}$. 
It would become relevant, however, if we were to consider a frame that is {\bf literally coincident} with particle $I$, for all $t$, as proposed in the relational \cite{Giacomini:2017zju} and perspective-neutral formalisms \cite{Vanrietvelde:2018pgb}. 
But what would the system look like from such a frame?

Naively, we may expect to obtain the relevant Hamiltonian simply by setting ${\bf x}_{I}(t) = {\bf 0}$ and ${\bf p}_{I}(t) = m_I\dot{\bf x}_I(t) = {\bf 0}$, for all $t$, which automatically ensures that $\dot{{\bf p}}_{I}(t) = m_I\ddot{\bf x}_I(t) = {\bf 0}$, and likewise for all higher-order time-derivatives, giving ${\bf p}_{J:I}(t) = {\bf p}_{J}(t) = m_J\dot{{\bf x}}_J(t)$ and $\dot{{\bf p}}_{J:I}(t) = \dot{{\bf p}}_{J}(t) = m_J\ddot{{\bf x}}_J(t)$, etc. 
Substituting these expressions into (\ref{A.1}) yields a Hamiltonian that is formally equivalent to that for a system of $N-1$ particles, in the presence of {\bf external potential} $V_{\rm ext}({\bf x}) := \sum_{J \neq I}V_{IJ}(|{\bf x}_J|)$ \cite{Kibble&Berkshire:2004,McCall:2011}. 
But this procedure is valid if and only if $V_{IJ}(|{\bf x}_{J}|) = 0$ for all $J \neq I$, since only under these conditions is the frame defined by the motion of particle $I$ inertial. 
(Recall that we require $I (=O)$ to be inertial, in order for the inertial phase space coordinates, in which the Hamiltonian takes the form (\ref{A.1}), to be well defined to begin with \cite{Baszo:2016}.)

Alternatively, if $I$ interacts with even a single other particle then $V_{\rm ext}({\bf x}) \neq 0$ and the frame defined by its motion is accelerated, according to the law of action and reaction \cite{Kibble&Berkshire:2004,McCall:2011}. 
If, {\it despite this}, we continue to insist that substituting ${\bf x}_{I}(t) = {\bf 0}$ and ${\bf p}_{I}(t) = m_I\dot{\bf x}_I(t) = {\bf 0}$ into (\ref{A.1}) gives the correct Hamiltonian for the ``relational'' $N$-particle state \cite{Giacomini:2017zju,Vanrietvelde:2018pgb}, we are, in effect, assuming one of two equally strange possibilities: either (i) that, contrary to the law of action and reaction, $I$'s frame is {\bf not} accelerating, or (ii) that, {\` a} la Mach's principle \cite{Mach's_Principle:Book,Barbour:Mach's_Principle}, acceleration is a {\bf relational} quantity. 

If we assume the former then the closed system contains an ``unmoved mover'' \cite{Aristotle,Aristotle:SEP}, namely, particle $I$, which acts on the remaining $N-1$ particles without itself being back-reacted upon. 
Since this violates the law of action and reaction, it also violates both the conservation energy and the conservation of linear momentum -- and, hence, the conservation of angular momentum \cite{Kibble&Berkshire:2004,McCall:2011,Arnold:1978,Marsden&Ratiu:1999}. 
It therefore violates all conservation laws required by the centrally-extended Galilean group \cite{Bargmann:1954gh,Levy-Leblond:1963qdx,Levy-Leblond:1967eic,Horzela:1991pa,Giulini:1995te}.
As this does not seem like a very promising basis for a theory, in which material systems are embedded within the Newtonian spacetime \cite{SEP:Newtonian_Space-time}, we are left with the second option.

In fact, the second assumption is precisely that adopted in the relational QRF literature \cite{Giacomini:2017zju,Castro-Ruiz:2019nnl,delaHamette:2020dyi,Krumm:2020fws,Giacomini:2021gei,Castro-Ruiz:2021vnq,delaHamette:2022cka,Kabel:2024lzr,Giacomini:2020ahk,Giacomini:2021aof,Cepollaro:2021ccc,Giacomini:2018gxh,Mikusch:2021kro,Ballesteros:2020lgl,Ballesteros:2025ypr,Vanrietvelde:2018pgb,Hohn:2018iwn,Hohn:2018toe,delaHamette:2021oex,Hoehn:2019fsy,AliAhmad:2021adn,Vanrietvelde:2018dit,Hoehn:2023ehz,Hoehn:2023axh,DeVuyst:2025ezt,Hoehn:2020epv,Apadula:2022pxk,Kabel:2022cje,Carrozza:2021gju,Goeller:2022rsx,Kabel:2023jve,Chen:2026kui,Hoehn-Presentation:CERN-2019,Hoehn-Presentation:HK-2020}. 
In \cite{Giacomini:2017zju}, the canonical Hamiltonian for a two-particle system, in the presence of an external potential, is written down and immediately reinterpreted as the Hamiltonian for the ``relational'' three-particle state.
\footnote{This is the ``initial Hamiltonian'' given just above Eq. (10), in that text, where particle $C$ is taken as the ``reference system''.}
In \cite{Vanrietvelde:2018pgb}, a similar expression is derived, again as a ``relational'' three-particle Hamiltonian, and, allegedly, as a necessary consequence of spatial translation invariance. 
If this derivation is valid then Galilean invariance must {\bf imply} Mach's Principle \cite{Mach's_Principle:Book,Barbour:Mach's_Principle}, in the non-relativistic regime. 
 
This is precisely the argument made in \cite{Vanrietvelde:2018pgb,Vanrietvelde:2018dit}. 
Although, therein, the authors describe their theory as a ``toy model for Mach's Principle'', it is clear that applying the same logic to an arbitrary Hamiltonian results in a purely relational description of the Universe, irrespective of the system considered. 
Moreover, these models are closely related to those presented in \cite{Giacomini:2017zju,delaHamette:2020dyi,Krumm:2020fws,delaHamette:2021oex}, a fact which has been commented on extensively in the relational literature itself \cite{Castro-Ruiz:2021vnq,Garmier:2025soc,DeVuyst:2025ezt}.
The resulting formalisms, such as \cite{Giacomini:2017zju,Vanrietvelde:2018pgb}, contain only {\bf relational variables}, which means that even the acceleration of the reference particle must be written in terms of these quantities. 

In an {\bf inertial frame}, this can be done using the conservation of the net momentum -- or, equivalently, the fact that the sum of all forces in an isolated system is zero -- since the identity
\begin{eqnarray} \label{A.17*}
\ddot{{\bf x}}_I(t) = -\frac{1}{M}\sum_{J \neq I}m_J\ddot{{\bf x}}_{J|I}(t) 
\end{eqnarray}
follows from the condition $\ddot{{\bf X}} = 0$, together with the definition of the relative displacements (\ref{A.13}). 
In the canonical theory, however, this does not make $\ddot{{\bf x}}_I(t)$ a relational quantity, since we may obtain its value by means of a single direct measurement, without specifying the values of any of the individual $\ddot{{\bf x}}_{J|I}$'s \cite{Kibble&Berkshire:2004,McCall:2011,Arnold:1978,Marsden&Ratiu:1999}. 

Our point is that, formally, (\ref{A.17*}) always holds, in both the canonical theory \cite{Rae:2002,Isham:1995,Dirac:1958,vonNeumann:1955} and the relational models \cite{Vanrietvelde:2018pgb,Vanrietvelde:2018dit}, since both claim to describe a frame in which the net momentum for an isolated system is conserved \cite{Kibble&Berkshire:2004,McCall:2011}. 
But its implications are clearly very different, within the two paradigms. 
Models that are ``relational by construction'' \cite{Giacomini:2017zju,Vanrietvelde:2018pgb} require the coordinate origin of the reference frame to coincide, quite literally, with the trajectory of the reference particle, $I$. 
In these models, it is not only permissible but {\bf necessary} to set $\ddot{{\bf x}}_I(t) = -\frac{1}{M}\sum_{J \neq I}m_J\ddot{{\bf x}}_{J|I}(t) = -\frac{1}{M}\sum_{J \neq I}m_J\ddot{{\bf x}}_{J}(t) = {\bf 0}$, even when $I$ interacts with other subsystems, and is therefore accelerated \cite{Giacomini:2017zju,Vanrietvelde:2018pgb}. 
This is {\bf not} equivalent to the condition $m_I\ddot{{\bf x}}_I + {\bf F}_{I:I} = {\bf 0}$, where ${\bf F}_{I:I}$ represents the relevant pseudo-force, and nor are the remaining pseudo-forces, ${\bf F}_{J:I} := -m_J\ddot{{\bf x}}_I$ for $J \neq I$, required by canonical classical mechanics \cite{Kibble&Berkshire:2004,McCall:2011}, introduced at the level of the classical Hamiltonian. 
Thus, the pseudo-forces associated with particle $I$'s acceleration are completely ignored, in the relational and perspective-neutral frameworks \cite{Giacomini:2017zju,Vanrietvelde:2018pgb}. 
This is not regarded as problematic, in these theories, because {\bf all} dynamical quantities, including accelerations, are defined only {\bf relative} to an embodied frame. 

For the remainder of the section, however, we will consider only canonical classical mechanics. 
We repeat that such relativism is unacceptable in the canonical theory because acceleration is absolute. 
The physical interpretation of setting ${\bf x}_I(t) = {\bf 0}$ and ${\bf p}_I(t) = m_I\dot{{\bf x}}_I(t) = {\bf 0}$ in (\ref{A.1}), for some $I$, and for all $t$, even when ${\bf F}_{I} := -\sum_{J \neq I}\partial V_{IJ}(|{\bf x}_J-{\bf x}_I|)/\partial {\bf x}_I \neq {\bf 0}$, is that the law of action and reaction is violated and the system contains ``unmoved movers''. 
While these would not be out of place in a Aristotelean Universe \cite{Aristotle,Aristotle:SEP}, they are decidedly at odds with a Newtonian spacetime \cite{SEP:Newtonian_Space-time}, in which they can be tolerated only as approximations, corresponding to the infinite-mass limit of the interacting reference system, $m_I \rightarrow \infty$. 

The considerations above show that, within the context of canonical Newtonian mechanics, we cannot jump {\bf literally} into the perspective of an accelerated particle, in a closed system, without introducing pseudo-forces to describe inertial effects \cite{Kibble&Berkshire:2004,McCall:2011,Arnold:1978,Marsden&Ratiu:1999}. 
The best we can do, instead, is to perform the coordinate transformation 
\begin{eqnarray} \label{A.17**}
\left\{{\bf x}_J,{\bf p}_J\right\}_{J=1}^{N} \mapsto \left\{{\bf X},{\bf P};{\bf x}_{J|I},{\bf p}_{J|I}\right\}_{J \neq I} \, , 
\end{eqnarray}
and to work with the Hamiltonian (\ref{A.16}). 
We cannot stress, strongly enough, that this is not the same as starting with the Hamiltonian for an open system of $N-1$ particles, subject to an {\bf external potential}, then re-interpreting the variables $\left\{{\bf x}_J,{\bf p}_J\right\}_{J=1}^{N-1}$ as the  ``relational'' description of a closed $N$-particle state. 
This is the approach adopted in \cite{Giacomini:2017zju,Vanrietvelde:2018pgb,Vanrietvelde:2018dit}, in the quantum regime, and we have shown that it leads to a loss of {\bf physical information}, even in the classical limit. 

Nonetheless, it is important to note that the relational variables (\ref{A.13}) and (\ref{A.14}) are the only measurable, and, hence, the only physical displacements and momenta, in (\ref{A.16}) \cite{Kibble&Berkshire:2004,McCall:2011,Arnold:1978,Marsden&Ratiu:1999}. 
This does not obviate our previous criticisms as it is only ${\bf X}$ and ${\bf P}$ that can be legitimately dispensed with, in {\it any} closed classical system, by setting {\bf both} equal to zero. 
In the classical regime, we may therefore impose the conditions ${\bf X} = {\bf 0}$ and ${\bf P} \approx {\bf 0}$, without loss of generality, where the weak equality ``$\approx$'' indicates that we can set ${\bf P} = {\bf 0}$ only after all the relevant Poisson brackets, involving this quantity, have been defined \cite{Dirac-CHS:1958,Dirac-CHS:1964,Henneaux-Teitelboim:1992,Rothe-RotheCHS:2010}. 
(See Appendix \ref{App.C} for details.)
 
In addition, we note that the relational variables (\ref{A.13})-(\ref{A.14}), though physical, do not form a canonically conjugate set for systems with three or more particles \cite{Angelo:2011}, 
\begin{eqnarray} \label{A.17***}
\left\{({\bf x}_{J|I})^i,({\bf p}_{K|I})_j\right\}_{\rm PB} \neq \delta_{JK}\delta^{i}{}_{j} \, , \quad {\rm if} \, \,  N \geq 3 \, . 
\end{eqnarray}
They are, therefore, not always the most convenient choice of phase space coordinates, for many practical purposes, such as the construction of Hamilton's equations. 

Taking our position-type variables as the set $\left\{{\bf X};{\bf x}_{J|I}\right\}_{J \neq I}$, the canonically conjugate momenta are denoted as $\left\{{\bf P};\boldsymbol{\pi}_{J|I}\right\}_{J \neq I}$, yielding the Poisson bracket algebra
\begin{eqnarray} \label{A.18}
\left\{X^{i},P_j\right\}_{\rm PB} = \delta^{i}{}_{j} \, , \quad \left\{({\bf x}_{J|I})^i,(\boldsymbol{\pi}_{K|I})_j\right\}_{\rm PB} = \delta_{JK}\delta^{i}{}_{j} \, , 
\nonumber\\
\left\{X^i,(\boldsymbol{\pi}_{K|I})_j\right\}_{\rm PB} = 0 \, , \quad \left\{({\bf x}_{J|I})^i,P_j\right\}_{\rm PB} = 0 \, , 
\nonumber\\
\left\{({\bf x}_{J|I})^i,({\bf x}_{K|I})^j\right\}_{\rm PB} = 0 \, , \, 
\left\{(\boldsymbol{\pi}_{J|I})_i,(\boldsymbol{\pi}_{K|I})_j\right\}_{\rm PB} = 0 \, , 
\end{eqnarray}
and the explicit expression for $\boldsymbol{\pi}_{J|I}$ is 
\begin{eqnarray} \label{A.19}
\boldsymbol{\pi}_{J|I} := {\bf p}_J - \frac{m_J}{M}{\bf P} \, ,
\end{eqnarray}
whatever the value of $I$. 
It is straightforward to see that both the relative displacements ${\bf x}_{J|I}$ and their conjugate momenta $\boldsymbol{\pi}_{J|I}$ are observable, because they are invariant under translations and Galilean boosts. 

Using these variables, the Hamiltonian (\ref{A.1}) takes the form
\begin{eqnarray} \label{A.20}
H({\bf x}_{\rm rel},\boldsymbol{\pi}_{\rm rel}) &=& \frac{P^2}{2M} + \sum_{J \neq I}\frac{\pi_{J|I}^2}{2m_J} + \frac{1}{2m_I}\left(\sum_{J \neq I}\boldsymbol{\pi}_{J|I}\right)^2 
\nonumber\\
&+& \sum_{J \neq I}V_{IJ}(|{\bf x}_{J|I}|)  + \sum_{K < L} V_{KL}(|{\bf x}_{L|I} - {\bf x}_{K|I}|) \, ,  
\end{eqnarray}
with $K,L \neq I$, and Hamilton's equations are
\begin{eqnarray} \label{A.21}
{\bf P} = M{\dot{\bf X}} \, , \quad \dot{{\bf P}} = {\bf 0} \, ,
\nonumber\\
\dot{{\bf x}}_{J|I} = \frac{\boldsymbol{\pi}_{J|I}}{m_J} + \frac{1}{m_I}\sum_{K \neq I}\boldsymbol{\pi}_{K|I}
\, , \quad
\dot{\boldsymbol{\pi}}_{J|I} = -\left\{\sum_{K \neq I}\frac{\partial V_{IK}}{\partial {\bf x}_{J|I}} + \sum_{K < L} \frac{\partial V_{KL}}{\partial {\bf x}_{J|I}}\right\} \, .
\end{eqnarray} 
The upper equations are those for the centre of mass (\ref{A.10}), which were previously derived from (\ref{A.4}), whereas the lower set give the second time-derivatives of the relative displacements, $\ddot{{\bf x}}_{J|I}$. 
While these are observable, we have no way to determine, from a measurement of $\ddot{{\bf x}}_{J|I}$ alone, what the physical accelerations of the individual particles $I$ and $J$ are. 
This does not matter, however, because $\ddot{{\bf x}}_{J}$ and $\ddot{{\bf x}}_{I}$ are also invariant under translations and boosts, and are therefore measurable, individually. 

Equations (\ref{A.21}), which can be formulated purely in terms of the derivatives of the inter-particle potentials as
\begin{eqnarray} \label{A.21*}
{\ddot{\bf X}} = {\bf 0} \, ,
\nonumber\\
\ddot{{\bf x}}_{J|I} = -\frac{1}{m_J}\left\{\sum_{K \neq I}\frac{\partial V_{IK}}{\partial {\bf x}_{J|I}} + \sum_{K < L} \frac{\partial V_{KL}}{\partial {\bf x}_{J|I}}\right\} - \frac{1}{m_I}\sum_{K \neq I}\left\{\sum_{L \neq I}\frac{\partial V_{IL}}{\partial {\bf x}_{K|I}} + \sum_{M < L} \frac{\partial V_{LM}}{\partial {\bf x}_{K|I}}\right\} \, ,
\end{eqnarray} 
with index $M \neq I$, must give exactly the same {\bf physical information} as (\ref{A.4}), because the coordinate transformation 
\begin{eqnarray} \label{A.22}
\left\{{\bf x}_J,{\bf p}_J\right\}_{J = 1}^{N} \mapsto \left\{{\bf X},{\bf P};{\bf x}_{J|I},\boldsymbol{\pi}_{J|I}\right\}_{J \neq I}
\end{eqnarray}
is {\bf invertible}. 
In order to perform the inversion, we must use the relations
\begin{eqnarray} \label{A.23}
{\bf x}_I = {\bf X} - \frac{1}{M}\sum_{J \neq I}m_J {\bf x}_{J|I} \, , \quad {\bf p}_I = \frac{m_I}{M}{\bf P} - \sum_{J \neq I}\boldsymbol{\pi}_{J|I} \, ,
\end{eqnarray}
which follow from the definitions already provided.
We also note that, although we are free to choose the inertial frame $O$ to coincide with the trajectory of the centre of mass, we are still required to `keep' both ${\bf X}$ and ${\bf P}$, as unspecified variables, in (\ref{A.22}). 
Were we to set ${\bf X} = {\bf 0}$ and ${\bf P} = {\bf 0}$, {\bf identically}, the transformation (\ref{A.22}) would no longer be invertible and physical information would be lost. 
This caveat is compatible with the stipulation that we must set ${\bf X} = {\bf 0}$ and ${\bf P} \approx {\bf 0}$, instead, as this allows all the relevant Poisson brackets (\ref{A.18}), required for Hamilton's equations (\ref{A.21}), to be formed \cite{Dirac-CHS:1958,Dirac-CHS:1964,Henneaux-Teitelboim:1992,Rothe-RotheCHS:2010}.

Lastly, we consider the conjugate variables $\left\{{\bf X},{\bf P};{\bf q}_{J|I},{\bf p}_{J|I}\right\}_{J \neq I}$, satisfying the algebra
\begin{eqnarray} \label{A.24}
\left\{X^{i},P_j\right\}_{\rm PB} = \delta^{i}{}_{j} \, , \quad \left\{({\bf q}_{J|I})^i,({\bf p}_{K|I})_j\right\}_{\rm PB} = \delta_{JK}\delta^{i}{}_{j} \, , 
\nonumber\\
\left\{X^i,({\bf p}_{K|I})_j\right\}_{\rm PB} = 0 \, , \quad \left\{({\bf q}_{J|I})^i,P_j\right\}_{\rm PB} = 0 \, , 
\nonumber\\
\left\{({\bf q}_{J|I})^i,({\bf q}_{K|I})^j\right\}_{\rm PB} = 0 \, , \, 
\left\{({\bf p}_{J|I})_i,({\bf p}_{K|I})_j\right\}_{\rm PB} = 0 \, . 
\end{eqnarray}
The explicit expression for ${\bf q}_{J|I}$ is 
\begin{eqnarray} \label{A.25}
{\bf q}_{J|I} := \frac{m_J}{\mu_{IJ}}\left({\bf x}_J - {\bf X}\right) \, , 
\end{eqnarray}
and the coordinate transformation 
\begin{eqnarray} \label{A.26}
\left\{{\bf x}_J,{\bf p}_J\right\}_{J = 1}^{N} \mapsto \left\{{\bf X},{\bf P};{\bf q}_{J|I},{\bf p}_{J|I}\right\}_{J \neq I} 
\end{eqnarray}
is also invertible. 
In order to perform the inversion, we must use the relations
\begin{eqnarray} \label{A.27}
{\bf x}_I = {\bf X} - \frac{1}{m_I}\sum_{J \neq I}\mu_{IJ}{\bf q}_{J|I} \, , 
\quad
{\bf p}_I = \frac{m_I}{M}\left({\bf P} - \sum_{J \neq I}\frac{m_J}{\mu_{IJ}}{\bf p}_{J|I}\right) \, , 
\end{eqnarray}
which follow from our previous definitions. 
In these coordinates, the canonical Hamiltonian takes the form
\begin{eqnarray} \label{A.28}
H({\bf q}_{\rm rel},{\bf p}_{\rm rel}) 
&=& \frac{P^2}{2M} + \sum_{J \neq I}^{N}\frac{m_J}{\mu_{IJ}}\frac{p_{J|I}^2}{2\mu_{IJ}} - \frac{1}{2M}\left(\sum_{J \neq I}^{N}\frac{m_J}{\mu_{IJ}}{\bf p}_{J|I}\right)^2 
\nonumber\\
&+& \sum_{J \neq I}V_{IJ}\left(\bigg|\frac{\mu_{IJ}}{m_J}{\bf q}_{J|I} + \frac{1}{m_I}\sum_{K \neq I}\mu_{IK}{\bf q}_{K|I}\bigg|\right) 
\nonumber\\
&+& \sum_{K < L} V_{KL}\left(\bigg|\frac{\mu_{IL}}{m_L}{\bf q}_{L|I} - \frac{\mu_{IK}}{m_K}{\bf q}_{K|I}\bigg|\right) \, ,  
\end{eqnarray}
and Hamilton's equations are
\begin{eqnarray} \label{A.27*}
{\bf P} = M\dot{{\bf X}} \, , \quad \dot{{\bf P}} = 0 \, ,
\nonumber\\
\dot{{\bf q}}_{J|I} = \frac{m_J}{\mu_{IJ}}\left(\frac{{\bf p}_{J|I}}{\mu_{IJ}} - \frac{1}{M}\sum_{K \neq I}\frac{m_K}{\mu_{IK}}{\bf p}_{K|I}\right)
\, , \quad
\dot{{\bf p}}_{J|I} = -\left\{\sum_{K \neq I}\frac{\partial V_{IK}}{\partial {\bf q}_{J|I}} + \sum_{K < L} \frac{\partial V_{KL}}{\partial {\bf q}_{J|I}}\right\} \, .
\end{eqnarray} 
Again, these equations contain exactly the same physical information as (\ref{A.21}) and (\ref{A.4}), and it is straightforward to see that both the ${\bf q}_{J|I}$'s and the ${\bf p}_{J|I}$'s are observable, because they are invariant under translations and boosts. 
Equations (\ref{A.27*}) can also be reformulated as
\begin{eqnarray} \label{A.27**}
\ddot{{\bf X}} = {\bf 0} \, , 
\nonumber\\
\ddot{{\bf q}}_{J|I} = -\frac{m_J}{\mu_{IJ}^2}\left\{\sum_{K \neq I}\frac{\partial V_{IK}}{\partial {\bf q}_{J|I}} + \sum_{K < L} \frac{\partial V_{KL}}{\partial {\bf q}_{J|I}}\right\} 
+ \frac{1}{M}\sum_{K \neq I}\frac{m_K}{\mu_{IK}} \left\{\sum_{L \neq I}\frac{\partial V_{IL}}{\partial {\bf q}_{K|I}} + \sum_{M < L} \frac{\partial V_{LM}}{\partial {\bf q}_{K|I}}\right\} \, . 
\end{eqnarray} 

Finally, we note that all three coordinates systems, $\left\{{\bf x}_{J},{\bf p}_{J}\right\}_{J = 1}^{N}$, $\left\{{\bf X},{\bf P};{\bf x}_{J|I},\boldsymbol{\pi}_{J|I}\right\}_{J \neq I}$ and $\left\{{\bf X},{\bf P};{\bf q}_{J|I},{\bf p}_{J|I}\right\}_{J \neq I}$, are {\bf inertial coordinate systems}. 
That is, they represent different ways of expressing the same information, which corresponds to the `view' of the $N$-particle system, from the perspective of the external, abstract, inertial frame $O$.  

\subsection{Galilean symmetries in classical mechanics} \label{App.A.3}

When we talk about the ``Galilean invariance'' of systems in classical mechanics \cite{Kibble&Berkshire:2004,McCall:2011,Arnold:1978,Marsden&Ratiu:1999}, we actually mean invariance under the {\bf Poisson bracket representation} of the {\bf centrally-extended Galilean algebra},
\begin{eqnarray} \label{PB_Gal_Alg}
\left\{J_i,J_j\right\}_{\rm PB} = \epsilon_{ij}{}^{k}J_k \, , \quad \left\{J_i,G_j\right\}_{\rm PB} = \epsilon_{ij}{}^{k}G_k  \, , \quad \left\{J_i,P_j\right\}_{\rm PB} = \epsilon_{ij}{}^{k}P_k \, , 
\nonumber\\
\left\{J_i,H\right\}_{\rm PB} = 0 \, , \quad \left\{G_i,G_j\right\}_{\rm PB} = 0  \, , \quad \left\{G_i,P_j\right\}_{\rm PB} = M\delta_{ij} \, , 
\nonumber\\
\left\{G_i,H\right\}_{\rm PB} = P_i \, , \quad \left\{P_i,P_j\right\}_{\rm PB} = 0  \, , \quad \left\{P_i,H\right\}_{\rm PB} = 0 \, , 
\end{eqnarray}
where 
\begin{eqnarray} \label{PB_Gal_Alg_gens}
H := \frac{P^2}{2M} + U \, , \quad {\bf P} := \sum_{J = 1}^{N}{\bf p}_J \, , \quad 
{\bf J} := {\bf L} + {\bf S} = \sum_{J = 1}^{N} {\bf x}_J \times {\bf p}_J + \sum_{J = 1}^{N} {\bf s}_J \, , 
\nonumber\\
{\rm and} \, \, {\bf G} := M{\bf X} - t{\bf P} \, , \, {\rm with} \, \, {\bf X} := \frac{1}{M}\sum_{J = 1}^{N} m_J{\bf x}_J \, ,
\end{eqnarray}
and ${\bf s}_J$ represents the internal `spin' of the $J^{\rm th}$ particle. 
All non-internal quantities are again defined with respect to the external inertial frame $O$. 

The components of the net momentum, $P_i$, generate spatial translations, the $J_i$ are generalised rotation generators, and the $G_i$ are the generators of Galilean velocity boosts. 
It may also be shown that $H_{\rm CoM} := P^2/(2M)$ generates global time translations, whereas $H := P^2/(2M) + U$ generates non-trivial time-evolution of the closed system \cite{Rae:2002,Isham:1995,Dirac:1958,vonNeumann:1955,Geometric-CM-QM}, and it is a remarkable fact that {\bf both} these operators satisfy (\ref{PB_Gal_Alg}). 
This should be compared with the case of the canonical Galilean group, in which {\bf only} the generator of global time translations, $\mathcal{H} := \partial/\partial t$, satisfies the canonical (unextended) Galilean algebra \cite{Bargmann:1954gh,Levy-Leblond:1963qdx,Levy-Leblond:1967eic,Horzela:1991pa,Giulini:1995te,Csillag:2003}. 

Both $H_{\rm CoM}$ and $H := H_{\rm CoM} + U$ satisfy (\ref{PB_Gal_Alg}) because the internal energy $U$ is a {\bf Casimir invariant} of this algebra. 
There are two more independent invariants, namely, the central extension itself -- that is, the total mass $M$ -- and the non-relativistic analogue of the Pauli-Lubansky pseudo-vector in classical field theory. 
Hence, we have
\begin{eqnarray} \label{PB_Gal_Alg_Casimirs}
C_1 := M \, , \quad C_2 := U = H - \frac{P^2}{2M}  \, , \quad 
C_3 := \big|{\bf J} - \frac{1}{M}{\bf G} \times {\bf P} \big|^2 \, . 
\end{eqnarray}

Strictly, the symmetry generators in (\ref{PB_Gal_Alg}) are identified with the dynamical quantities (\ref{PB_Gal_Alg_gens}) up to a multiplicative factor with dimensions of inverse action. 
The latter is not explicitly shown and may be considered as a unit numerical factor. 
This ensures that the product of a generator and its respective parameter remains dimensionless, and can be exponentiated to yield the required transformation. 
Thus, the operators
\begin{eqnarray} \label{PB_Gal_Alg_Ops}
S_{\rm T}({\bf x}_0) := \exp\left({\bf P}.{\bf x_0}\right) \, , \quad
S_{\rm B}({\bf v}) := \exp\left({\bf G}.{\bf v}\right) \, , 
\nonumber\\
S_{\rm R}(\boldsymbol{\theta}) := \exp\left({\bf J}.\boldsymbol{\theta}\right) \, , \quad
S_{0}(t) := \exp\left(-H_{\rm CoM}t\right) \, , \,
\end{eqnarray}
enact spatial translations by ${\bf x}_0$, boosts by velocity $-{\bf v}$, rotations by $\theta$ about the unit vector ${\bf n}$, where $\boldsymbol{\theta} := {\bf n}\theta$, and temporal translations by time $t$, respectively. 
A general element of the centrally-extended Galilean group is represented by the operator \cite{Arnold:1978,Marsden&Ratiu:1999}
\begin{eqnarray} \label{PB_Gal_Alg_G}
S_{\rm Gal}({\bf x}_0,{\bf v},\boldsymbol{\theta},t) := \exp\left({\bf P}.{\bf x}_0 + {\bf G}.{\bf v} + {\bf J}.\boldsymbol{\theta} - H_{\rm CoM}t\right) \, .
\end{eqnarray}

\section{The closed N-particle system in quantum mechanics} \label{App.B}

\subsection{Quantum dynamics in an external inertial frame} \label{App.B.1}

The quantum mechanical analogue of (\ref{A.1}) is
\begin{eqnarray} \label{B.1}
\widehat{H}({\bf x},{\bf p}) = \sum_{J=1}^{N}\frac{\widehat{p}_J^2}{2m_J} + \sum_{K < L}\widehat{V}_{KL}(|{\bf x}_L-{\bf x}_K|) \, . 
\end{eqnarray}
This is the canonical Hamiltonian of a system of $N$ quantum particles, or, in other words, the operator whose eigenvalues correspond to the possible total energies of the system, as seen from an arbitrary inertial and {\bf classical} frame of reference, $O$. 
As before, if the system is interpreted as closed, $O$’s frame defines an abstract inertial coordinate system on the Newtonian spacetime manifold \cite{SEP:Newtonian_Space-time}.  

The quantum mechanical analogues of (\ref{A.2}) are
\begin{eqnarray} \label{B.2}
[(\widehat{{\bf x}}_I)^i,(\widehat{{\bf p}}_J)_j]= i\hbar\delta_{IJ}\delta^{i}{}_{j}\widehat{\mathbb{I}} \, , \quad 
[(\widehat{{\bf x}}_I)^i,(\widehat{{\bf x}}_J)^j]= 0 \, , \quad 
[(\widehat{{\bf p}}_I)_i,(\widehat{{\bf p}}_J)_j]= 0 \, , 
\end{eqnarray}
and the analogues of (\ref{A.3})-(\ref{A.4}) are derived from the Heisenberg equation,
\begin{eqnarray} \label{B.3}
i\hbar \frac{d\widehat{O}(t)}{dt} = [\widehat{O}(t),\widehat{H}]  \, , 
\end{eqnarray}
giving
\begin{eqnarray} \label{B.4}
\widehat{{\bf p}}_{J} = m_J\widehat{\dot{{\bf x}}}_J \, , \quad 
\widehat{\dot{{\bf p}}}_J = -\sum_{K < L}\frac{\partial \widehat{V}_{KL}}{\partial {\bf x}_{J}} \, , 
\end{eqnarray}
and, hence,
\begin{eqnarray} \label{B.5} 
\widehat{\ddot{{\bf x}}}_J = -\frac{1}{m_J}\sum_{K < L}\frac{\partial \widehat{V}_{KL}}{\partial {\bf x}_{J}} \, .
\end{eqnarray}
From here on, we deal only with the closed-system scenario so that no operators carry an explicit time-dependence in the Schr{\" o}dinger picture, $\partial \widehat{O}/\partial t = 0$.

The operator-analogues of the net momentum and classical centre-of-mass coordinate are
\begin{eqnarray} \label{B.6}
\widehat{{\bf P}} := \sum_{J=1}^{N}\widehat{{\bf p}}_J \, , 
\end{eqnarray}
and
\begin{eqnarray} \label{B.7}
\widehat{{\bf X}} := \frac{1}{M}\sum_{J=1}^{N}m_J\widehat{{\bf x}}_J \, , 
\end{eqnarray}
yielding
\begin{eqnarray} \label{B.8}
[\widehat{{\bf X}},\widehat{{\bf P}}] = i\hbar\widehat{\mathbb{I}} \, . 
\end{eqnarray}
Substituting (\ref{B.7}) and (\ref{B.6}) into (\ref{B.3}) gives 
\begin{eqnarray} \label{B.9}
\widehat{{\bf P}} = M\widehat{\dot{{\bf X}}} \, , \quad \widehat{\dot{{\bf P}}} = {\bf 0} \, , 
\end{eqnarray}
or, equivalently, 
\begin{eqnarray} \label{B.10}
\widehat{{\bf P}} = M\widehat{{\bf v}} \, , \quad \widehat{\ddot{{\bf X}}} = {\bf 0} \, , 
\end{eqnarray}
and, hence,
\begin{eqnarray} \label{B.11}
\widehat{{\bf X}}(t) = \widehat{{\bf v}}t + \widehat{{\bf x}}_0 \, , 
\end{eqnarray}
where $\widehat{{\bf x}}_0$ and $\widehat{{\bf v}}$ are {\bf time-independent} operators whose individual components satisfy the commutation relations
\begin{eqnarray} \label{B.12}
[\widehat{x}_0^i,\widehat{v}_j] = \frac{i\hbar}{M}\delta^{i}{}_{j}\widehat{\mathbb{I}} \, .
\end{eqnarray}

While (\ref{B.9})-(\ref{B.11}) are the obvious quantum analogues of (\ref{A.10})-(\ref{A.12}), it is due to the structure of the operators $\widehat{{\bf P}}$ and $\widehat{{\bf X}}(t)$ that the seemingly tight analogy between the classical and quantum regimes begins to break down. 
The superficial correspondence between (\ref{A.10})-(\ref{A.12}) and (\ref{B.9})-(\ref{B.11}) is misleading because, although the classical coordinates $X^i(t)$ and their conjugate momenta $P_i$ are {\bf ignorable} -- that is, unphysical -- this is not the case for $\widehat{X}^i(t)$ and $\widehat{P}_i$. 
We now show that it is impossible to set $\widehat{P}_i|\Psi\rangle \approx 0$, for an arbitrary physical state $|\Psi\rangle$ -- by analogy with the classical conditions $P_i \approx 0$, which we are free to impose for an arbitrary classical state by jumping to the centre-of-mass frame -- without losing {\bf physical information}. 

To see why this is the case, we need only note that the general solution of the Heisenberg equation, for an arbitrary operator $\widehat{O}$ (with no explicit time-dependence), is
\begin{eqnarray} \label{B.13}
\widehat{O}(t) = \exp\left(\frac{i}{\hbar}\widehat{H}t\right) \widehat{O} \, \exp\left(-\frac{i}{\hbar}\widehat{H}t\right) \, .
\end{eqnarray}
This implies that {\bf any} operator which commutes with the Hamiltonian does not depend on time, even in the Heisenberg picture:
\begin{eqnarray} \label{B.14}
[\widehat{O}(t),\widehat{H}] = 0 \implies \widehat{\dot{O}}(t) = 0  \implies \widehat{O}(t) = \widehat{O} \, . 
\end{eqnarray} 
But the condition (\ref{B.14}) does {\bf not} imply that $\widehat{O}$ is capable of taking only a {\bf single value} (such as zero). 
Nor does it imply that the eigenstates of such an operator -- for example, those with eigenvalue zero -- correspond to the physical states of the theory. 
In fact, if $\widehat{O}$ possesses a continuous eigenvalue spectrum, its eigenstates cannot be {\bf canonical physical states} \cite{Rae:2002,Isham:1995,Dirac:1958,vonNeumann:1955}, as they are non-normalisable. 
For this reason, we strongly disagree with the claim, made in \cite{Vanrietvelde:2018pgb}, that the condition $\widehat{{\bf P}}|\Psi\rangle \approx {\bf 0}$ {\bf selects} the subset of physical (i.e., Galilean-invariant) states, from the kinematical Hilbert space of Galilean-symmetric quantum mechanics. 
The canonical theory is already Galilean-symmetric \cite{Bargmann:1954gh,Levy-Leblond:1963qdx,Levy-Leblond:1967eic,Horzela:1991pa,Giulini:1995te,Csillag:2003}, and does not permit the existence of physical states with $P_i = {\rm const}$. 

Following standard terminology \cite{Rae:2002,Isham:1995,Dirac:1958,vonNeumann:1955}, we refer to operators satisfying the condition (\ref{B.14}) as either {\bf stationary operators} or {\bf time-independent operators}. 
Operators of this kind are also often referred to as ``conserved quantities” or “conserved charges'', in the quantum mechanical literature, paralleling the standard terminology of classical mechanics \cite{Kibble&Berkshire:2004,McCall:2011,Arnold:1978,Marsden&Ratiu:1999,Noether'sTheorems}. 
This terminology is potentially misleading, however, as it appears to suggest that they represent only a single ``conserved'' value of a given quantity. 
We have been at pains to stress that this is not the case, and opt to use the terms ``stationary'' or ``time-independent'', instead, to avoid confusion. 
Throughout the rest of this paper, we denote such operators as $\widehat{O}$, rather than $\widehat{O}(0)$, for the sake of notational elegance, and use parentheses to include the time parameter, $\widehat{O}(t)$, only when explicit time-dependence exists. 

It is important to understand that stationary operators are not constants, and that they still possess standard {\bf spectral representations} \cite{Rae:2002,Isham:1995,Dirac:1958,vonNeumann:1955}. 
These representations sum over a discrete (or continuous) range of eigenvalues and their associated proper (or improper) projectors. 
In particular, we have that
\begin{eqnarray} \label{B.16}
\widehat{{\bf P}} := \int {\bf P} \bigotimes_{J=1}^{N}|{\bf p}_J \rangle\langle {\bf p}_J|_{J} {\rm d}^3p_J \, , \, \, {\rm with} \, \, {\bf P} := \sum_{J=1}^{N} {\bf p}_J \, \, {\rm and} \, \, \dot{\bf P} = {\bf 0} \, , 
\end{eqnarray}
in the canonical momentum basis. 
The net momentum operator may also be rewritten as
\begin{eqnarray} \label{B.17}
\widehat{{\bf P}} := M\widehat{{\bf v}} := M\int {\bf v} \bigotimes_{J=1}^{N}|{\bf v}_J \rangle\langle {\bf v}_J|_{J} {\rm d}^3v_J \, , \, \, {\rm with} \, \, {\bf v} := \frac{1}{M}\sum_{J=1}^{N} m_J{\bf v}_J\, \, {\rm and} \, \, \dot{\bf v} = {\bf 0} \, , 
\end{eqnarray}
which implicitly defines the operator $\widehat{{\bf v}}$, and we may define the operator $\widehat{{\bf x}}_0 = \widehat{{\bf X}}(0)$ as
\begin{eqnarray} \label{B.18}
\widehat{{\bf x}}_0 := \frac{1}{M}\sum_{J=1}^{N}m_J\widehat{{\bf x}}_J(0)  \, , \, \, {\rm with} \, \, {\bf x}_0 := \frac{1}{M}\sum_{J=1}^N m_J{\bf x}_J(0) \, \, {\rm and} \, \, \dot{\bf x}_0 = {\bf 0} \, .
\end{eqnarray}

We cannot stress, strongly enough, that the physical implications of permitting a spectrum of values for the centre-of-mass coordinate, and the net momentum, as required by (\ref{B.18}) and (\ref{B.17}), are {\bf not} the same as those obtained by imposing the conditions 
\begin{eqnarray} \label{B.19}
\widehat{P}_i|\Psi\rangle \approx 0 \, .
\end{eqnarray}
These conditions are the {\bf literal analogues} of the classical constraints $P_i \approx 0$, which are obtained from the general constraints $P_i \approx Mv_i = {\rm const.}$ by setting $v_i = 0$ for all $i$. 
In classical physics, we can choose these specific values without loss of generality. 
This choice simply fixes the specific values of three quantities that are already known to be constants. 

But, in quantum physics, this is no longer the case. 
The correct analogue of $P_i \approx Mv_i$, in the quantum regime, is $\widehat{P}_i \approx M\widehat{v}_i$, where $\widehat{v}_i$ is a {\bf time-independent operator}. 
However, this condition is already strongly, and not just weakly satisfied, by $\widehat{{\bf P}} = M\widehat{{\bf v}}$ (\ref{B.10}). 
{\bf This means that it does not act as a constraint} \cite{Dirac-CHS:1958,Dirac-CHS:1964,Henneaux-Teitelboim:1992,Rothe-RotheCHS:2010}.
In particular, we cannot impose the weak equality $\widehat{P}_i|\Psi\rangle \approx Mv_i|\Psi\rangle$, for any {\bf fixed} value of $v_i$, without imposing the strong equality $\widehat{P}_i|\Psi\rangle = Mv_i|\Psi\rangle$. 

This fixes $|\Psi\rangle$ as an eigenstate of $\widehat{P}_i$, but such a state is non-normalisable, and therefore {\bf unphysical}, using the standard inner product of canonical quantum mechanics \cite{Rae:2002,Isham:1995,Dirac:1958,vonNeumann:1955}. 
In this sense, the canonical theory predicts the very opposite result of the perspective-neutral formalism \cite{Vanrietvelde:2018pgb}, in which the condition (\ref{B.19}) is said to select the subset of physical states from the kinematical Hilbert space. 
While it is possible to introduce a different, non-canonical, norm -- as proposed in the perspective-neutral framework, which appeals to the notions of ``group averaging'' and ``refined algebraic quantisation'' as motivations for its construction \cite{Vanrietvelde:2018pgb} -- this is by no means required, in order to enforce Galilean invariance. 
The canonical norm of the canonical theory, which is already Galilean-invariant, works just fine \cite{Bargmann:1954gh,Levy-Leblond:1963qdx,Levy-Leblond:1967eic,Horzela:1991pa,Giulini:1995te,Csillag:2003}, and this theory does not permit the net momentum to be ``gauge fixed'', due to canonical quantum uncertainty.

In summary, we have shown that the imposition of (\ref{B.19}) is not justified, by the imposition of Galilean invariance (specifically, translation invariance) for quantum systems, as claimed in \cite{Vanrietvelde:2018pgb}. 
The dynamics implied by the Schr{\" o}dinger and Heisenberg equations, with Hamiltonian (\ref{B.1}), is already invariant under arbitrary transformations of the centrally-extended Galilean group \cite{Bargmann:1954gh,Levy-Leblond:1963qdx,Levy-Leblond:1967eic,Horzela:1991pa,Giulini:1995te,Csillag:2003}. 
Hence, we cannot reduce the spectrum of values, permitted by $\widehat{{\bf P}}$ (\ref{B.10}), to a {\bf single value}, without {\bf artificially} restricting the dynamics of the system, {\it beyond} the requirements imposed by Galilean symmetry. 

In the next section, App. \ref{App.B.2}, we consider the quantum mechanical analogue of ``jumping'' to an internal frame, associated with one of the particles in the closed system. 
As in the classical scenario, we will see that, if the reference particle interacts with other subsystems but no pseudo-forces are introduced to describe inertial effects, this ``jump'' should not be interpreted too literally.

\subsection{Quantum dynamics in an `internal' frame} \label{App.B.2}

The Hamiltonian (\ref{B.1}) can also be rewritten as
\begin{eqnarray} \label{B.20}
\widehat{H}({\bf x}_{\rm rel},{\bf p}_{\rm rel}) 
&=& \frac{\widehat{P}^2}{2M} + \sum_{J \neq I}^{N}\frac{m_J}{\mu_{IJ}}\frac{\widehat{p}_{J|I}^2}{2\mu_{IJ}} - \frac{1}{2M}\left(\sum_{J \neq I}^{N}\frac{m_J}{\mu_{IJ}}\widehat{{\bf p}}_{J|I}\right)^2 
\nonumber\\
&+& \sum_{J \neq I}\widehat{V}_{IJ}(|{\bf x}_{J|I}|) + \sum_{K<L} \widehat{V}_{KL}(|{\bf x}_{L|I}-{\bf x}_{K|I}|) \, , 
\end{eqnarray}
where $K, L \neq I$,
\begin{eqnarray} \label{B.21}
\widehat{{\bf x}}_{J|I} := \widehat{{\bf x}}_{J} - \widehat{{\bf x}}_{I} \, , 
\end{eqnarray}
and
\begin{eqnarray} \label{B.22}
\widehat{{\bf p}}_{J|I} := \mu_{IJ}\left(\frac{\widehat{{\bf p}}_J}{m_J}-\frac{\widehat{{\bf p}}_I}{m_I}\right)
= \mu_{IJ}(\widehat{\dot{{\bf x}}}_J-\widehat{\dot{{\bf x}}}_I)
= \mu_{IJ}\widehat{\dot{{\bf x}}}_{J|I} \, . 
\end{eqnarray}
As in the classical scenario, we must draw a distinction between $\widehat{{\bf p}}_{J|I}$ (\ref{B.22}) and the momentum of particle $J$ `seen'  by particle $I$,
\begin{eqnarray} \label{B.23}
\widehat{{\bf p}}_{J:I} := m_J\widehat{{\bf v}}_{J|I} = \frac{m_J}{\mu_{IJ}}\widehat{{\bf p}}_{J|I} \, . 
\end{eqnarray}
The quantum analogue of (\ref{A.17*}),
\begin{eqnarray} \label{B.24}
\widehat{\ddot{{\bf x}}}_I(t) = -\frac{1}{M}\sum_{J \neq I}m_J \widehat{\ddot{{\bf x}}}_{J|I}(t) \, ,
\end{eqnarray}
also holds identically. 

In the remainder of this section, we construct the quantum analogues of (\ref{A.17**})-(\ref{A.27*}) as well as the concrete unitary operators that give rise to the relevant (quantum) coordinate transformations. 
In fact, the quantum counterparts of (\ref{A.17**})-(\ref{A.27*}) are straightforward and we list them here only for completeness. 
However, to the best of our knowledge, the relevant unitaries have not appeared before in the literature. 
Hence, in the quantum regime -- in which no particle has a well-defined trajectory, including the chosen reference particle, $I$ \cite{Rae:2002,Isham:1995,Dirac:1958,vonNeumann:1955} -- the formal structure of the relevant equations still parallels that of the classical theory. 
In this case, the closest we can get to a ``jump'' from the external inertial frame $O$, to the `view' of $I$, without introducing (quantised) pseudo-forces, is the transformation  
\begin{eqnarray} \label{B.25}
\left\{\widehat{{\bf x}}_J,\widehat{{\bf p}}_J\right\}_{J=1}^{N} \mapsto \left\{\widehat{{\bf X}},\widehat{{\bf P}};\widehat{{\bf x}}_{J|I},\widehat{{\bf p}}_{J|I}\right\}_{J \neq I} \, . 
\end{eqnarray}

By analogy with the classical case, we have that
\begin{eqnarray} \label{B.26}
[(\widehat{{\bf x}}_{J|I})^i,(\widehat{{\bf p}}_{K|I})_j] \neq i\hbar \, \delta_{JK}\delta^{i}{}_{j}\widehat{\mathbb{I}} \, , \quad {\rm if} \, \,  N \geq 3 \, ,
\end{eqnarray}
which prompts us to search for an alternative set of canonically conjugate variables, satisfying the algebra
\begin{eqnarray} \label{B.26}
\left[\widehat{X}^{i},\widehat{P}_j\right] = i\hbar\delta^{i}{}_{j} \, \widehat{\mathbb{I}} \, , \quad \left[(\widehat{{\bf x}}_{J|I})^i,(\widehat{\boldsymbol{\pi}}_{K|I})_j\right] = i\hbar\delta_{JK}\delta^{i}{}_{j} \, \widehat{\mathbb{I}} \, , 
\nonumber\\
\left[\widehat{X}^i,(\widehat{\boldsymbol{\pi}}_{K|I})_j\right] = 0 \, , \quad \left[(\widehat{{\bf x}}_{J|I})^i,\widehat{P}_j\right] = 0 \, , 
\nonumber\\
\left[(\widehat{{\bf x}}_{J|I})^i,(\widehat{{\bf x}}_{K|I})^j\right] = 0 \, , \, \quad
\left[(\widehat{\boldsymbol{\pi}}_{J|I})_i,(\widehat{\boldsymbol{\pi}}_{K|I})_j\right] = 0 \, , 
\end{eqnarray}
and the explicit expression for $\widehat{\boldsymbol{\pi}}_{J|I}$ is 
\begin{eqnarray} \label{B.27}
\widehat{\boldsymbol{\pi}}_{J|I} := \widehat{{\bf p}}_J - \frac{m_J}{M}\widehat{{\bf P}} \, . 
\end{eqnarray}
Using these variables, the Hamiltonian (\ref{B.20}) takes the form
\begin{eqnarray} \label{B.28}
\widehat{H}({\bf x}_{\rm rel},\boldsymbol{\pi}_{\rm rel}) &=& \frac{\widehat{P}^2}{2M} + \sum_{J \neq I}\frac{\widehat{\pi}_{J|I}^2}{2m_J} + \frac{1}{2m_I}\left(\sum_{J \neq I}\widehat{\boldsymbol{\pi}}_{J|I}\right)^2 
\nonumber\\
&+& \sum_{J \neq I}\widehat{V}_{IJ}(|{\bf x}_{J|I}|)  + \sum_{K < L} \widehat{V}_{KL}(|{\bf x}_{L|I} - {\bf x}_{K|I}|) \, ,  
\end{eqnarray}
with $K,L \neq I$, and the Heisenberg equation yields 
\begin{eqnarray} \label{B.29}
\widehat{{\bf P}} = M\widehat{\dot{{\bf X}}} \, , \quad \widehat{\dot{{\bf P}}}= 0 \, ,
\nonumber\\
\widehat{\dot{{\bf x}}}_{J|I} = \frac{\widehat{\boldsymbol{\pi}}_{J|I} }{m_J}+ \frac{1}{m_I}\sum_{K \neq I}\widehat{\boldsymbol{\pi}}_{K|I}
\, , \quad
\widehat{\dot{\boldsymbol{\pi}}}_{J|I} = -\left\{\sum_{K \neq I}\frac{\partial \widehat{V}_{IK}}{\partial {\bf x}_{J|I}} + \sum_{K < L} \frac{\partial \widehat{V}_{KL}}{\partial {\bf x}_{J|I}}\right\} \, ,
\end{eqnarray}
also by analogy with the classical case. 
Equations (\ref{B.29}) can then be combined to give
\begin{eqnarray} \label{B.29*}
\widehat{\ddot{{\bf X}}} = {\bf 0} \, ,
\nonumber\\
\widehat{\ddot{{\bf x}}}_{J|I} = -\frac{1}{m_J}\left\{\sum_{K \neq I}\frac{\partial \widehat{V}_{IK}}{\partial {\bf x}_{J|I}} + \sum_{K < L} \frac{\partial \widehat{V}_{KL}}{\partial {\bf x}_{J|I}}\right\} - \frac{1}{m_I}\sum_{K \neq I}\left\{\sum_{L \neq I}\frac{\partial \widehat{V}_{IL}}{\partial {\bf x}_{K|I}} + \sum_{M < L} \frac{\partial \widehat{V}_{LM}}{\partial {\bf x}_{K|I}}\right\} \, .
\end{eqnarray}  

If the new variables are expanded in terms of the canonical bases $\left\{\bigotimes_{J=1}^{N}|{\bf x}_J\rangle_J ; \bigotimes_{J=1}^{N}|{\bf p}_J\rangle_J \right\}$, then the change of quantum coordinates 
\begin{eqnarray} \label{B.30}
\left\{\widehat{{\bf x}}_J,\widehat{{\bf p}}_J\right\}_{J = 1}^{N} \mapsto \left\{\widehat{{\bf X}},\widehat{{\bf P}};\widehat{{\bf x}}_{J|I},\widehat{\boldsymbol{\pi}}_{J|I}\right\}_{J \neq I}
\end{eqnarray}
is enacted in the same way as its classical counterpart (\ref{A.22}), but with hats on the relevant quantities. 
The additional relations, required to perform the inverse transformation, are 
\begin{eqnarray} \label{B.31}
\widehat{{\bf x}}_I = \widehat{{\bf X}} - \frac{1}{M}\sum_{J \neq I}m_J \widehat{{\bf x}}_{J|I} \, , \quad {\bf p}_I = \frac{m_I}{M}\widehat{{\bf P}} - \sum_{J \neq I}\widehat{\boldsymbol{\pi}}_{J|I} \, .
\end{eqnarray}

If, however, we wish to obtain expressions for our new operators in terms of their natural bases, $\left\{\bigotimes_{J = 1}^{I-1}|{\bf x}_{J|I}\rangle_{J|I}|{\bf X}\rangle_{\rm CoM}\bigotimes_{J = I+1}^{N}|{\bf x}_{J|I}\rangle_{J|I} ; \bigotimes_{J = 1}^{I-1}|\boldsymbol{\pi}_{J|I}\rangle_{J|I} |{\bf P}\rangle_{\rm CoM}\bigotimes_{J = I+1}^{N}|\boldsymbol{\pi}_{J|I}\rangle_{J|I} \right\}$, we must perform an {\bf active unitary transformation}, given by the operator
\begin{eqnarray} \label{B.32}
\widehat{B}_{\rm O \rightarrow I} := \widehat{S}_{I \rightarrow {\rm CoM}}\prod_{J \neq I}\widehat{S}_{J \rightarrow J|I}
\times
\exp\left[-\frac{i}{\hbar}\widehat{{\bf p}}_I. \left(\widehat{{\bf X}} - \frac{m_I}{M}\widehat{{\bf x}}_I\right)\right]
\exp\left[\frac{i}{\hbar}(\widehat{{\bf P}} - \widehat{{\bf p}}_I).\widehat{{\bf x}}_I \right] \, ,
\end{eqnarray}
where $\widehat{S}_{A \rightarrow B}$ is the label-swap operator, which acts on the ket $| \dots \rangle_A$ such that $\widehat{S}_{A \rightarrow B}| \dots \rangle_A = | \dots \rangle_B$. 
If $I$ interacts with other particles, the action of $\widehat{B}_{\rm O \rightarrow I}$ gives the closest description we can obtain, to what it `sees' in real space, without introducing quantised pseudo-forces via the maps $\widehat{{\bf p}}_J \mapsto \widehat{{\bf p}}_J - m_J\widehat{\dot{{\bf x}}}_I$ for $J \neq I$. 
(See Sec. \ref{Sec.3.1}.)
This operator acts on the canonical position and momentum spaces bases such that
\begin{eqnarray} \label{B.33}
\widehat{B}_{\rm O \rightarrow I} \bigotimes_{J = 1}^{N} |{\bf x}_J\rangle_{J} 
= \left(\, \, \bigotimes_{J = 1}^{I-1} |{\bf x}_{J|I}\rangle_{J|I}\right) |{\bf X}\rangle_{\rm CoM} \left(\bigotimes_{J = I+1}^{N} |{\bf x}_{J|I}\rangle_{J|I}\right) \, , 
\nonumber\\
\widehat{B}_{\rm O \rightarrow I} \bigotimes_{J = 1}^{N} |{\bf p}_J\rangle_{J} =  
\left(\, \, \bigotimes_{J = 1}^{I-1} |\boldsymbol{\pi}_{J|I}\rangle_{J|I}\right) |{\bf P}\rangle_{\rm CoM} \left(\bigotimes_{J = I+1}^{N} |\boldsymbol{\pi}_{J|I}\rangle_{J|I}\right) \, ,
\end{eqnarray}
and the Jacobian for the corresponding classical coordinate transformation (\ref{A.22}) is unity, so that  
\begin{eqnarray} \label{B.34}
\prod_{J = 1}^{N} {\rm d}^3{\bf x}_J = {\rm d}^3{\bf X}  \prod_{J \neq I} {\rm d}^3{\bf x}_{J|I} \, , \quad
\prod_{J = 1}^{N} {\rm d}^3{\bf p}_J = {\rm d}^3{\bf P} \prod_{J \neq I} {\rm d}^3\boldsymbol{\pi}_{J|I} \, .
\end{eqnarray}
Using (\ref{B.33})-(\ref{B.34}) and (\ref{A.22}), together, we then have
\begin{eqnarray} \label{B.35}
&&\widehat{B}_{\rm O \rightarrow I} \, \widehat{{\bf x}}_I \, \widehat{B}_{\rm O \rightarrow I}^{\dagger} 
= \widehat{{\bf X}} - \frac{1}{M}\sum_{K \neq I}m_{K}\widehat{{\bf x}}_{K|I} \, ,  
\nonumber\\ 
&&\widehat{B}_{\rm O \rightarrow I} \, \widehat{{\bf x}}_{J} \, \widehat{B}_{\rm O \rightarrow I}^{\dagger} 
= \widehat{{\bf x}}_{J|I} + \widehat{{\bf X}} - \frac{1}{M}\sum_{K \neq I}m_{K}\widehat{{\bf x}}_{K|I} \, ,  
\nonumber\\
&&\widehat{B}_{\rm O \rightarrow I} \, \widehat{{\bf p}}_I \, \widehat{B}_{\rm O \rightarrow I}^{\dagger} 
= \frac{m_I}{M}\widehat{{\bf P}} - \sum_{K \neq I}\widehat{\boldsymbol{\pi}}_{K|I} \, ,  
\nonumber\\ 
&&\widehat{B}_{\rm O \rightarrow I} \, \widehat{{\bf p}}_{J} \, \widehat{B}_{\rm O \rightarrow I}^{\dagger} 
= \widehat{\boldsymbol{\pi}}_{J|I} + \frac{m_J}{M}\widehat{{\bf P}} \, , \, \,  (J \neq I) \, . 
\end{eqnarray}
Note that, strictly, the operators $\widehat{{\bf X}}$, $\widehat{{\bf P}}$ and $\widehat{{\bf x}}_{J|I}$, $\widehat{\boldsymbol{\pi}}_{J|I}$, on the right-hand sides of (\ref{B.35}), are not those defined previously in (\ref{B.7}), (\ref{B.6}) and (\ref{B.21}), (\ref{B.27}). 
But they are unitarily equivalent to these operators, respectively, and belong to the same equivalence classes \cite{Rae:2002,Isham:1995,Dirac:1958,vonNeumann:1955}.
They represent the same set of observables, expressed in terms of a different basis. 
The Hamiltonians $\widehat{H}({\bf x},{\bf p})$ (\ref{B.1}) and $\widehat{H}({\bf x}_{\rm rel},\boldsymbol{\pi}_{\rm rel})$ (\ref{B.28}) are also unitarily equivalent, in this sense:
\begin{eqnarray} \label{B.35*}
\widehat{H}({\bf x}_{\rm rel},\boldsymbol{\pi}_{\rm rel}) = \widehat{B}_{\rm O \rightarrow I} \widehat{H}({\bf x},{\bf p}) \widehat{B}_{\rm O \rightarrow I}^{\dagger} \, . 
\end{eqnarray}

We may also transform the non-spin part of the canonical $N$-particle wave function,
\begin{eqnarray} \label{B.36}
|\Psi\rangle &=& \int \Psi\left(\left\{{\bf x}_J\right\}_{J=1}^{N}\right)\bigotimes_{J = 1}^{N} |{\bf x}_J\rangle_{J}\prod_{J = 1}^{N} {\rm d}^3{\bf x}_J 
\nonumber\\
&=& \int \tilde{\Psi}\left(\left\{{\bf p}_J\right\}_{J=1}^{N}\right)\bigotimes_{J = 1}^{N} |{\bf p}_J\rangle_{J}\prod_{J = 1}^{N} {\rm d}^3{\bf p}_J \, , 
\end{eqnarray}
by acting with $\widehat{B}_{\rm O \rightarrow I}$ (\ref{B.32}), yielding a {\bf passive unitary transformation}. 
In this case, $|\Psi\rangle \mapsto |\Psi'\rangle := \widehat{B}_{\rm O \rightarrow I}|\Psi\rangle$, where
\begin{eqnarray} \label{B.37}
|\Psi'\rangle &=& \int \Psi'\left(\left\{{\bf X},{\bf x}_{J|I}\right\}_{J \neq I}\right) \bigotimes_{J = 1}^{I-1} |{\bf x}_{J|I}\rangle_{J|I} |{\bf X}\rangle_{\rm CoM} \bigotimes_{J = I+1}^{N} |{\bf x}_{J|I}\rangle_{J|I} \, {\rm d}^3{\bf X}\prod_{J \neq I} {\rm d}^3{\bf x}_{J|I} 
\nonumber\\
&=& \int \tilde{\Psi}'\left(\left\{{\bf P},\boldsymbol{\pi}_{J|I}\right\}_{J \neq I}\right) \bigotimes_{J = 1}^{I-1} |\boldsymbol{\pi}_{J|I}\rangle_{J|I} |{\bf P}\rangle_{\rm CoM} \bigotimes_{J = I+1}^{N} |\boldsymbol{\pi}_{J|I}\rangle_{J|I} \, {\rm d}^3{\bf P} \prod_{J \neq I} {\rm d}^3\boldsymbol{\pi}_{J|I}  \, ,
\end{eqnarray}
and the functions $\Psi'$ and $\tilde{\Psi}'$ are defined as
\begin{eqnarray} \label{B.38}
\Psi'\left(\left\{{\bf X},{\bf x}_{J|I}\right\}_{J \neq I}\right) := \Psi\left(\left\{{\bf x}_J\right\}_{J=1}^{N}\right) \, , \quad
\tilde{\Psi}'\left(\left\{{\bf P},\boldsymbol{\pi}_{J|I}\right\}_{J \neq I}\right) := \tilde{\Psi}\left(\left\{{\bf p}_J\right\}_{J=1}^{N}\right) \, ,
\end{eqnarray}
using (\ref{A.22}) to perform the relevant substitutions. 

Finally, we consider the alternative set of canonically conjugate variables, satisfying the algebra
\begin{eqnarray} \label{B.45}
\left[\widehat{X}^{i},\widehat{P}_j\right] = i\hbar\delta^{i}{}_{j} \, \widehat{\mathbb{I}} \, , \quad \left[(\widehat{{\bf q}}_{J|I})^i,(\widehat{{\bf p}}_{K|I})_j\right] = i\hbar \delta_{JK}\delta^{i}{}_{j} \, \widehat{\mathbb{I}} \, , 
\nonumber\\
\left[\widehat{X}^i,(\widehat{{\bf p}}_{K|I})_j\right] = 0 \, , \quad \left[(\widehat{{\bf q}}_{J|I})^i,\widehat{P}_j\right] = 0 \, , 
\nonumber\\
\left[(\widehat{{\bf q}}_{J|I})^i,(\widehat{{\bf q}}_{K|I})^j\right] = 0 \, , \, 
\left[(\widehat{{\bf p}}_{J|I})_i,(\widehat{{\bf p}}_{K|I})_j\right] = 0 \, . 
\end{eqnarray}
The explicit expression for $\widehat{{\bf q}}_{J|I}$ is 
\begin{eqnarray} \label{B.46}
\widehat{{\bf q}}_{J|I} := \frac{m_J}{\mu_{IJ}}(\widehat{{\bf x}}_J - \widehat{{\bf X}}) \, , 
\end{eqnarray}
and the transformation 
\begin{eqnarray} \label{B.47}
\left\{\widehat{{\bf x}}_J,\widehat{{\bf p}}_J\right\}_{J = 1}^{N} \mapsto \left\{\widehat{{\bf X}},\widehat{{\bf P}};\widehat{{\bf q}}_{J|I},\widehat{{\bf p}}_{J|I}\right\}_{J \neq I} 
\end{eqnarray}
is also invertible. 
The additional relations we need, in order to perform the inversion, are
\begin{eqnarray} \label{B.48}
\widehat{{\bf x}}_I = \widehat{{\bf X}} - \frac{1}{m_I}\sum_{J \neq I}\mu_{IJ}\widehat{{\bf q}}_{J|I} \, , 
\quad
\widehat{{\bf p}}_I = \frac{m_I}{M}\left(\widehat{{\bf P}} - \sum_{J \neq I}\frac{m_J}{\mu_{IJ}}\widehat{{\bf p}}_{J|I}\right) \, . 
\end{eqnarray}
In this coordinate system, the canonical Hamiltonian is written as
\begin{eqnarray} \label{B.48*}
\widehat{H}({\bf q}_{\rm rel},{\bf p}_{\rm rel}) 
&=&  \frac{\widehat{P}^2}{2M} + \sum_{J \neq I}^{N}\frac{m_J}{\mu_{IJ}}\frac{\widehat{p}_{J|I}^2}{2\mu_{IJ}} - \frac{1}{2M}\left(\sum_{J \neq I}^{N}\frac{m_J}{\mu_{IJ}}\widehat{{\bf p}}_{J|I}\right)^2 
\nonumber\\
&+& \sum_{J \neq I}\widehat{V}_{IJ}\left(\bigg|\frac{\mu_{IJ}}{m_J}{\bf q}_{J|I} + \frac{1}{m_I}\sum_{K \neq I}\mu_{IK}{\bf q}_{K|I}\bigg|\right) 
\nonumber\\
&+& \sum_{K < L} \widehat{V}_{KL}\left(\bigg|\frac{\mu_{IL}}{m_L}{\bf q}_{L|I} - \frac{\mu_{IK}}{m_K}{\bf q}_{K|I}\bigg|\right) \, ,  
\end{eqnarray}
and the Heisenberg equation yields the equations of motion
\begin{eqnarray} \label{B.48**}
\widehat{{\bf P}} = M\widehat{\dot{{\bf X}}} \, , \quad \widehat{\dot{{\bf P}}} = 0 \, ,
\nonumber\\
\widehat{\dot{{\bf q}}}_{J|I} = \frac{m_J}{\mu_{IJ}}\left(\frac{\widehat{{\bf p}}_{J|I}}{\mu_{IJ}} - \frac{1}{M}\sum_{K \neq I}\frac{m_K}{\mu_{IK}}\widehat{{\bf p}}_{K|I}\right)
\, , \quad
\widehat{\dot{{\bf p}}}_{J|I} = -\left\{\sum_{K \neq I}\frac{\partial \widehat{V}_{IK}}{\partial {\bf q}_{J|I}} + \sum_{K < L} \frac{\partial \widehat{V}_{KL}}{\partial {\bf q}_{J|I}}\right\} \, .
\end{eqnarray} 
by complete analogy with the classical case. 
Equations (\ref{B.48*}) can also be reformulated as
\begin{eqnarray} \label{B.48**}
\widehat{\ddot{{\bf X}}} = {\bf 0} \, , 
\nonumber\\
\widehat{\ddot{{\bf q}}}_{J|I} = -\frac{m_J}{\mu_{IJ}^2}\left\{\sum_{K \neq I}\frac{\partial \widehat{V}_{IK}}{\partial {\bf q}_{J|I}} + \sum_{K < L} \frac{\partial \widehat{V}_{KL}}{\partial {\bf q}_{J|I}}\right\} 
+ \frac{1}{M}\sum_{K \neq I}\frac{m_K}{\mu_{IK}} \left\{\sum_{L \neq I}\frac{\partial \widehat{V}_{IL}}{\partial {\bf q}_{K|I}} + \sum_{M < L} \frac{\partial \widehat{V}_{LM}}{\partial {\bf q}_{K|I}}\right\} \, . 
\end{eqnarray} 

Again, if the new quantum coordinates are expanded in terms of the canonical bases, the transformation (\ref{B.47}) proceeds as in the classical case, but with hats on the relevant quantities. 
If, however, we wish to obtain expressions for these operators in terms of their natural bases, $\left\{\bigotimes_{J = 1}^{I-1}|{\bf q}_{J|I}\rangle_{J|I}|{\bf X}\rangle_{\rm CoM}\bigotimes_{J = I+1}^{N}|{\bf q}_{J|I}\rangle_{J|I} ; \bigotimes_{J = 1}^{I-1}|{\bf p}_{J|I}\rangle_{J|I}|{\bf P}\rangle_{\rm CoM}\bigotimes_{J = I+1}^{N}|{\bf p}_{J|I}\rangle_{J|I} \right\}$, we must perform an {\bf active unitary transformation}, given by the operator
\begin{eqnarray} \label{B.49}
\widehat{\tilde{B}}_{\rm O \rightarrow I} &:=& \widehat{S}_{I \rightarrow {\rm CoM}}\prod_{J \neq I}\widehat{S}_{J \rightarrow J|I}  
\times 
\exp\left[-\frac{i}{\hbar} \boldsymbol{\gamma}.\sum_{J \neq I} \ln\left(\frac{m_J}{\mu_{IJ}}\right)\widehat{{\bf D}}_J\right]
\exp\left[\frac{i}{\hbar}(\widehat{{\bf P}} - \widehat{{\bf p}}_I).\widehat{{\bf x}}_I\right]
\nonumber\\
&\times& 
\exp\left[-\frac{i}{\hbar}\widehat{{\bf p}}_I.\left(\widehat{{\bf X}} - \frac{m_I}{M}\widehat{{\bf x}}_I\right)\right]
\exp\left[-\frac{i}{\hbar}\ln\left(\frac{m_I}{M}\right) \boldsymbol{\gamma}.\widehat{{\bf D}}_I\right] \, ,
\end{eqnarray}
where ${\bf \gamma} := [1,1,1]^{\rm T}$ and the individual components of the vector operator $\widehat{{\bf D}}_J$ are defined as the anti-commutators 
\begin{eqnarray} \label{B.49*}
\widehat{D}_J^i := \frac{1}{2}\left\{\widehat{x}_J^i,\widehat{p}_J^i \right\}_{+} \, .
\end{eqnarray}

If $I$ interacts with other particles the action of $\widehat{\tilde{B}}_{\rm O \rightarrow I}$ is the closest we can get, to a description of what it `sees' in velocity space, without again introducing quantised pseudo-forces. 
This operator acts on the canonical bases such that 
\begin{eqnarray} \label{B.50}
\widehat{\tilde{B}}_{\rm O \rightarrow I} \bigotimes_{J = 1}^{N} |{\bf x}_J\rangle_{J} 
= \mathcal{J}^{\frac{3}{2}}\left(\, \, \bigotimes_{J = 1}^{I-1} |{\bf q}_{J|I}\rangle_{J|I}\right)  |{\bf X}\rangle_{\rm CoM} \left(\bigotimes_{J = I+1}^{N} |{\bf q}_{J|I}\rangle_{J|I}\right) \, , 
\nonumber\\
\widehat{\tilde{B}}_{\rm O \rightarrow I} \bigotimes_{J = 1}^{N} |{\bf p}_J\rangle_{J} 
= \mathcal{J}^{-\frac{3}{2}}\left(\, \, \bigotimes_{J = 1}^{I-1} |{\bf p}_{J|I}\rangle_{J|I}\right)  |{\bf P}\rangle_{\rm CoM} \left(\bigotimes_{J = I+1}^{N} |{\bf p}_{J|I}\rangle_{J|I}\right) \, .
\end{eqnarray}
where
\begin{eqnarray} \label{B.50*}
\mathcal{J} = {\rm det}\left[\frac{\partial({\bf X},{\bf q}_{J|I})}{\partial {\bf x}}\right] = \frac{m_I}{M} \prod_{J \neq I}\frac{m_J}{\mu_{IJ}} \, , 
\nonumber\\
\mathcal{J}^{-1} = {\rm det}\left[\frac{\partial({\bf P},{\bf p}_{J|I})}{\partial {\bf p}}\right] = \left(\frac{m_I}{M} \prod_{J \neq I}\frac{m_J}{\mu_{IJ}}\right)^{-1} \, , 
\end{eqnarray}
from (\ref{A.26}), giving
\begin{eqnarray} \label{B.51}
\prod_{J = 1}^{N} {\rm d}^3{\bf x}_J = \mathcal{J}^{-1}{\rm d}^3{\bf X} \prod_{J \neq I} {\rm d}^3{\bf q}_{J|I} \, , \quad
\prod_{J = 1}^{N} {\rm d}^3{\bf p}_J = \mathcal{J}^3{\rm d}^3{\bf P} \prod_{J \neq I} {\rm d}^3{\bf p}_{J|I} \, .
\end{eqnarray}
Using (\ref{B.49})-(\ref{B.50}) and (\ref{A.26}), together, we have
\begin{eqnarray} \label{B.52}
&&\widehat{\tilde{B}}_{\rm O \rightarrow I} \, \widehat{{\bf x}}_I \, \widehat{\tilde{B}}_{\rm O \rightarrow I}^{\dagger} 
= \widehat{{\bf X}} - \frac{1}{m_I}\sum_{K \neq I}\mu_{IK}\widehat{{\bf q}}_{K|I} \, ,  
\nonumber\\ 
&&\widehat{\tilde{B}}_{\rm O \rightarrow I} \, \widehat{{\bf x}}_{J} \, \widehat{\tilde{B}}_{\rm O \rightarrow I}^{\dagger} 
= \frac{\mu_{IJ}}{m_J}\widehat{{\bf q}}_{J|I} + \widehat{{\bf X}} \, ,  
\nonumber\\
&&\widehat{\tilde{B}}_{\rm O \rightarrow I} \, \widehat{{\bf p}}_I \, \widehat{\tilde{B}}_{\rm O \rightarrow I}^{\dagger} 
= \frac{m_I}{M}\left(\widehat{{\bf P}} - \sum_{K \neq I}\frac{m_K}{\mu_{IK}}\widehat{{\bf p}}_{K|I}\right) \, ,  
\nonumber\\ 
&&\widehat{\tilde{B}}_{\rm O \rightarrow I} \, \widehat{{\bf p}}_{J} \, \widehat{\tilde{B}}_{\rm O \rightarrow I}^{\dagger} 
= \frac{m_J}{\mu_{IJ}}\widehat{{\bf p}}_{J|I} + \frac{m_J}{M}\left(\widehat{{\bf P}} - \sum_{K \neq I}\frac{m_K}{\mu_{IK}}\widehat{{\bf p}}_{K|I}\right)\, , \, \, (J \neq I) \, . 
\end{eqnarray}
The operators $\widehat{{\bf X}}$, $\widehat{{\bf P}}$ and $\widehat{{\bf q}}_{J|I}$, $\widehat{{\bf p}}_{J|I}$ are not the same as those defined previously, in (\ref{B.7}), (\ref{B.6}) and (\ref{B.46}), (\ref{B.22}). 
But they are unitarily equivalent to these operators, respectively, and belong to the same equivalence classes \cite{Rae:2002,Isham:1995,Dirac:1958,vonNeumann:1955}. 
Yet again, they represent the same set of observables, expressed in terms of a different basis. 
The Hamiltonians $\widehat{H}({\bf x},{\bf p})$ (\ref{B.1}) and $\widehat{H}({\bf q}_{\rm rel},{\bf p}_{\rm rel})$ (\ref{B.48*}) are also unitarily equivalent, in the same sense:
\begin{eqnarray} \label{B.52*}
\widehat{H}({\bf q}_{\rm rel},{\bf p}_{\rm rel}) = \widehat{\tilde{B}}_{\rm O \rightarrow I} \widehat{H}({\bf x},{\bf p}) \widehat{\tilde{B}}_{\rm O \rightarrow I}^{\dagger}
\end{eqnarray}

We may also act with $\widehat{\tilde{B}}_{\rm O \rightarrow I}$ (\ref{B.49}) on the non-spin part of the canonical $N$-particle wave function, (\ref{B.36}), yielding a {\bf passive unitary transformation}. 
In this case, $|\Psi\rangle \mapsto |\Psi''\rangle := \widehat{\tilde{B}}_{\rm O \rightarrow I}|\Psi\rangle$, where
\begin{eqnarray} \label{B.53}
|\Psi''\rangle &=& \int \Psi''\left(\left\{{\bf X},{\bf q}_{J|I}\right\}_{J \neq I}\right) 
\bigotimes_{J = 1}^{I-1} |{\bf q}_{J|I}\rangle_{J|I}  |{\bf X}\rangle_{\rm CoM} \bigotimes_{J = I+1}^{N} |{\bf q}_{J|I}\rangle_{J|I} \, {\rm d}^3{\bf X}\prod_{J \neq I} {\rm d}^3{\bf q}_{J|I} 
\nonumber\\
&=& \int \tilde{\Psi}''\left(\left\{{\bf P},{\bf p}_{J|I}\right\}_{J \neq I}\right) 
\bigotimes_{J = 1}^{I-1} |{\bf p}_{J|I}\rangle_{J|I} |{\bf P}\rangle_{\rm CoM} \bigotimes_{J = I+1}^{N} |{\bf p}_{J|I}\rangle_{J|I} \, {\rm d}^3{\bf P}\prod_{J \neq I} {\rm d}^3{\bf p}_{J|I} \, .
\end{eqnarray}
and the functions $\Psi''$ and $\tilde{\Psi}''$ are defined by the relations
\begin{eqnarray} \label{B.54}
\Psi''\left(\left\{{\bf X},{\bf q}_{J|I}\right\}_{J \neq I}\right) := \mathcal{J}^{-\frac{3}{2}}\Psi\left(\left\{{\bf x}_J\right\}_{J=1}^{N}\right) \, , \quad
\tilde{\Psi}''\left(\left\{{\bf P},{\bf p}_{J|I}\right\}_{J \neq I}\right) := \mathcal{J}^{\frac{3}{2}}\tilde{\Psi}\left(\left\{{\bf p}_J\right\}_{J=1}^{N}\right) \, ,
\end{eqnarray}
using (\ref{A.26}) to perform the relevant substitutions. 

We stress, once more, that neither $\widehat{B}_{\rm O \rightarrow I}$ nor $\widehat{\tilde{B}}_{\rm O \rightarrow I}$ represent a {\bf literal} ``jump'' from the external inertial frame $O$ to the accelerated frame of particle $I$, when the latter is interacting. 
Nonetheless, they have physical significance, since they map {\bf unobservable} quantities to {\bf observable} ones -- whilst, strictly, remaining `within the frame' of $O$. 
It is straightforward to show that neither $\left\{\widehat{{\bf x}}_J,\widehat{{\bf p}}_J\right\}_{J=1}^{N}$ nor $\left\{\widehat{{\bf X}},\widehat{{\bf P}}\right\}$ are observable, since they are not invariant under translations or Galilean boosts. 
Conversely, both $\left\{\widehat{{\bf x}}_{J|I},\widehat{\boldsymbol{\pi}}_{J|I}\right\}_{J \neq I}$ and $\left\{\widehat{{\bf q}}_{J|I},\widehat{{\bf p}}_{J|I}\right\}_{J \neq I}$ are invariant under these transformations, and, therefore, represent observables \cite{Bargmann:1954gh,Levy-Leblond:1963qdx,Levy-Leblond:1967eic,Horzela:1991pa,Giulini:1995te,Csillag:2003}. 

Both $\widehat{{\bf q}}_{J|I}$ and $\widehat{\boldsymbol{\pi}}_{J|I}$ are defined relationally, with respect to the centre of mass, whereas $\widehat{{\bf x}}_{J|I}$ and $\widehat{{\bf p}}_{J|I}$ are relational, with respect to particle $I$. 
In this sense, and in this sense only, do the latter represent the ``QRF associated with $I$''. 
The phase space coordinate operators, $\left\{\widehat{{\bf X}},\widehat{{\bf x}}_{J|I};\widehat{{\bf P}},\widehat{\boldsymbol{\pi}}_{J|I}\right\}_{J \neq I}$ and $\left\{\widehat{{\bf X}},\widehat{{\bf q}}_{J|I};\widehat{{\bf P}},\widehat{{\bf p}}_{J|I}\right\}_{J \neq I}$, are still {\bf inertial}. 
That is, they are defined relative to a purely abstract, and classical, inertial frame, {\bf external} to the closed system of material quantum particles.  

\subsection{Galilean symmetries in canonical quantum mechanics} \label{App.B.3}

When we talk about the ``Galilean invariance'' of canonical quantum systems, we actually mean invariance under the {\bf commutator representation} of the {\bf centrally-extended Galilean algebra},
\begin{eqnarray} \label{CC_Gal_Alg}
\left[\widehat{J}_i,\widehat{J}_j\right] = i\hbar\epsilon_{ij}{}^{k}\widehat{J}_k \, , \quad \left[\widehat{J}_i,\widehat{G}_j\right] = i\hbar\epsilon_{ij}{}^{k}\widehat{G}_k  \, , \quad \left[\widehat{J}_i,\widehat{P}_j\right] = i\hbar\epsilon_{ij}{}^{k}\widehat{P}_k \, , 
\nonumber\\
\left[\widehat{J}_i,\widehat{H}\right] = 0 \, , \quad \left[\widehat{G}_i,\widehat{G}_j\right] = 0  \, , \quad \left[\widehat{G}_i,\widehat{P}_j\right] = i\hbar M\delta_{ij} \, \widehat{\mathbb{I}} \, , 
\nonumber\\
\left[\widehat{G}_i,\widehat{H}\right] = i\hbar\widehat{P}_i \, , \quad \left[\widehat{P}_i,\widehat{P}_j\right] = 0  \, , \quad \left[\widehat{P}_i,\widehat{H}\right] = 0 \, , 
\end{eqnarray}
where
\begin{eqnarray} \label{CC_Gal_Alg_gen}
\widehat{H} := \frac{\widehat{P}^2}{2M} + \widehat{U} \, , \quad \widehat{{\bf P}} := \sum_{J = 1}^{N}\widehat{{\bf p}}_J \, , \quad 
\widehat{{\bf J}} := \widehat{{\bf L}} + \widehat{{\bf S}} = \sum_{J = 1}^{N}  \widehat{{\bf x}}_J \times \widehat{{\bf p}}_J + \sum_{J = 1}^{N} \widehat{{\bf s}}_J \, , 
\nonumber\\
{\rm and} \, \, \widehat{{\bf G}} := M\widehat{{\bf X}} - t\widehat{{\bf P}} \, , \, {\rm with} \, \, \widehat{{\bf X}} := \frac{1}{M}\sum_{J = 1}^{N} m_J\widehat{{\bf x}}_J \, .  
\end{eqnarray}
In the commutator representation, the independent {\bf Casimir operators} are \cite{Levy-Leblond:1963qdx,Levy-Leblond:1967eic}
\begin{eqnarray} \label{CC_Gal_Alg_Casimirs}
\widehat{C}_1 = \widehat{M} := M\widehat{\mathbb{I}} \, , \quad \widehat{C}_2 := \widehat{U} := \widehat{H} - \frac{\widehat{P}^2}{2M}  \, , \quad 
\widehat{C}_3 := \big|\widehat{{\bf J}} - \frac{1}{M}\widehat{{\bf G}} \times \widehat{{\bf P}} \big|^2 \, .
\end{eqnarray}
$M$ is again the total mass of the system, $\widehat{U}$ its internal energy, and the third invariant is the non-relativistic analogue of the Pauli-Lubansky pseudo-vector in canonical QFT \cite{Tong-QFT:2007,Padmanabhan-QFT:2016,Parker&Toms:2009,Wald-QFT:2004}. 

It may further be shown that the individual spin vector of the $J^{\rm th}$ particle is given by $\widehat{{\bf s}}_J = (\hbar/2)\sigma^{i}(s_J){\bf e}_i$, where $\sigma^{i}(s_J)$ is the $s_J$-representation of the $i^{\rm th}$ $SU(2)$ generator, which satisfies the algebra $[\sigma^{i},\sigma^{j}] = 2i\epsilon^{ij}{}_{k}\sigma^{k}$ with $s_J \in \mathbb{Z}^+/2$\cite{Rae:2002,Isham:1995,Dirac:1958,vonNeumann:1955}. 
The third invariant may be rewritten as $|\widehat{{\bf L}}_{\rm rel} + \widehat{{\bf S}}|^2$, where $\widehat{{\bf L}}_{\rm rel} := \widehat{{\bf L}} - \widehat{{\bf X}} \times \widehat{{\bf P}}$ is the relative angular momentum of the system and $\widehat{{\bf S}}$ is its total spin. 
This operator possesses eigenvectors with eigenvalues of the form $\lambda(\lambda + 1)\hbar^2$, with $\lambda \in \mathbb{Z}^+/2$ \cite{Levy-Leblond:1963qdx,Levy-Leblond:1967eic}.

In the commutator representation, there is no need for unit numerical factors, and the constant with units of inverse action may be written explicitly as $\hbar^{-1}$. 
The operators
\begin{eqnarray} \label{PB_Gal_Alg_Ops}
\widehat{S}_{\rm T}({\bf x}_0) := \exp\left(\frac{i}{\hbar}\widehat{{\bf P}}.{\bf x}_0\right) \, , \quad
\widehat{S}_{\rm B}({\bf v}) := \exp\left(\frac{i}{\hbar}\widehat{{\bf G}}.{\bf v}\right) \, , 
\nonumber\\
\widehat{S}_{\rm R}(\boldsymbol{\theta}) := \exp\left(\frac{i}{\hbar}\widehat{{\bf J}}.\boldsymbol{\theta}\right) \, , \quad
\widehat{S}_{0}(t) := \exp\left(-\frac{i}{\hbar}\widehat{H}_{\rm CoM}t\right) \, , \,
\end{eqnarray}
enact spatial translations by ${\bf x}_0$, boosts by velocity $-{\bf v}$, rotations by $\theta$ about the unit vector ${\bf n}$, where $\boldsymbol{\theta} := {\bf n}\theta$, and temporal translations by time $t$, respectively. 
A general element of the centrally-extended Galilean group is represented by the operator \cite{Bargmann:1954gh,Levy-Leblond:1963qdx,Levy-Leblond:1967eic,Horzela:1991pa,Giulini:1995te,Csillag:2003}
\begin{eqnarray} \label{CC_Gal_Alg_G}
S_{\rm Gal}({\bf x}_0,{\bf v},\boldsymbol{\theta},t) := \exp\left[\frac{i}{\hbar}\left({\bf P}.{\bf x}_0 + {\bf G}.{\bf v} + {\bf J}.\boldsymbol{\theta} - H_{\rm CoM}t\right)\right] \, .
\end{eqnarray}
This representation is obtained from (\ref{PB_Gal_Alg_G}) by performing the {\bf canonical quantisation},
\begin{eqnarray} \label{Canon_Quant}
\left\{O_1,O_2\right\}_{\rm PB} \mapsto \frac{1}{i\hbar}[\widehat{O}_1,\widehat{O}_2] \, , \, \, {\rm where} \, \, \left\{O_1,O_2\right\}_{\rm PB} = \lim_{\hbar \rightarrow 0}\frac{1}{i\hbar}[\widehat{O}_1,\widehat{O}_2] \, .
\end{eqnarray}

\section{Quantisation schemes for constrained Hamiltonian systems} \label{App.C}

\subsection{Primary versus secondary and first-class versus second-class constraints} \label{App.C.1}

A {\bf primary constraint} is a constraint that holds irrespective of whether the equations of motion for the classical system are satisfied. 
A {\bf secondary constraint} is one that is required, in order to ensure that the primary constraints are preserved in time \cite{Dirac-CHS:1958,Dirac-CHS:1964,Henneaux-Teitelboim:1992,Rothe-RotheCHS:2010}. 

This terminology should not be confused with that for first- and second-class constraints. 
A {\bf first-class constraint}, $\phi_a \approx {\rm const}.$, is one that weakly commutes with all other constraints ($\left\{\phi_a,\phi_b\right\}_{\rm PB} \approx 0$ for all $b \in \left\{1,2,\dots\right\}$), whereas a {\bf second-class constraint} does not. 
Here, the notion of a {\bf weak equality}, denoted ``$\approx$'', indicates that equality holds only on the relevant {\bf constraint surface} in the classical phase space.
In the {\bf classical} theory of {\bf constrained Hamiltonian systems}, the number of physical degrees of freedom in a system is given by
\begin{eqnarray} \label{count}
N_{\rm phys} = \frac{1}{2}(N_{\rm var} - 2N_{\rm first-class} - N_{\rm second-class}) \, , 
\end{eqnarray}
where the quantities on the right-hand side refer to the number of canonical variables, the number of first-class constraints, and the number of second-class constraints, respectively. 
This count is taken over the full first-class set, including both primary and secondary constraints \cite{Dirac-CHS:1958,Dirac-CHS:1964,Henneaux-Teitelboim:1992,Rothe-RotheCHS:2010}.

It is important to understand that, in such systems, the constraints may come from two qualitatively different sources. 
In {\bf open systems}, they may be imposed by an ``external system''. 
For example, one may consider charged particles in three-dimensional space, that are constrained to move within a one-, two-, or three-dimensional submanifold, by the action of an external potential. 
In this case, the back-reaction of the particles on the potential is neglected and the latter does not form part of the dynamical description. 
\footnote{Practically all systems considered in introductory quantum mechanics courses arise from classical systems of this kind. Hence, these models represent approximate open-system descriptions of exact closed-system physics \cite{Rae:2002,Isham:1995,Dirac:1958,vonNeumann:1955}.} 
Or one may consider a ball rolling down a slope without sliding. 
In this case, the slope forms part of the ``external system'' and the constraint represents a restriction, reflecting a physical assumption, to a subset of all possible solutions to the equation of motion \cite{Dirac-CHS:1958,Dirac-CHS:1964,Henneaux-Teitelboim:1992,Rothe-RotheCHS:2010}.

In {\bf closed} or {\bf isolated systems}, however, there is no ``external system'', capable of generating constraints. 
In this case, the only possible sources of constraints are certain symmetry groups, known as ``gauge groups'' \cite{GaugeTheory:Healy2007}, under which the dynamics of the system must remain invariant. 
But what makes a group ``gauge''? 

A standard distinction is to split the total symmetry group of a system into two subgroups: the {\bf spacetime symmetry group} and the {\bf gauge group} proper. 
This is a good working definition, and a standard one in the particle physics literature; for example, it is well-known that the spacetime symmetry group of the Standard Model is the Poincar{\' e} group, $\mathbb{R}^{1,3} \rtimes SO^{+}(1,3)$, whereas its gauge group is given by the direct product $U(1) \times SU(2) \times SU(3)$ \cite{SM}.
But one may reasonably ask if there is any fundamental difference -- stemming from the nature of the symmetries themselves, and how they appear in the Lagrangian and Hamiltonian descriptions of a classical theory  -- which permits us to make such a distinction? 
Put another way, one may ask: what is the {\it mathematical} difference between a spacetime symmetry and a ``gauge'' symmetry? 
 
To answer this question, we refer to the words of Henneaux and Teitelboim \cite{Henneaux-Teitelboim:1992}, who state that:
``A transformation of the variables induced by a change in the arbitrary reference frame is called a gauge transformation. 
Physical variables (``observables'') are then said to be gauge invariant. 
{\bf In a gauge theory, one cannot expect that the equations of motion will determine all the dynamical variables for all times if the initial conditions are given because one can always change the reference frame in the future, say, while keeping the initial conditions fixed. 
A different time evolution will then stem from the same initial conditions. 
Thus, it is a key property of a gauge theory that the general solution of the equations of motion contains arbitrary functions of time} \dots 
{\it Thus, a gauge system is always a constrained Hamiltonian system}. 
The converse, however, is not true. 
Not all conceivable constraints of a Hamiltonian system arise from a gauge invariance.'' 
In this quotation, the italics are original, but the boldface type is ours. 

We quote these passages at length, and verbatim, because we believe them to be the best and most succinct description of what a gauge theory actually is. 
The last two sentences refer to what we have already said -- namely, that in open systems the constraints may be imposed by ``external'' conditions -- whereas the passage in boldface refers directly to the question raised above. 
On our reading of this passage, there is a clear mathematical difference between spacetime symmetries and gauge symmetries. 
Put simply, if a system can be described by means of a regular (non-singular) Lagrangian, then the time-evolution implied by the canonical Euler-Lagrange equations is unique, for a given set of initial conditions. 
That is, there is no ``gauge ambiguity'' that must be fixed, before physical information about the system can be extracted. 
In this case, the symmetry group of the system is not a gauge group. 

We may then ask whether the spacetime symmetry group of the theories we have considered -- namely, the centrally-extended Galilean group  \cite{Dirac-CHS:1958,Dirac-CHS:1964,Henneaux-Teitelboim:1992,Rothe-RotheCHS:2010} -- acts as a gauge group? 
(That is, do Galilean symmetries satisfy the Henneaux-Teitelboim definition of ``gauge symmetries''?) 
The fact that we obtain the same equations of motion, for the canonical displacements of the closed $N$-particle system, by either (a) varying the regular, canonical, and non-Galilean-invariant Lagrangian $L$ (\ref{X.1}), or (b) varying the singular, non-canonical, and Galilean-invariant Lagrangian $L^*$ (\ref{X.2}), subject to the constraints $P_i \approx 0$ (see Sec. \ref{Sec.X}), shows that it does not. 

The central issue is that, although the canonical theory can be recast as a {\it bona fide} gauge theory, according to the Henneaux-Teitelboim criterion \cite{Henneaux-Teitelboim:1992} -- namely, by mapping $L \mapsto L^*$ and setting $P_i \approx 0$, thereby ``gauging'' three of the canonical degrees of freedom --  it does not need to be formulated in this way, in order to impose Galilean invariance. 
This distinguishes Galilean-invariant theories from genuine gauge theories, in which the Lagrangian is automatically singular, and in which one {\it must} impose certain constraints, in order to obtain a self-consistent and physically acceptable description of the system. 
Hence, the regular $L$ is the physically preferred description, because its Euler-Lagrange dynamics is unique for given initial data, and satisfies the required symmetries. 
The ``gauge'' structure of the alternative description, whose Euler-Lagrange dynamics is not unique for given initial data, due to the fact that $L^*$ is invariant under the transformation ${\bf x}_J \mapsto {\bf x}_J + {\bf f}(t)$, and which leaves ${\bf X}(t)$ undetermined, is optional rather than mandatory.

This distinction between gauge symmetries and spacetime symmetries also generalises to the relativistic regime. 
For example, in classical electromagnetism, Amp{\`e}re's Law and Faraday's Law are the equations of motion for the electric and magnetic fields, whereas Gauss's Laws for magnetism and electricity represent constraints, which must hold at all times $t$ \cite{Jackson:EM}. 
Here, one does not have the option of imposing these constraints (or not): failure to do so results in inconsistencies. 
Conversely, no such mandatory constraints arise from the invariance of Maxwell's equations under the Poincar{\' e} group, which is now the relevant spacetime symmetry group. 

Another pertinent example is the non-relativistic free-particle \cite{Henneaux-Teitelboim:1992}. 
The dynamics of this system must, of course, be Galilean-invariant. 
However, if one {\it also} requires the particle world-line to be re-parameterisation-invariant then its dynamics is described by a gauge theory, whose ``gauge freedom'' is associated with the transformation of the affine parameter, $\tau \mapsto \tau'(\tau)$. 
If one does not require this freedom then a regular-Lagrangian description will suffice, for the extraction of all physical data. 
We mention this example because, in our view, it leads to the same conclusion as before: here, there is no way to obtain a re-parameterisation-invariant theory without ``gauging'' the regular Lagrangian. 
But such a procedure is not necessary, in order to impose Galilean-invariance. 
Hence, it is clear that the gauge freedom corresponds to invariance under one-dimensional local diffeomorphisms (re-parameterisations of the world-line), whereas the group of $(3+1)$-dimensional Galilean transformations is the spacetime symmetry group of the theory.
 
This is the reason why, in fact, we have not adopted the terminology of \cite{Vanrietvelde:2018pgb,Vanrietvelde:2018dit}, such as ``gauge degrees of freedom'', ``gauge-invariant Dirac observables'' and ``gauge-fixing'', to refer to Galilean-invariant systems. 
The careful reader will have noticed that, instead, we have paraphrased this terminology in phrases like ``{\it analogous} to a gauge-fixing procedure, which removes redundancy in the description of the system'' (see Abstract), or that we have quoted these terms directly, using quotation marks. 
This is because we do not agree with the use of these terms, to describe Galilean-invariance -- or, following Henneaux and Teitelboim's definition of a {\bf gauge symmetry}, to describe spacetime symmetries, more generally.
\footnote{In the case of general relativity, the gauge group and the spacetime symmetry group may, in fact, be one and the same. This is because the physics represented by an arbitrary solution to the Einstein field equations is invariant under active transformations of the diffeomorphism group \cite{Gaul:1999ys,Lusanna:2019}. This, in turn, implies that canonical GR can be recast as a constrained Hamiltonian system, subject to a topological restriction on the solutions to the field equations: topology $\mathbb{R} \times \Sigma_t$, with $\mathbb{R}$ a global time-like direction. (This is known as the Arnowitt-Deser-Misner, or ADM formulation of GR \cite{MTW:1973}). In this case, the local spacetime symmetry group equals the gauge group, which also equals the covariance group of the theory \cite{Norton:1993eqc}. But this need not concern us. In this critique, we consider only theories with global spacetime symmetries \cite{ErlangenProgram_Klein_1872,ErlangenProgram_EMS_2015,Kisil:2010,ErlangenProgram_Encycolpedia_Springer,Goenner:2015}  -- and, in fact, criticise the unwarranted extrapolations of these models into the regime of gravitational physics.}  

To be clear, we do not deny that both spacetime symmetries {\it and} gauge symmetries can give rise to the existence of unphysical degrees of freedom, in the canonical description of the relevant closed systems. 
Nevertheless, the number of constraints a closed system {\it must} be subject to is determined by its gauge group, {\bf not} its spacetime symmetry group. 
It is not necessary to ``gauge'' the spacetime symmetries, by altering the canonical Lagrangian, in order to identify the physical content of the theory.

Finally, another important distinction is between holonomic and non-holonomic constraints. 
A {\bf holonomic constraint} is one that can be expressed as a function of the configuration space variables, whereas a {\bf nonholonomic constraint} cannot be expressed in this way \cite{Neimark-Fufaev:1972,Cushman:2010}. 
Constraints that can be expressed as functions of both the configuration space variables and their generalised velocities, but which remain integrable -- that is, reducible to holonomic form -- are called {\bf semi-holonomic}. 
For the systems considered in this paper, all relevant constraints can be expressed as semi-holonomic constraints, {\it if} one makes use of the equations of motion, to rewrite the momenta as time derivatives of the position variables. 
Specifically, we have 
\begin{eqnarray} \label{semi-holonomic}
P_i \approx 0 \implies \dot{X}^{i} \approx 0 \implies X^{i} \approx {\rm const.}
\end{eqnarray}
However, since the $P_i \approx 0$ represent primary constraints, this step is not necessary, as the equations of motion are not required in order for them to hold \cite{Dirac-CHS:1958,Dirac-CHS:1964,Henneaux-Teitelboim:1992,Rothe-RotheCHS:2010}. 

Nonetheless, it is interesting to consider the Dirac-quantisation analogue of (\ref{semi-holonomic}), which then reads
\begin{eqnarray} \label{semi-holonomic*}
\widehat{P}_i \approx 0 \implies \widehat{\dot{X}}^{i} \approx 0 \implies \widehat{X}^{i} \approx {\rm const.} \, ,
\end{eqnarray}
or, equivalently, 
\begin{eqnarray} \label{semi-holonomic**}
\widehat{P}_i|\Psi\rangle^{\rm phys} \overset{!}{=} 0 \implies \widehat{\dot{X}}^{i}|\Psi\rangle^{\rm phys} \overset{!}{=}  0 
\implies \widehat{X}^{i}|\Psi\rangle^{\rm phys} \overset{!}{=}  {\rm const.} \times |\Psi\rangle^{\rm phys} \, , 
\end{eqnarray}
using the notation of \cite{Vanrietvelde:2018pgb,Vanrietvelde:2018dit}. 
This is clearly incompatible with the requirement that $[\widehat{X}^i,\widehat{P}_j] = i\hbar\delta^{i}{}_{j}\widehat{\mathbb{I}}$, which gives rise to the canonical uncertainty relation $\Delta_{\Psi}X^i\Delta_{\Psi}P_j \geq \hbar/2$. 
We comment on the relevance of this result, for the Dirac quantisation procedure advocated in the perspective-neutral formalism \cite{Vanrietvelde:2018pgb,Vanrietvelde:2018dit}, in App. \ref{App.C.4}.

\subsection{Canonical quantisation} \label{App.C.2}

For {\bf unconstrained systems}, {\bf canonical quantisation} is implemented by a map from the canonical {\bf Poisson bracket} to the canonical commutator \cite{Rae:2002,Isham:1995,Dirac:1958,vonNeumann:1955}. 
However, for {\bf constrained systems}, the {\bf Dirac bracket} must be mapped to the commutator instead:
\begin{eqnarray} \label{C.0}
\left\{O_1,O_2\right\}_{\rm DB} \mapsto \frac{1}{i\hbar}[\widehat{O}_1,\widehat{O}_2] \, , 
\, {\rm where} \, \left\{O_1,O_2\right\}_{\rm DB} = \lim_{\hbar \rightarrow 0} \frac{1}{i\hbar}[\widehat{O}_1,\widehat{O}_2] \, .
\end{eqnarray}

For systems with only first-class constraints, the Dirac bracket between any two functions of the canonical phase space variables, $F({\bf x},{\bf p})$ and $G({\bf x},{\bf p})$, always coincides with the Poisson bracket. 
For systems with second-class constraints, the Dirac bracket is given by \cite{Dirac-CHS:1958,Dirac-CHS:1964,Henneaux-Teitelboim:1992,Rothe-RotheCHS:2010}
\begin{eqnarray} \label{C.1}
\left\{F,G\right\}_{\rm DB} := \left\{F,G\right\}_{\rm PB} - \sum_{a}\left\{F,\phi_a\right\}_{\rm PB} M_{ab}^{-1} \sum_{b}\left\{\phi_b,G\right\}_{\rm PB} \, , 
\end{eqnarray}
where
\begin{eqnarray} \label{C.2}
M_{ab} := \left\{\phi_a,\phi_b\right\}_{\rm PB} 
\end{eqnarray}
is the {\bf constraint matrix}, with  
\begin{eqnarray} \label{C.3}
\phi_a \approx {\rm const} \, .
\end{eqnarray}
Here, the $\phi_a$'s also denote functions of the canonical phase space variables, $\left\{{\bf x}_J,{\bf p}_J\right\}_{J=1}^{N}$, and the {\bf weak equality} ``$\approx$'' indicates that the condition $\phi_a({\bf x},{\bf p}) = {\rm const.}$ cannot be implemented, until all the relevant Poisson brackets have been formed \cite{Dirac-CHS:1958,Dirac-CHS:1964,Henneaux-Teitelboim:1992,Rothe-RotheCHS:2010}. 

For our purposes, this is all we need to know. 
We do not attempt to outline the derivation of the expression (\ref{C.1}), or to explain how, or why, the `naive' Poisson bracket quantisation fails, when the system is constrained.
The reader is referred to \cite{Dirac-CHS:1958,Dirac-CHS:1964,Henneaux-Teitelboim:1992,Rothe-RotheCHS:2010}, and references therein, for further details. 

\subsection{Reduced quantisation} \label{App.C.3}

{\bf Reduced quantisation} can be applied both to gauge systems, in which $N_{\rm phys} \neq \frac{1}{2}N_{\rm var}$, $N_{\rm first-class} > 0$, and to non-gauge systems, with $N_{\rm phys} \neq \frac{1}{2}N_{\rm var}$, $N_{\rm first-class} = 0$, in which spacetime symmetries render some of the canonical degrees of freedom {\bf unphysical} at the classical level. 
In this procedure, the classically-unphysical variables are simply removed, or set equal to zero, at the level of the classical Hamiltonian \cite{Giacomini:2017zju,Castro-Ruiz:2019nnl,delaHamette:2020dyi,Krumm:2020fws,Giacomini:2021gei,Castro-Ruiz:2021vnq,Vanrietvelde:2018pgb,Hohn:2018iwn,Hohn:2018toe,delaHamette:2021oex,Hoehn:2019fsy,AliAhmad:2021adn,Vanrietvelde:2018dit,Hoehn:2023ehz,Hoehn:2023axh,DeVuyst:2025ezt}. 
The remaining phase space variables are then promoted to quantum operators. 

In the case of genuine {\bf gauge theories} \cite{GaugeTheory:Healy2007}, this does not represent a systematic treatment of the constraints, which is given, instead, by canonical quantisation \cite{Dirac-CHS:1958,Dirac-CHS:1964,Henneaux-Teitelboim:1992,Rothe-RotheCHS:2010}. 
For {\bf non-gauge theories}, simple reduction of the classical Hamiltonian is equivalent to removing any unphysical degrees of freedom that arise in the canonical description, due to spacetime symmetries. 
It is important to understand, however, that according to the formal theory of constrained Hamiltonian systems, such theories are {\bf not constrained}; only when a theory possesses genuine gauge symmetries, which meet the Henneaux-Teitelboim criteria \cite{Henneaux-Teitelboim:1992}, quoted in App. \ref{App.C.1}, is the imposition of the constraints {\bf mandatory}, in order to maintain the invariance of the equations of motion, under the relevant group \cite{Dirac-CHS:1958,Dirac-CHS:1964,Henneaux-Teitelboim:1992,Rothe-RotheCHS:2010}.

The most pertinent example, for our present work, of a system in which spacetime symmetries imply the existence of classically-unphysical degrees of freedom, but in which the equations of motion remain fully ``'invariant'', with or without the imposition of constraints, is the closed Galilean system. 
For this system, imposing the conditions $P_i \approx 0$ removes the unphysical degrees of freedom associated with the classical centre of mass, and is equivalent to performing the map $H = P^2/(2M) + U \mapsto U$. 
Nevertheless, in the canonical description, the closed Galilean-invariant system is not a constrained Hamiltonian system, and does not possess gauge symmetries, according to the formal meanings of these terms \cite{Henneaux-Teitelboim:1992}. 
The ``invariance'' referred to above means invariance under the spacetime symmetry group (i.e., the centrally-extended Galilean group \cite{Bargmann:1954gh,Levy-Leblond:1963qdx,Levy-Leblond:1967eic,Horzela:1991pa,Giulini:1995te,Csillag:2003}), which is not a gauge group. 

For our purposes, it is important to understand that classically-unphysical degrees of freedom can `survive' the canonical quantisation process, {\it if} one chooses not to implement the constraints that kill them at the classical level \cite{Rae:2002,Isham:1995,Dirac:1958,vonNeumann:1955}. 
In principle, removing these degrees of freedom by imposing the constraints {\it could} still give the correct results in the quantum regime,  if one is lucky, but it cannot give the correct results in general. 
More specifically, the correct results can be obtained if and only if one is interested in a specific state of some system, which corresponds to the zero eigenvalue(s) of the operator(s) one removes. 

This means that it does not make physical sense to remove operators with continuous eigenspectra -- unless one is prepared to change the canonical norm, and to invoke a superselection rule for the relevant continuous quantity. 
This is precisely the kind of removal, and alteration of the canonical theory, that the perspective-neutral formalism advocates, in order to impose Galilean invariance \cite{Vanrietvelde:2018pgb,Vanrietvelde:2018dit}. 
Unfortunately, the resulting theory actually violates the requirements of Galilean symmetry. 
(See Sec. \ref{Sec.X}.)

\subsection{Dirac quantisation} \label{App.C.4}

In {\bf Dirac quantisation}, the classical constraints are promoted to operators, and are imposed directly in the quantum regime \cite{Kunstatter:1991ds,Schleich:1990gd,Plyushchay:1994pk,Barvinsky:1996cg,Shimizu:1996vf}. 
Specifically, one implements the map
\begin{eqnarray} \label{C.4}
\phi_a({\bf x},{\bf p}) \approx 0 \mapsto \widehat{\phi}_a({\bf x},{\bf p}) \approx 0 \, , 
\end{eqnarray}
where the expression on the right-hand side of this equation is interpreted as restriction that defines a {\bf reduced Hilbert space}, spanned by the {\bf physical states} of the theory:
\begin{eqnarray} \label{C.5}
\widehat{\phi}_a({\bf x},{\bf p})|\Psi\rangle^{\rm phys} \overset{!}{=} 0 \, . 
\end{eqnarray}
Here, the notation ``$\overset{!}{=}$'', which is defined in \cite{Vanrietvelde:2018pgb,Vanrietvelde:2018dit}, indicates that $|\Psi\rangle^{\rm phys}$ represents a ``physical state'', if and only if it satisfies the quantised constraint equations. 

This procedure clearly differs from reduced quantisation, but it also differs from canonical quantisation \cite{Ita:2021cak,BarroseSa:2023rvz,Juhasz:2024twu}, in a subtle way. 
In the latter one may find that, after promoting the classical constraint functions to quantum operators,
\begin{eqnarray} \label{C.4}
\phi_a({\bf x},{\bf p}) \mapsto \widehat{\phi}_a({\bf x},{\bf p}) \, , 
\end{eqnarray}
they no longer function {\it as constraints} on the system \cite{Bargmann:1954gh,Levy-Leblond:1963qdx,Levy-Leblond:1967eic,Horzela:1991pa,Giulini:1995te,Csillag:2003}.
(Note that the Dirac bracket quantisation, (\ref{C.0}), in no way precludes the possibility of quantising the functions $\phi_a({\bf x},{\bf p})$.) 
In other words, just because we may set $\phi_a({\bf x},{\bf p}) \approx {\rm const}$, does not mean that we may set $\widehat{\phi}_a({\bf x},{\bf p}) \approx {\rm const}$ \cite{Dirac-CHS:1958,Dirac-CHS:1964,Henneaux-Teitelboim:1992,Rothe-RotheCHS:2010}. 
In general, a classical function with a specific constant value -- say, zero -- will not map to a quantum operator with the same specific value, or even any specific value at all. 
Instead, $\widehat{\phi}_a({\bf x},{\bf p})$ will be {\bf time-independent}, but with a standard spectral decomposition \cite{Rae:2002,Isham:1995,Dirac:1958,vonNeumann:1955}.

As an example, consider again the case of a closed $N$-particle system in the Newtonian spacetime \cite{SEP:Newtonian_Space-time}. 
Here, the relevant symmetry group is the centrally-extended Galilean group \cite{Bargmann:1954gh,Levy-Leblond:1963qdx,Levy-Leblond:1967eic,Horzela:1991pa,Giulini:1995te}. 
The classical (Poisson bracket) representation of its Lie algebra is compatible with the implementation of three primary first-class constraints, and imposing $P_i \approx 0$ ensures that the external inertial frame $O$ comoves with the centre of mass. 
In the classical regime, these constraints are well-defined, and unambiguous, because the trajectory of the centre of mass also defines a classical inertial frame. 
Like the centre of $O$'s frame, the centre-of-mass frame is sharply defined in the classical phase space -- it is not `smeared' over a finite region, in either real space or momentum space, due to canonical quantum uncertainty. 

However, under canonical quantisation, in which we choose {\it not} to impose $P_i \approx 0$ at the classical level, we have $P_i \mapsto \widehat{P}_i$ and $\left\{P_i,P_j\right\}_{\rm DB} = \left\{P_i,P_j\right\}_{\rm PB} = 0 \mapsto (i\hbar)^{-1}[\widehat{P}_i,\widehat{P}_j] = 0$, where $\widehat{P}_i$ is a normal time-independent operator ($\widehat{\dot{P}}_i = 0$) \cite{Rae:2002,Isham:1995,Dirac:1958,vonNeumann:1955,Dirac-CHS:1958,Dirac-CHS:1964,Henneaux-Teitelboim:1992,Rothe-RotheCHS:2010}. 
It does not, therefore, make sense to impose $\widehat{P}_i = 0$, at {\bf any} stage, since this would entail a loss of physical information. 
(For example, regarding the Galilean-invariant momentum operator, $\widehat{\boldsymbol{\pi}}_{J|I} := \widehat{{\bf p}}_J - (m_J/M)\widehat{{\bf P}}$, which is canonically conjugate to the relative displacement, $\widehat{{\bf x}}_{J|I} := \widehat{{\bf x}}_J - \widehat{{\bf x}}_I$. See App. \ref{App.B.2}.) 
The physical issue is that the quantised centre-of-mass coordinate, $\widehat{\bf X}(t)$, can now exist in a superposition of states, relative to $O$'s sharp frame. 
What has `gone wrong' with the process of Dirac quantisation, here, is that it requires us to impose a {\bf superselection rule} (SSR), for the net momentum of a closed quantum system, whereas no such rule exists in canonical quantum mechanics \cite{SSR:QM-Compendium,Giulini:2007fn}.
\footnote{Strictly, the requirement that $\widehat{P}_i|\Psi\rangle^{\rm phys} \overset{!}{=} 0$, which limits ``physical states'' to only a single $P_i$-eigenvalue (zero), is a stronger condition than the imposition of an SSR. The latter is implied by the former, but the converse is not true.} 

In the context of the perspective-neutral formalism \cite{Vanrietvelde:2018pgb}, this leads to immediate contradictions. 
For example, it requires that $\Delta_{\Psi}P_i = 0$, which is compatible with the assertion that $[\widehat{x}_A^i,\widehat{P}_j] = 0$ -- and, hence, that $\Delta_{\Psi}x_A^i\Delta_{\Psi}P_j = 0$ -- for the chosen ``reference particle'' $A$. 
But the same condition is {\bf incompatible} with the requirement that $[\widehat{X}^i,\widehat{P}_j] = i\hbar\left(1 - \frac{m_A}{M}\right)\delta^{i}{}_{j}\widehat{\mathbb{I}}$ -- and, hence, that $\Delta_{\Psi}X^i\Delta_{\Psi}P_j \geq \frac{\hbar}{2}\left(1 - \frac{m_A}{M}\right)$ -- where, here, $X^i$ and $\widehat{P}_i$ denote the position and momentum of the centre of mass, relative to particle $A$. 
Nevertheless, {\bf both} these conditions are required to hold in the perspective-neutral formalism \cite{Vanrietvelde:2018pgb,Vanrietvelde:2018dit}.
\footnote{The relations $[\widehat{x}_A^i,\widehat{P}_j] = 0$ and $[\widehat{X}^i,\widehat{P}_j] = i\hbar\left(1 - \frac{m_A}{M}\right)\delta^{i}{}_{j}\widehat{\mathbb{I}}$ follow from the three-dimensional generalisation of Eqs. (12) in \cite{Vanrietvelde:2018pgb}, with arbitrary particle masses in place of the unit masses considered therein. Strictly, Eqs. (12) give the classical Dirac brackets used in the perspective-neutral formalism -- which, we argued in Sec. \ref{Sec.3.2}, are wrongly constructed. Upon quantisation, these yield the commutators $[\widehat{x}_A^i,\widehat{p}_{Aj}] = 0$ and $[\widehat{x}_I^i,\widehat{p}_{Jj}] = i\hbar\delta_{IJ}\delta^{i}{}_{j}\widehat{\mathbb{I}}$, for $I,J \neq A$, in our notation.}


In summary, the imposition of a superselection rule for the net momentum of a closed system is not justified, by invariance under the centrally-extended Galilean group \cite{Bargmann:1954gh,Levy-Leblond:1963qdx,Levy-Leblond:1967eic,Horzela:1991pa,Giulini:1995te}, and, moreover, leads to internal contradictions within the framework proposed in \cite{Vanrietvelde:2018pgb,Vanrietvelde:2018dit}.


\end{document}